# ECONOMICS AND HUMAN DIMENSION OF ACTIVE MANAGEMENT OF FOREST-GRASSLAND ECOTONE IN SOUTH-CENTRAL USA UNDER CHANGING CLIMATE

By

BIJESH MISHRA

Master of Science in Environmental Studies
Kentucky State University
Frankfort, KY
2017

B. Sc. (Agriculture)
Tribhuvan University
Kathmandu, Nepal
2013

Submitted to the Faculty of the
Graduate College of the
Oklahoma State University
in partial fulfillment of
the requirements for
the Degree of
DOCTOR OF PHILOSOPHY
July, 2022

ECONOMICS AND HUMAN DIMENSION OF

ACTIVE MANAGEMENT OF FOREST-GRASSLAND

ECOTONE IN SOUTH-CENTRAL USA UNDER

CHANGING CLIMATE

Dissertation Approved:

Omkar Joshi, Ph.D.

Dissertation Adviser

Rodney E. Will, Ph.D.

Binod Chapagain, Ph.D.

Lixia He Lambert, Ph.D.

ACKNOWLEDGEMENTS

I would like to acknowledge my parents Bishnu Prasad Mishra and Juna Bhattarai, spouse Nisha Poudel, sister Deepa Mishra, and other family members for their continuous motivation, support, and believing on my almost three decades of academic journey.

I acknowledge my major advisor Dr. Omkar Joshi, committee members Dr. Rodney E. Will (PI), Dr. Binod Chapagain, and Dr. Lixia Lambert for their guidance, and support during this rigorous journey of accomplishing a terminal degree.

I further acknowledge USDA NIFA Foundational Knowledge of Agricultural Production Systems, Grant No: 2018-67014-27504 for providing funding support and the Department of Natural Resource Ecology and Management at Oklahoma State University for providing the opportunity to pursue this academic journey.

Finally, I would like to remember friends, lab mates, departmental fellows, faculties, and staffs who supported me emotionally to cope with the COVID-19 pandemic seeing family and friends being infected from COVID-19 yet, unable to be with them physically.

Name: BIJESH MISHRA

Date of Degree: JULY, 2022

Title of Study: ECONOMICS AND HUMAN DIMENSION OF ACTIVE

MANAGEMENT OF FOREST-GRASSLAND ECOTONE IN SOUTH-CENTRAL USA UNDER CHANGING CLIMATE

Major Field: NATURAL RESOURCE ECOLOGY AND MANAGEMENT

Abstract:

The south-central ecotone of the USA, characterized by a mix of forest, savanna, and grasslands, previously maintained by fire, is changing towards closed-canopy forest due to exclusion of fire. The management is complicated by the encroachment of species such as eastern redcedar (*Juniperus virginiana)* and changing climate such as drought and more sporadic rainfall. Active management using prescribed fire and thinning can restore the ecosystem services, generate revenue, and motivate landowners to manage their land, but the cost is a major barrier. However, economic benefit from actively managed ecosystem under changing climate is not well known in this region. I estimated benefits of various ecosystems maintained using prescribed fire and thinning under changing climate scenarios. I further studied the willingness to pay (WTP), and intentions of landowners to actively manage their land for deer habitat management. I found that sawlog generally increased with the increase in rainfall, but pulpwood growth varied by management regime. The change in net future value (NFV) from timber was relatively stable in non-burned stands compared to frequently burned stands with increasing rainfall. The change in timber volume was greater in harvested and thinned stands. Stands burned in two- and three-year intervals the supported the greatest number of cattle and deer. The willingness to pay for deer hunting was higher in hunting sites with opportunities to observe more deer per visit. The WTP to observe 10 and 6 deers per visit instead of 1 deer per visit are about $11 and $9, respectively. The WTP for deer hunting was higher for deer habitats with food plots and deer sanctuaries. Forest canopy cover had non-significant impact on WTP for deer hunting, providing flexibility for landowners to change canopy and manage land for multiple objectives such as hunting, wildlife management, cattle grazing, and timber production. Landowners had positive intentions, social pressure, but the negative attitude toward actively managing their land. Financial burden, potential reasons for negative attitude, can be offset by revenue generated by managing land for multiple objectives.

TABLE OF CONTENTS

Chapter Page

LIST OF TABLES

LIST OF FIGURES

# CHAPTER I: INTRODUCTION

The south-central transitional ecoregion of the USA contains a mix of forest, grassland, and savanna from southern Illinois to northern Texas (Hallgren et al., 2012; Joshi et al., 2019b). This is a dynamic region positioned between evergreen and dense forest on the east and Great Plains on the west. The transitional and dynamic nature of the ecoregion was historically maintained by frequent fire (Stambaugh et al., 2014). The fire was largely excluded in the early 20$^{th}$ century, which resulted in the encroachment of fire sensitive species such as eastern redcedar (ERC, *Juniperus virginiana*) (Kaur et al., 2020) and mesic hardwood (Joshi et al., 2019b). The ecoregion is further affected by prolonged drought (Tian et al., 2018), erratic rainfall (Otkin et al., 2019), and unpredictable wildfire (Clark et al., 2007; Hallgren et al., 2012). The prolonged drought coupled with fuel load from fire-intolerant ERC increase the risk of crown fire (Hoff et al., 2018a).

Active management tools such prescribed burning and thinning are important in restoring characteristic ecosystem in this region (Feltrin et al., 2016), reducing the fuel load (Starr et al., 2019), and maintaining its transitional nature. Active management can enhance resiliency against changing climate and the adaptive capacity to mitigate the negative effects of climate change (Campbell and Ager, 2013; Loudermilk et al., 2017). These management practices, however, bring additional financial burdens (Starr et al., 2019) which could be compensated by the economic return from actively managed ecosystem. The compensation may be obtained in different forms such as hunting leases (Martínez-Jauregui et al., 2016), timber (Brodrechtova, 2015), wildlife and livestock management (Loomis et al., 1989), and recreational benefits (Bertram and Larondelle, 2017).

The adoption of active management tools such as prescribed fire depends upon intention of landowners to implement these tools on their property (Elmore et al., 2010; Harr et al., 2014; Joshi et al., 2019a; Morton et al., 2010). Despite proven benefits, limited inquiry has been conducted in this region to understand economic returns from actively managed ecosystem under changing climate. To this end,  Elmore et al. (2010) found that farmers and the public of Oklahoma are supportive of using prescribed fire in their land. Starr et al. (2019) identified financial burden and liabilities as potential barriers for the adoption of active management tools. Joshi et al. (2019a) identified sociodemographic factors affecting active management in this region. The economic returns from actively managed ecosystem, which is considered as important motivation for landowners to active manage their land (Starr et al., 2019), has yet to be recognized in this region. Despite positive support for the adoption of prescribed fire among farmers and public (Elmore et al., 2010), intentions of forest and rangeland landowners to actively manage their lands using these tools is not explored yet.

This research aimed to fulfill the knowledge gap by studying how active management of ecosystem may affect its productivity and economic return under changing climate in the south-central transitional ecoregion of the USA. This aim was fulfilled by answering three specific research questions, which were articulated as three objectives of the research. The first objective of this research was to estimate economic valuation of actively managed forest, savanna, and grassland using prescribed fire and thinning for timber, cattle, and deer management under varying rainfall patterns. The second objective was to study how different habitat characteristics impact willingness to pay (WTP) to lease a deer habitat using the best worse choice modeling approach. The third objective of this research was to study intentions of landowners to actively manage their lands for deer habitat. This research was framed within the premises of two prominent social psychological theories, the theory of reasoned action (TRA) and the theory of planned behavior (TPB), which were used to study landowners' intentions towards active management. This research further expanded these theories using moral norms.

This research is divided into five chapters. The first chapter introduces the relevance of research. Three subsequent chapters answer three objectives in the form of three independent but interrelated manuscripts. The fifth chapter concludes and summarizes this research.

# CHAPTER II: QUANTIFYING ECOSYSTEM BENEFITS OF ACTIVE MANAGEMENT IN SOUTH-CENTRAL TRANSITIONAL ECOREGION, USA UNDER CHANGING CLIMATE

## Abstract

Due largely to fire exclusion, the south-central ecoregion of the USA is transitioning from a mosaic grassland, savanna, and forest ecosystem towards closed-canopy forest, and losing important ecosystem benefits. Prescribed fire and thinning are vital in restoring woodlands, savannas, and grasslands. The resulting benefits from restoration may result in a more sustainable and resilient system in the face of changing climate. Managing the ecosystem for combinations of deer habitat, cattle grazing, and timber production can offset the cost of management, which is a major barrier to active management in this region. However, the economic return of actively managed forest, grassland, and savanna managed for multiple objectives is unknown in this region. We conducted benefit-cost analysis from actively managed forest, savanna, and grassland for 40 years using prescribed fire and thinning for deer habitat management, cattle grazing, and timber production. The data came from a long-term study in southeastern Oklahoma that used pine harvesting, hardwood thinning, and prescribed fire at various return intervals to produce different ecosystems ranging from closed-canopy shortleaf pine (*Pinus echinata*) – post oak (*Quercus stellata*) forest to savanna. This research further quantified the effect of changing rainfall on timber production and total economic return. This research found that increase in the sawlog weight (92.4 ton $ha^{-1}$) and the pulpwood weight (98.4 ton $ha^{-1}$) due to increase in rainfall was highest in closed canopy forest either not burned or burned every four years. Sawlog weight increased with an increase in rainfall in all treatments. Sawlog grew by 10.11% with 10% to increasing rainfall in the harvested pine and thinned hardwood treatment which is most responsive to change in rainfall. Pulpwood weight growth showed mixed responses to changes in rainfall among treatments. The land area required is lowest for deer (1.85 ha $year^{-1}$ $animal^{-1}$) and cattle (1.85 ha $year^{-1}$ $animal^{-1}$) in savanna ecosystems burned every two or three years. The net present value ($8,367 $ha^{-1}$), annual equivalent income ($460), and benefit-cost ratios (10.06) were the highest in non-burned forest stands primarily due to timber value. isThis research results could help landowners in the south-central US to make an informed decision to optimize their economic returns for deer habitat management, cattle grazing, and timber production based on their specific management objectives.

## 1. Introduction

The south-central transitional ecoregion is a dynamic region consisting of tallgrass prairie, savanna, and upland forest, sandwiched between the grasslands of Great Plains and eastern forest, occurring from Southern Illinois to Northern Texas (Hallgren et al., 2012; Hoff et al., 2018a; Joshi et al., 2019b). Historically, the transitional nature of the region was maintained by frequent burning (Hoff et al., 2018a). Fire was largely excluded following European settlement which led to the growth of fire-intolerant, mesic hardwood forest, and increased encroachment of species such as eastern redcedar (ERC, *Juniperus virginiana*) (Hoff et al., 2018a; Kaur et al., 2020). The ecoregion is further affected by prolonged droughts (Tian et al., 2018), erratic rainfall (Otkin et al., 2019), and unpredictable wildfire (Clark et al., 2007; Hallgren et al., 2012). These encroachment of redcedar further worsen negative consequences of drought by depleting the soil moisture due reduced surface runoff and potential deep water recharge (Zou et al., 2018) thus increasing risk of wildfire  (Hoff et al., 2018a).

Prescribed burning and thinning are crucial in restoring the restoring and maintaining  open woodlands, savannas, and grasslands which provide increasing the sustainability and resilience in the face of future climate change. For example, management using prescribed fire at a return interval of three years or less can maintain savannas while longer fire intervals or no fire result in uneven-aged or even-aged forests respectively (Adhikari et al., 2021a; Feltrin et al., 2016; Masters et al., 2006) improving yield. The evapotranspiration is higher for forested land compared to mixed cover and non-forested land (Zhang et al., 2001). Changing cover from forested to grassland reduce water losses (Cardella Dammeyer et al., 2016; Zhang et al., 2001) thus, increasing the sustainability and resilience in the face of future climate change. Prescribed fire and thinning further reduces wildfire risk by reducing fuel load (Starr et al., 2019), improve timber, reduce ERC encroachment, and improve wildlife habitat (Harper and Johnson, 2008). Active management using prescribed fire and thinning is, however, limited by economic and financial burdens and liabilities associated with prescribed fire to landowners (Starr et al., 2019). Previous research in this region suggests that healthy and resilient forests provide an opportunity to increase revenue, which in turn drives active management (Joshi et al., 2019b; Starr et al., 2019).

The ecosystem benefits from the forest, savanna, and grassland in this region are vital for reducing the financial and economic burden to landowners. The inherent economic incentive is an important motivation for landowners to manage and open their land for recreational hunting (Mozumder et al., 2007). The economic return from actively managed ecosystems, however, are unrealized in this region. This research, thus, quantifies the economic return of the forest, grassland, and savanna resulting from different combinations of prescribed fire return interval, hardwood thinning, and pine harvesting. This research quantified timber, cattle forage, and deer forage values from the forest, savanna, and grassland ecosystems in the experimental research plots at Pushmataha Wildlife Management Area in South-eastern Oklahoma over the nearly four decades. This research further evaluated the economic feasibility of ecosystem management based on financial indicators including net present value (NPV), land expectation value (LEV), and equivalent annual income (EAI). We further simulated change in rainfall, its impact on timber production, and economic returns from the forest, savanna, and grassland of the south-central ecoregion of the USA.

Timber production, deer habitat management, and cattle management are important rangeland and forest management objectives in the USA (Hines et al., 2021). South-central transitional region forests are dominated by shortleaf pine (*Pinus echinata*), hickory (*Carya* spp.), and oak (*Quercus* spp.) (Adhikari et al., 2021a; Johnson et al., 2010). In Oklahoma, forest covers about 23% of the landscape, is largely owned by private landowners (Johnson et al., 2010). Forest sector contributed more than 5 billion in industry output directly and indirectly generating more than 19,300 jobs and 1.20 jobs in other sector with 1 addition job creation in forestry sector in Oklahoma in 2016 (Starr et al., 2018). In Oklahoma, the forestry and logging industry accounts about 13% of total employment, generated $575 million in payroll, and the state received $1.4 billion directly through payroll, employee compensation, and property tax (Gore et al., 2022). Forest area and harvested timber volumes are growing in this region with the socio-economic development and increase in demand for forest products (Harper and Johnson, 2008). Besides producing timber and forest vegetation protects deer (*Odocoileus virginiana*) from extreme weather and predators (Hines et al., 2021).

Deer hunting is an important wildlife management tool (Arnett and Southwick, 2015; Hines et al., 2021), an avenue for social interaction, and holds cultural and traditional values in the southern United States (Arnett and Southwick, 2015). Annually, 13.7 million deer hunters contribute $38.3 billion to the US economy, create several employment opportunities, and contribute to taxes and revenues (Arnett and Southwick, 2015). Hunters spend more than one billion dollars in Texas and Oklahoma in 2011 (Poudel et al., 2016).

Texas (15.3%), Oklahoma (3.4%), Kansas (1.7%), and Missouri (1.1%), four states lying in transitional ecoregion of the USA, have estimated 21.5% of private and federal rangeland and pasture grazed by cattle (Maher et al., 2021). Oklahoma contributes estimated $722.4 million in total annual ecosystem service from cattle grazing grazed in 43,837 acres of federal and 15.6 million acres of private rangeland (Maher et al., 2020). While the number of beef cattle increased by 40% in 2017 from 2012 in Oklahoma (Maher et al., 2020), cattle and deer often co-occur in the North American rangelands (Hines et al., 2021). Rangelands provide forage for cattle grazing and habitat for wildlife such as deer (Maher et al., 2021). Cattle and deer are compatible when the stocking rate of cattle is maintained below 0.12 - 0.17 animal units per year per hectare (AUY $ha^{-1}$) (Hines et al., 2021). They often do not compete with each other because cattle graze on grasses and deer browse on leaves and woody plants (Hintze et al., 2021). Integrating grazers and browsers often improves carrying capacity (Hintze et al., 2021; Walker, 1994), increases ecosystem stability, and helps with uniform utilization of vegetation (Walker, 1994).

## 2. Methods

### 2.1 Study Site

The study was conducted on about 53-hectare (ha) (130 acres) forest habitat research area (FHRA) located in Pushmataha Wildlife Management Area (WMA) in southeastern Oklahoma. The research area was initiated in 1982 and subsequently developed as a long-term ecological research site (Adhikari et al., 2021a; Masters et al., 2006). The overstory canopy of the research site is dominated by post oak (*Quercus* *stellata*), hickory *spp.*), shortleaf pine (*Pinus echinate)*, and blackjack oak (*Q. marilandica*) (Adhikari et al., 2021a). The understory vegetation is composed of shrubs, grass, forbs, legumes, panicum (*Dichanthelium spp., Panicum spp.*), and sedges (*Carex* spp.) (Adhikari et al., 2022). Some of the common understory plants are grapes (*Vitis spp.*), greenbriers (*Smilax* spp.), winged Sumac (*Rhus copallinum*), little bluestem (*Schizachyrium scoparium*), and poison ivy (*Toxicodendron radicans*).

The research site was located in the semi-arid region of Oklahoma (Adhikari et al., 2021a; Feltrin et al., 2016) with hot summers and moderate winters (Adhikari et al., 2021a). This region experienced prolonged droughts from 2003 to 2006, and from 2009 to 2013 (Adhikari et al., 2021a). South-central USA experienced an unusual sequence of extreme drought in the summer of 2015 followed by heavy rainfall in October, eliminating the extreme drought within two weeks (Otkin et al., 2019). The study site received the lowest average annual precipitation in August, which is also the hottest month of the year leading to late summer water stress (Adhikari et al., 2021a). This unpredictable and largely varying precipitation and temperature make the study region vulnerable to changing climate. The growing season rainfall (April to September) in millimeter (mm) from 1986 to 2017 is plotted in Figure 1.

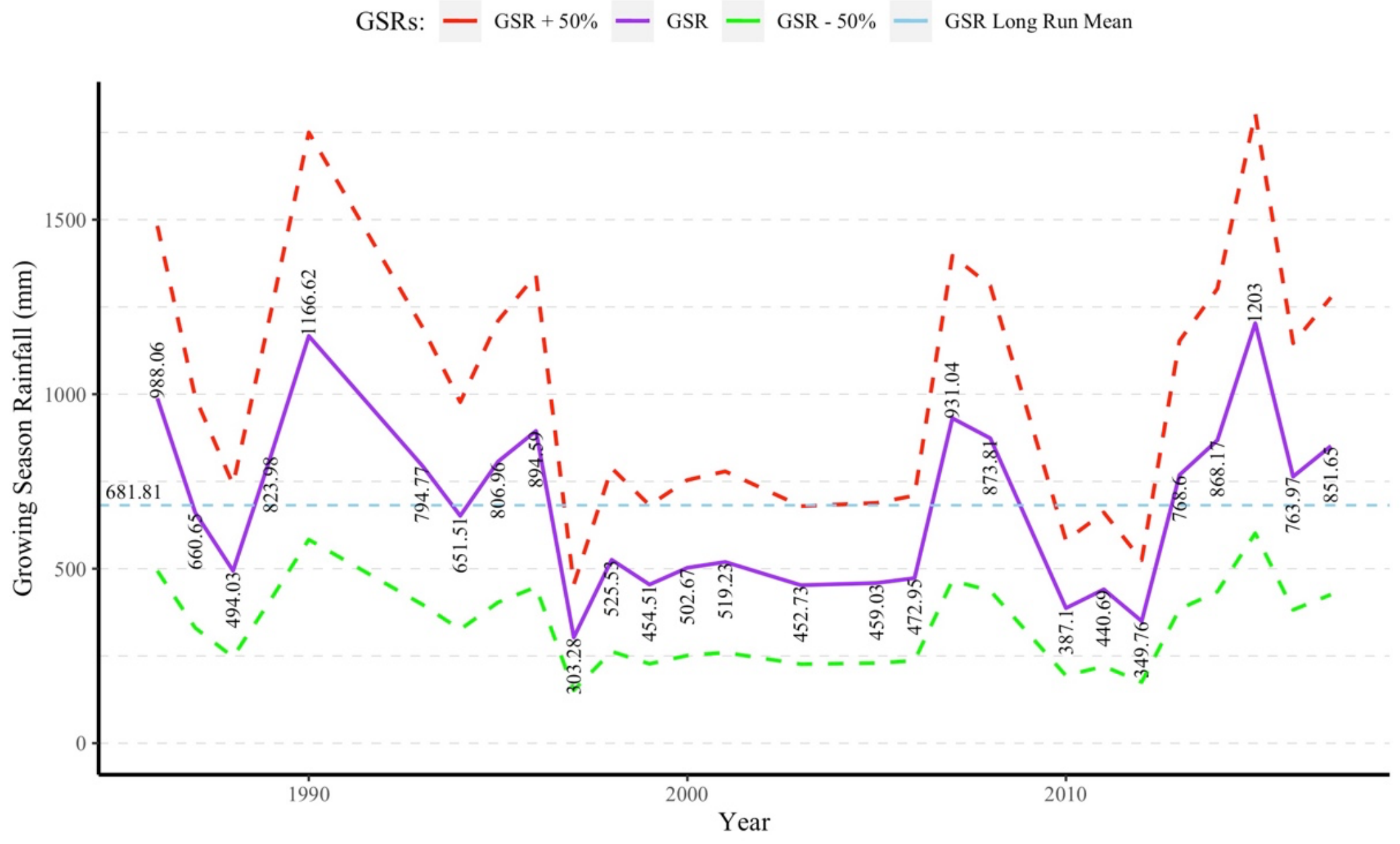

Figure 1: Growing seasons rainfall (GSR) from 1986 to 2017, and simulation range of +50% and -50%.

2.2 Treatments

The plots were burned in the late dormant season at an interval of 0 to 4 years starting in 1985 (Adhikari et al., 2021a) following the harvesting (H) of shortleaf pine greater than 11.4 cm diameter at breast height (DBH = 1.4 m) meter and thinning (T) of hardwoods such as hickory and post oak in 1984 to a basal area of approximately 9 $m^2$ $ha^{-1}$ using single stem injection of herbicide (Masters et al., 2006). Control stands were untreated, mature closed-canopied stands consisting of shortleaf pine, loblolly pine (*P. taeda*), post oak, elm, hickory, black gum (*Nyssa sylvatica*) and various other few unidentified hardwood. The major disturbance was observed in the 1930s (Adhikari et al., 2021a). HT stands were harvested and thinned but never burned. The HT2, HT3, and HT4 stands were harvested, thinned in 1984, and burned at 2, 3, and 4 years intervals from 1985 (Masters et al., 2006). Research plots were established using a random experiment design with 3 replications of Control, HT, HT2, and HT4, and 2 replications of HT3 (Adhikari et al., 2021a; Masters et al., 2006). Thinned and harvested stands were characterized as grassland and savanna in 1987. By 2018, Control, HT, and HT4 transformed into a forest; HT2 and HT3 remained as woodland and savanna (Adhikari et al., 2021a). The characteristics of stands are described in detail in Adhikari et al. (2021a), Adhikari et al. (2021b), and Adhikari et al. (2022).

Tree diameter at breast height (DBH, 1.37 m) and the species of trees were recorded in 2014 growing season (around October) from Control stands. DBH was measured using diameter tape. The inventory of tree species and DBH of individual trees from all three control stands were developed in Microsoft Access to input into FVS.

### 2.3 Data and Analysis

#### 2.3.1 Timber Growth Simulation using Forest Vegetation Simulator (FVS)

Control stands were naturally established stands with matured trees (Adhikari et al., 2021a). Pine trees were older than 61 to 111 years and hardwood trees were older than 73 to 177 years (Masters et al., 1993) by 2014. Give the age of trees, and the growth of existing trees limited by space and resource competition in control plots, this research used DBH of mature trees from control plots from 2014 as the initial condition of research site in 1984 to simulate timber volume growth for all treatments. This research simulated timber volume in FVS (Figure 2) from 1984 (FVS Key: INVYEAR and TIMEINT) to 2024 for each treatment assuming 40 years (FVS Key: NUMCYCLE) period for the timber management. Timber growth and yield models were simulated in the Forest Vegetation Simulator (FVS) growth and yield model for the southern variant using the locational code for the research area as 80906 (Dixon, 2022).

Treatments were simulated using various combinations of harvest (FVS Key: THINBBA), thinning (FVS Key: THINBBA), prescribed burn (FVS Key: SIMFIRE), annual growth (FVS Key: FIXDG), natural regeneration (FVS Key: FIXDG), and mortality modifiers (FIXMORT) in FVS to imitate treatments and vegetation dynamics in the research field. Post-oak and blackjack oak mortalities were taken from Masters et al. (2006) to account for tree mortality.

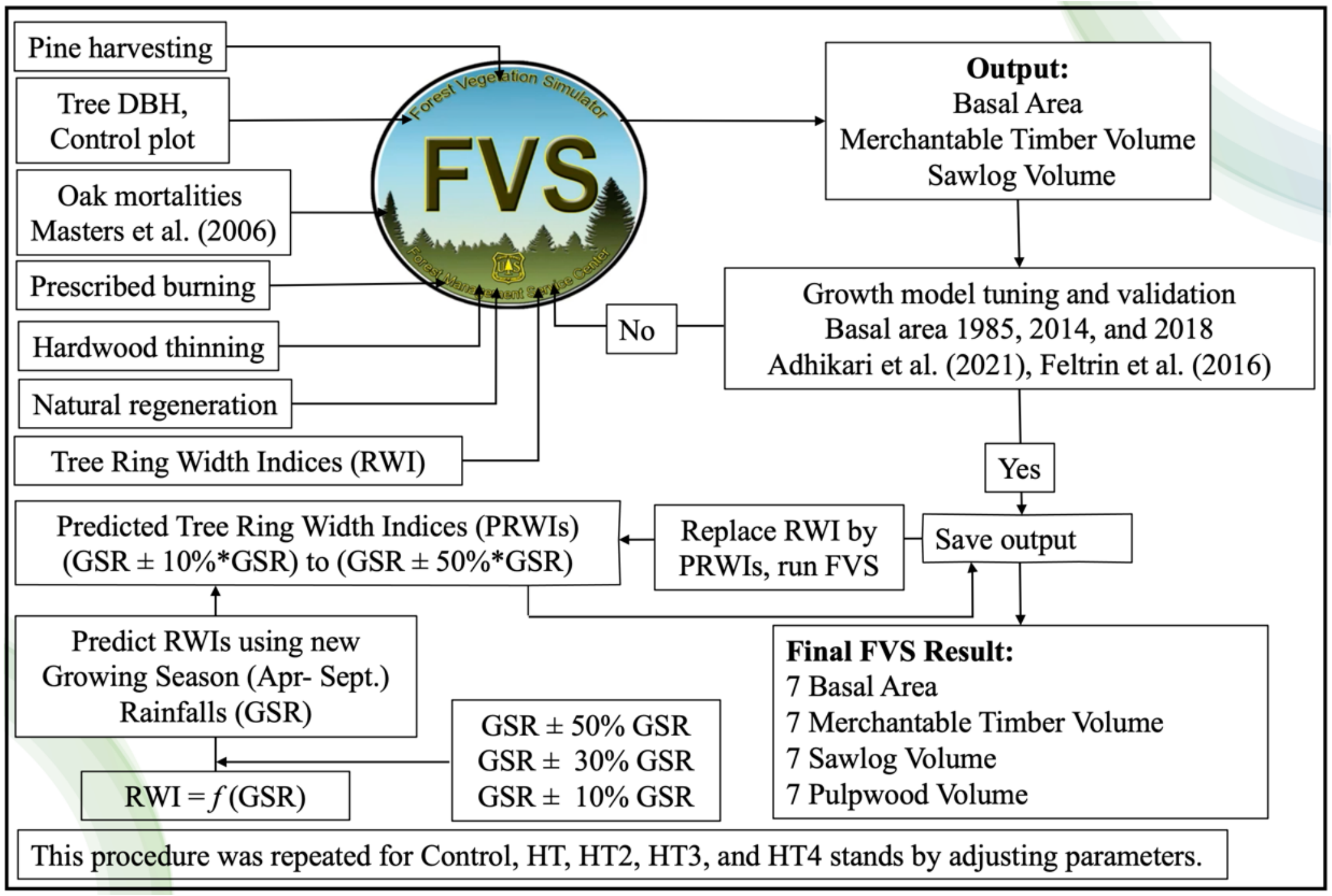

Figure 2: Flowchart of Timber Growth Simulation in FVS.

Tree-ring width indices for annual wood (RWI) for shortleaf pine were used as diameter growth multiplier values in the FVS simulation for annual growth and natural regeneration. The RWI for each treatment from 1984 to 2018 was calculated as described by Adhikari et al. (2021a). The FVS simulations were calibrated by matching simulated average basal area in FVS with basal area as reported in Adhikari et al. (2021a) and Feltrin et al. (2016) for the years 1984, 2014, and 2018. Basal area ($m^2$ $ha^{-1}$), merchantable timber volume ($ft^3$), and sawlog volume ($ft^3$) from 1984 to 2024 were obtained from FVS results for each treatment after calibrating volumetric growth. Pulpwood volume ($ft^3$) was calculated as the difference between merchantable timber and sawlog timber volume. The sawlog and pulpwood volume ($ft^3$) were converted into weight (ton $ha^{-1}$) using conversion factor of 2.7 divided by 128 (Self, 2020).

### 2.3.2 Timber Stumpage Price

The average annual stumpage prices of sawlog and pulpwood in north-eastern Texas were obtained from Texas A & M Forest Service annual summary report (Texas A & M Forest Service, 2021). The stumpage price of sawlog and pulpwood for 2024 was calculated by averaging the timber stumpage price for the last five years—from 2016 to 2021. The sawlog and pulpwood prices were multiplied with sawlog and pulpwood weight respectively to obtain respective market values for the year 2024, the end of the 40 year cycle. The stumpage of price of sawlog and pulpwood in 2024 were considered as \$27.69 $ton^{-1}$ and \$7.87 $ton^{-1}$ for the analysis.

### 2.3.3: Cattle and Deer Forage Classification

The aboveground net primary production (ANPP) of understory vegetation was collected annually from the research plots in Pushmataha Habitat Research Area from 1987 to 2017 immediately following the growing season (usually October) using a quadrant of 0.5 m x 0.5 m (0.25 $m^2$). Forage data collection methods are discussed in detail in Adhikari et al. (2021b). Understory vegetation collected were categorized into grass, sedge, panicum, legume, non-legume forb (henceforth, forb), and woody (current year woody and leaf). Understory vegetation samples were bagged, dried, and weighed. The dry mean weight of forage categories was calculated for each treatment and converted into average dry forage produced in kg $ha^{-1}$ $year^{-1}$. We assumed that cattle and deer don't compete for forage and thus, classified legume, woody, forbs, (Johnson et al., 1995), and panicum (Gee et al., 2011) as deer forages and grass and sedge (Sedivec and Printz, 2014) as cattle forage.

The total forage for cattle and deer were multiplied by 0.25 (25% of ANPP) to account for grazing and browsing efficiency (Redfearn and Bidwell, 2017). Total forage after the multiplication represents the total forage available for consumption for cattle and deer per hectare of land under carrying capacity. Carrying capacity is defined as number of individual animal supported by the ecosystem in given area (Chapman and Byron, 2018).

2.3.4: Valuation of Cattle Forage

The stocking rate of cattle (Meehan et al., 2018) under carrying capacity i.e. the total number of cattle supported by dry forage produced in one hectare land is calculated using Eq (1) assuming 10.89 kg $cattle^{-1}$ $day^{-1}$ (24 pound $cattle^{-1}$ $day^{-1}$) (Rasby, 2013) and 1.93 kg $deer^{-1}$ $day^{-1}$ (Fulbright and Ortega-S, 2013) dry matter intake by cattle. The total land required for a cattle is the inverse of cattle stocking rate.

$$\text{Cattle Stocking Rate} = \frac{\text{Cattle forage dry mass( Kg)/hectare/day}*365*0.25}{\text{Dry forage consumed by a cattle}*365\text{/day}} \qquad \text{Eq (1)}$$

The market price of cattle, 226.80 to 272.16 Kg (500 to 600 pounds) steer, was obtained from Oklahoma National Stockyards (Oklahoma National Stockyard, 2022) (Figure 3). The monthly average steer auction prices from 1984 to 2016 and weekly average steer auction prices from 2017 to 2021 were converted into the annual average price. Cattle auction market prices from 2022 to 2024 were predicted using calculated annual average cattle auction market price data from 1984 to 2021. Annual market prices of cattle were adjusted against inflation by multiplying the current market value with the producer price index (PPI) ratio of slaughter cattle (U.S. Bureau of Labor Statistics, 2022). The PPI ratio was calculated by dividing the PPI of individual year by that of 1984 assuming the PPI of 1894 is 100.

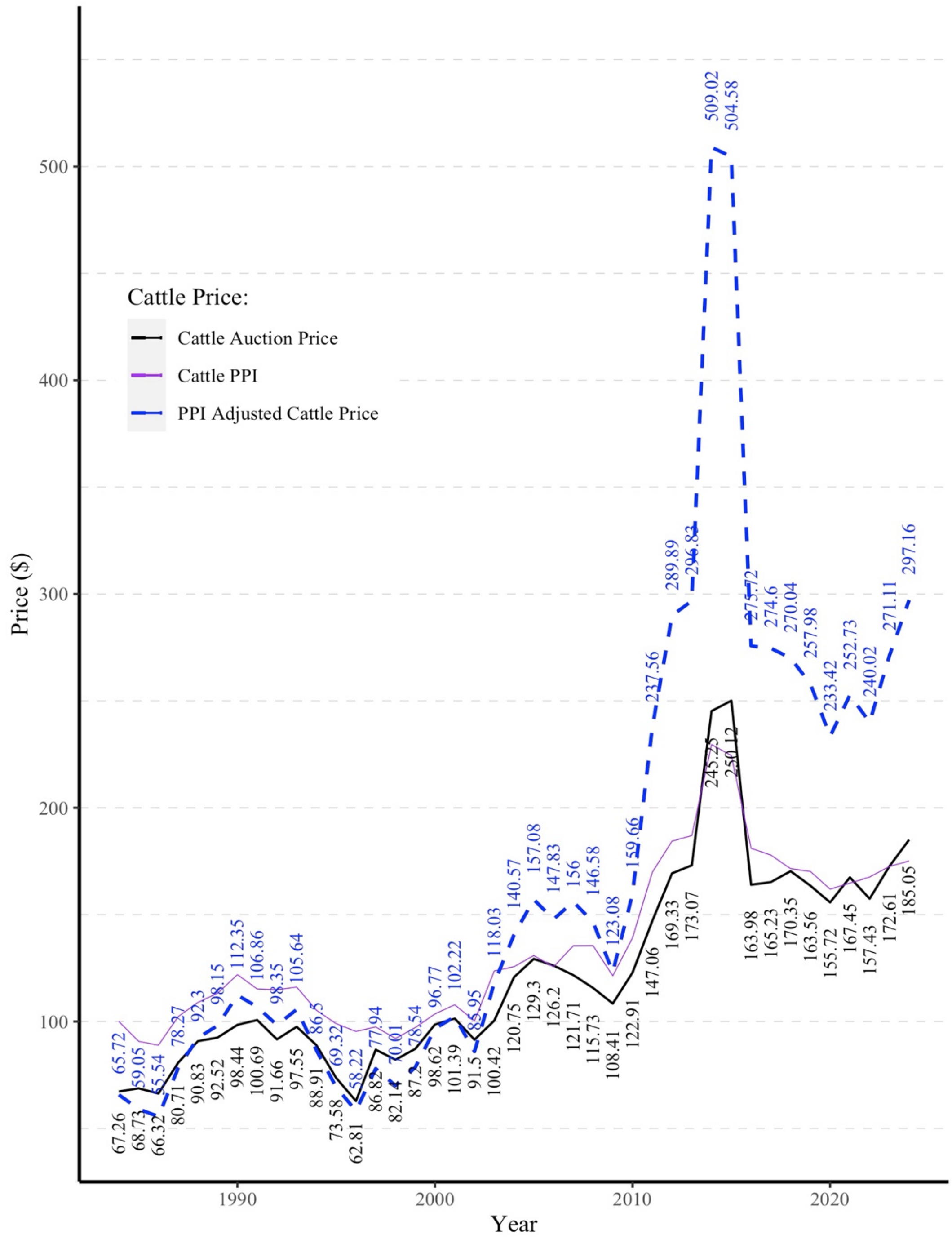

Figure 3: Cattle auction market price in Oklahoma, producer price index, and Producer Price index adjusted cattle price.

The annual economic value of cattle forage was obtained by multiplying annual auction price of cattle, PPI ratio, and stocking rate of cattle after adjusting for grazing efficiency (Meehan et al., 2018; Redfearn and Bidwell, 2017; Sedivec and Printz, 2014) using Eq. (2).

$$\text{Market Value}_{\text{cattle forage}} = \text{Cattle market price} * \text{PPI ratio} * \text{cattle stocking rate} \quad \text{Eq (2)}$$

2.3.5: Valuation of Deer Forage

Contingent valuation approach was used to elicit the value of a deer in respondent's regular deer hunting site using two questions of a questionnaire designed for this research. Survey responses and respondent details are discussed in detail in other chapters of this dissertation. The first question asked respondent to reveal total willingness to pay (WTP) to maintain the current population of deer observed at their regular deer hunting. The second question asked respondents to report total number of deer observed per visit in their regular hunting site. The reported average WTP in first question was divided by the average number of deer observed per visit in second question to calculate WTP per deer which is the economic value of a deer. Total economic return calculation in this research is consistent with the common practice of adding provisioning (timber and cattle) with cultural (WTP based on deer habitat management) ecosystem services (Costanza et al., 1997).

The stocking rate of deer was calculated using Eq. (3) which is a modification of cattle stocking rate formula. The total land required to support one deer from the dry forage production is the inverse of deer stocking rate.

$$\text{Deer Stocking Rate} = \frac{\text{Deer forage dry mass Kg/hectare/day} * 365 * 0.25}{\text{Dry forage consumed by a deer/day} * 365} \quad \text{Eq (3)}$$

The annual economic value of deer was obtained by multiplying WTP ($) deer$^{-1}$ ($12.96) with the deer stocking rate adjusted for grazing efficiency (Eq. 4). The WTP deer$^{-1}$ was assumed to remain constant over time.

$$\text{Market Value}_{\text{deer forage}} = \text{WTP}(\$)/\text{deer} * \text{Deer Stocking Rate} \quad \text{Eq (4)}$$

2.3.6: Cost of Timber and Pasture Management

The cost of prescribed burning, tax, and annual maintenance costs for were included in the benefit-cost analysis. Prescribed burning cost was considered as $61.75 ha$^{-1}$ ($25 acre$^{-1}$) (Bidwell et al., 2021) for 2022. Prescribed burning costs were discounted to the year 1984 and compounded for every burned year until the end of the 40 year cycle. Annual maintenance cost for timber including property tax was considered as $12.35 ha$^{-1}$ ($5 acre$^{-1}$) in 2024, discounted for every year until 1984 and summed as total annual maintenance cost for 40 years period. The total annual maintenance cost was $198.17 ha$^{-1}$ ($80.23 acre$^{-1}$) over 40 years. The interest rate varies from 1% to 7% (Hardaker and Healey, 2021) in the benefit cost analysis of timber valuation [Also see: (Bettinger et al., 2017; Chang, 2014; Donis et al., 2020; Hardaker and Healey, 2021)]. The real annual compounding and discounting interest rate was assumed to be 5.5% following previous literature published in this region (Chang, 2014).

Present value of future sum:

$$V_o = \frac{V_n}{(1+r)^n} \quad \text{Eq (4)}$$

Future value of present sum:

$$V_n = V_0(1 + r)^n \qquad \text{Eq (5)}$$

Where $V_o, V_n, r,$ and n are present \$ value, future \$ value, interest rate, and periods or years into the future respectively (Bullard and Straka, 2011).

The average pasture maintenance cost varies over 40 years period. Unlike timber, forage is harvested and utilized on annual basis. This research, thus, assumed average annual pasture maintenance cost as \$20 acre$^{-1}$ (\$49.42 ha$^{-1}$) (Halich et al., 2021) in 40 years period.

2.3.7: Financial Criteria for Benefit cost Analysis of Timber

Net present value (NPV), land expectation value (LEV), equal annual equivalent (EAE), and were used as financial analysis criteria for timber. The financial analysis were based on timber growth in each treatments. Since, the valuation of timber was based on 40 years period (year 2024), NPV is refer to as net future values (NFV) in this research.

Net future value is the sum of the future value of the revenues subtracted from the future values of the costs. If NFV > 0, the investment will return the desired discount and a future value of additional net revenue (Bettinger et al., 2017; Wagner, 2012). The NFV is calculated using Eq (6) following Bettinger et al. (2017).

$$NFV_T = \sum_{t=1}^{T} \frac{R_t}{(1+r)^t} - \sum_{t=0}^{T} \frac{C_t}{(1+r)^t} \qquad \text{Eq (6)}$$

Land/Soil Expectation Value (LEV): The LEV is the net future value of an investment in an even aged stand from the time of planting through infinite rotations of the same management regime. LEV considers value of bare land into calculation but purchasing cost and revenue from selling the land are not accounted (Bettinger et al., 2017). LEV is calculated following Eq. (7) (Bettinger et al., 2017; Bullard and Straka, 2011).

$$\mathrm{LEV} = \mathrm{NFV}\left(\frac{(1+r)^t}{(1+r)^t-1}\right) \qquad \text{Eq (7)}$$

Equivalent Annual Income (EAI) is the net revenue that can be obtained annually over the life of investment considering the discount rate applied (Bettinger et al., 2017). EAI is obtained from timber production assuming total cost of management for each treatment is allocated towards timber production. EAI is calculated using Eq. (8) (Bettinger et al., 2017; Bullard and Straka, 2011).

$$\mathrm{EAI} = \mathrm{NFV}\left(\frac{r}{(1+r)^t-1}\right) \qquad \text{Eq (8)}$$

2.3.8: Climatic Sensitivity on Timber Growth, and Economic Valuation

The sensitivity of timber production under changing climate was simulated by varying growing season (April to September) precipitation as a proxy of climate change because growing season rainfall was found to be most correlated with variation in annual ring width (Adhikari et al., 2021a). Growing season rainfall data of the study area from 1984 to 2018 were computed from daily Daymet data (Thornton et al., 2018). The RWI was regressed as a function of average growing season rainfall using ordinary least squared regression and new RWIs from 1984 to 2018 were predicted for ± 10%, ± 30%, and ± 50% change in rainfall. Predicted RWIs were used as a proxy of the effect of change in rainfall in the growth of timber volume during the rotation cycle. The growing season rainfall and the prediction range from 1986 to 2017 are presented in Figure 1.

The market value of timber calculated using the original RWI is the market value of timber under the current climatic scenario. After calibrating the timber growth model in FVS with original RWI, new timber growth models were simulated by replacing original RWI with predicted RWIs, *ceteris paribus*, for each year from 1984 to 2024 and each treatment. This gave rainfall sensitive predicted basal area, merchantable timber, sawlog, and pulpwood volumes from 1984 to 2024 for all treatments under the various rainfall scenarios. The predicted sawlog and pulpwood volumes were multiplied with timber market price to obtain predicted timber market values for the year 2024, the end of the rotation cycle. The predicted timber market values are market values of timber sensitive to rainfall conditions under changing climatic scenarios.

## 3. Result and Discussion

### 3.1: Stand Details and Timber Production

The basal area (BA), merchantable timber volume, and sawlog volume in control stands were the highest among all treatments in 1984 (Table 1). The change in basal area was smallest (3.14 $m^2$ $ha^{-1}$) in Control stands during the 40-year period but HT (34.58 $m^{-2}$ $ha^{-1}$) stand has the highest increase in BA followed by HT4 (19.63 $m^{-2}$ $ha^{-1}$). The Control plots were unmanaged stands with mostly mature trees standing at least for a century and a few younger trees. The growth of control plots is further limited by space and competition among new-old growths. The long term loblolly plantation established in 1975 in southeast Oklahoma attained maximum basal area up to 43.5 $m^2$ $ha^{-1}$ and started to decline (Hennessey et al., 2004). A synthesis study of intensively managed pine stands in the US south also found that the competition related mortality was initiated at basal area of 30-35 $m^2$ $ha^{-1}$ (Jokela et al., 2004). The simulated basal area of Control stand in this research plots by 2024 was 29.13 $m^2$ $ha^{-1}$. The basal area of the Control stand is comparatively low in our plots because of competition (Hennessey et al., 2004), competition related mortalities (Jokela et al., 2004; Masters et al., 2006) , and uneven aged mixed nature of the stand (Sterba and Monserud, 1993). Thinning and harvesting reduced competition and provided growing space for existing and new trees in HT stands allowing the stand to accumulate BA in the 40 years period.

Table 1: Basal area and timber production estimated using FVS growth and yield model simulation in southeastern Oklahoma, USA

| Treatments | Basal Area ($M^2$ $ha^{-1}$) | | | Timber Production (Ton $ha^{-1}$) | | | | | | | | |
|---|---|---|---|---|---|---|---|---|---|---|---|---|
| | | | | Merchantable | | | Saw Log | | | Pulp Wood | | |
| | 1984 | 2024 | Change | 1984 | 2024 | Change | 1984 | 2024 | Change | 1984 | 2024 | Change |
| **CONTROL** | 25.99 | 29.13 | 3.14 | 160.71 | 203.81 | 43.09 | 119.77 | 153.82 | 34.05 | 40.94 | 49.96 | 9.02 |
| **HT** | 4.14 | 38.72 | 34.58 | 33.28 | 178.04 | 144.75 | 32.84 | 79.22 | 46.38 | 0.44 | 98.82 | 98.37 |
| **HT2** | 5.06 | 7.36 | 2.30 | 40.85 | 65.73 | 24.88 | 40.30 | 65.36 | 25.06 | 0.54 | 0.37 | - 0.17 |
| **HT3** | 3.37 | 6.21 | 2.84 | 27.95 | 50.88 | 22.93 | 27.70 | 50.01 | 22.31 | 0.27 | 0.86 | 0.59 |
| **HT4** | 3.83 | 23.46 | 19.63 | 31.31 | 128.64 | 97.33 | 31.01 | 123.38 | 92.37 | 0.30 | 5.26 | 4.97 |

Note: The simulation based estimates using FVS growth and yield model were made in 1984 and 2024. The basal area of stands in 1985 and 2018 based on tree measurements are reported in Adhikari et al. (2021a).

The growth of sawlog timber was highest in HT4 (92.37 ton $ha^{-1}$) stand followed by HT (46.38 ton $ha^{-1}$), and control (34.05 ton $ha^{-1}$) (Table 1). Harvested and thinned (HT) stands had the highest weight pulpwood (98.37 ton $ha^{-1}$) growth. The rest of the stands added a very small weight of pulpwood and pulpwood weight decreased in HT2 stands. Thinning and harvesting created more open spaces for trees to grow (Harper and Johnson, 2008) in HT stand adding pulpwood volume. The lower stand density in HT4 resulted in larger trees that were predominantly sawtimber. The basal area of HT grew by 35.58 $m^2$ $ha^{-1}$ compared to 19.63 $m^2$ $ha^{-1}$ in HT4 which is about 76% faster growth in HT compared to that of HT4. But the merchantable timber grew 144.75 ton $ha^{-1}$ in HT compared to 97.33 ton $ha^{-1}$ in HT4 which is about 49% faster growth in HT compared to that of HT4. The growth of pulpwood is also higher in HT compared to HT4, but the growth of sawlog was higher in HT4 compared to HT. Most of the merchantable timbers produced in HT4 were sawlog timber.

The HT2 and, HT3 plots had the smallest growth in sawlog and pulpwood volume. The pulpwood volume decreased but sawlog volume increased in HT2 suggesting growth of small trees as bigger trees and conversion of pulpwood into sawlog. The HT2, and HT3 were disturbed more frequently by burning every two and three years. Frequent burning in HT2 and HT3 further suppressed and regeneration of new trees limiting the addition of basal area. New growths were further suppressed by frequent fire, thus, limiting their growth and development to become pulpwood and sawlog timber. Frequent burning created a favorable environment for the growth of understory vegetation.

3.2: Cattle and Deer Forage Production

The average ANPP of sedge was highest in HT4 stands (250 kg $ha^{-1}$ $yr^{-1}$) (Table 2). The average ANPP of forbs increased with the decrease in fire return interval in burned plots. Among burned plots, the highest average biomass of forbs is produced in HT2 (250.22 kg $ha^{-1}$ $yr^{-1}$). The highest average biomass of legume (356.92 kg $ha^{-1}$ $yr^{-1}$) and panicum (435.69 kg $ha^{-1}$ $yr^{-1}$), were produced in HT2 stands. The highest average biomass of sedge (250.22 kg $ha^{-1}$ $yr^{-1}$) was in HT4 stands. The average biomass of woody (2,930.81 kg $ha^{-1}$ $yr^{-1}$) and grasses (kg $ha^{-1}$ $yr^{-1}$) were highest in HT3 stands.

The production of cattle and deer forages were low in non-burned stands throughout the research fields in comparison to burned plots. The Control and HT plots produced the smallest average biomass volume of forbs, sedges, grasses, legume, panicum, and woody. The HT and Control stand formed a closed canopy dominant forest, which limited sunlight access thus limiting the growth of understory vegetation. Frequent fire provided a favorable environment for the growth of the understory by increasing organic matter (Pellegrini et al., 2021), mobilizing nutrient contents in soil, and suppressing the generation and growth of hardwood and pine. Relatively open canopy in HT2, HT3, and HT4 stands supported growth of understory vegetation compared to HT and Control.

Table 2: Annual average dry biomass of understory vegetation produced in the south-central transitional region of the USA.

| **Treatment** | **Cattle Forage (Kg ha$^{-1}$ year$^{-1}$)** | | | **Deer Forage (Kg ha$^{-1}$ year$^{-1}$)** | | | | |
|---|---|---|---|---|---|---|---|---|
| | **Sedge** | **Grass** | **Total** | **Legume** | **Woody** | **Forbs** | **Panicum** | **Total** |
| **CONTROL** | 23.03 | 434.27 | 457.30 | 31.48 | 432.67 | 44.61 | 58.97 | 567.74 |
| **HT** | 24.15 | 547.57 | 571.71 | 23.13 | 701.29 | 47.62 | 88.28 | 860.31 |
| **HT2** | 173.64 | 7,915.20 | 8,088.84 | 356.92 | 2,676.52 | 482.48 | 435.69 | 3,951.62 |
| **HT3** | 131.12 | 8,446.76 | 8,577.88 | 272.79 | 2,930.81 | 224.82 | 288.68 | 3,717.11 |
| **HT4** | 250.22 | 5,860.65 | 6,110.87 | 315.23 | 2,508.86 | 201.66 | 369.30 | 3,395.04 |

The total number of cattle and deer supported by Control and HT stands is very small compared to HT2, HT3, and HT4 stands (Table *3*). The total land required is the highest for Control followed by HT stands. The larger area of land required to raise per cattle and per deer in Control and HT stands are due to lower understory vegetation and forage production. The HT2, HT3, and HT4 produce a larger volume of cattle and deer forage thus, requiring comparatively smaller hectares of land to raise a deer or a cattle.

Table 3: Total number of cattle and deer supported by various treatments in south central transitional ecoregion of USA.

| **Treatments** | **Cattle Stocking Rate (Cattle ha$^{-1}$)** | **Deer Stocking Rate (Deer ha$^{-1}$)** |
|---|---|---|
| **CONTROL** | 0.20 | 0.03 |
| **HT** | 0.31 | 0.04 |
| **HT2** | 1.40 | 0.51 |
| **HT3** | 1.32 | 0.54 |
| **HT4** | 1.20 | 0.38 |

3.3: Timber Management Cost and Economic Return

The management cost of Control and HT stands were the smallest ($210.60) among all stands because management was not applied after the establishment of the research plots in both stands. The costs for both plots were from land property tax, and maintenance costs. The management cost was highest in HT2 ($1,588.84) stands followed by HT3 ($1,153.61). The cost of harvested and thinned plots originated from the land property tax, maintenance cost, and prescribed fire costs. The cost of management increased with the increase in fire frequency.

The economic return from sawlog timber growth in 40 years period was the highest in HT4 ($2,372.70) followed by HT ($1,191.43) (Table *4*). The return from sawlog timber in Control, HT2, and HT3 were relatively small. The return from pulpwood was the highest from HT ($773.98). The Control ($70.96) and HT4 (39.08) stands had relatively smaller return compared to HT. The return from pulpwood in HT2 was negative because HT2 losses pulpwood volume after the research plots were established.

Among five treatments, the highest total return from timber growth in 40 years period was obtained from HT4 ($ 2,411.78) followed by HT ($1,965.41). The higher weight of sawlog produced in the HT4 compared to HT in combination with the higher sawlog price generated higher revenue in HT4. The total economic return from HT and Control plots varies by more than $1,000 ha$^{-1}$ even though the sawlog weight growth is only 12.33 ton ha$^{-1}$. HT stand produced significantly more pulpwood (89.35 ton ha$^{-1}$) compared to Control stand increasing the economic return from HT significantly higher in comparison to Control stands.

The total economic return per hectare from timber after deducting timber management cost, also referred as NFV and LEV, in 40 years was highest for HT ($1,754.80) followed by HT4 ($1,548.32) and Control ($735.04). The economic return from stands based on LEV which accounts value of bare land also follows same pattern as NPV. The relatively small management cost but high economic return favored HT to became management regime generating highest return from timber management compared to HT4. HT2 and HT3 were unfavorable management options for landowners managing property for timber production due to the negative economic returns. The cost of management is higher compared to benefit in HT2 and HT3.

Table 4: Net change in revenue, timber management cost, and net future value from timber from actively managed stands in 40 years period in the south-central transitional region of the USA.

| **Treatment** | **Return from Timber ($ ha$^{-1}$) in 40 Years** | | | **Total Cost ($ ha$^{-1}$)** | **NFV ($ ha$^{-1}$)** | **LEV ($ ha$^{-1}$)** |
|---|---|---|---|---|---|---|
| | Sawlog | Pulpwood | Total | | | |
| **CONTROL** | 874.69 | 70.96 | 945.65 | 210.60 | 735.04 | 832.88 |
| **HT** | 1,191.43 | 773.98 | 1,965.41 | 210.60 | 1,754.80 | 1,988.36 |
| **HT2** | 643.64 | - 1.36 | 642.28 | 1,588.84 | - 946.57 | - 1,072.55 |
| **HT3** | 573.18 | 4.67 | 577.85 | 1,153.61 | - 575.76 | - 652.40 |
| **HT4** | 2,372.70 | 39.08 | 2,411.78 | 863.45 | 1,548.32 | 1,754.40 |

The highest annual economic return per hectare from timber, cattle forages, and deer forages were obtained from HT3 ($56.42) followed by HT2 ($49.48), and HT4 ($43.75) (Table 5). The higher HT3 provided highest annual return per hectare from cattle forage ($43.53) and comparatively lucrative return from deer forage ($17.10) in comparison to other stands. Though, the annual return from timber was negative, the loss was offset by the return from forages in HT3 stands. The HT2 provided highest annual economic return per hectare from deer forage ($18.18) and comparatively lucrative return from cattle forages ($38.23) which offset loss from timber production. The HT4, however, provided positive annual economic return per hectare from timber ($11.33), cattle forages ($16.80), and deer forages ($15.62).

The annual economic return analysis showed that landowners interested in managing their property for cattle and deer management will benefit by burning every 2 to 3 years creating woodland and savanna (Adhikari et al., 2021a) . Landowners interested to manage their land for timber production are benefitted by harvesting and thinning their stands. Periodic early and pre-commercial thinning and harvesting could provide additional source of income for timber producers if their goal is to maximize current year income (Hardaker and Healey, 2021).

Table 5: Net Annual Economic Return from timber and forage in actively managed ecosystems in south central transitional ecoregion, USA.

| Net Annual Economic Return ($ ha$^{-1}$) | | | | |
|---|---|---|---|---|
| Treatment | Timber | Cattle Forge | Deer Forage | Total |
| Control | 5.38 | - 44.46 | 2.61 | - 36.47 |
| HT | 12.85 | - 43.22 | 3.96 | - 26.42 |
| HT2 | - 6.93 | 38.23 | 18.18 | 49.48 |
| HT3 | - 4.21 | 43.53 | 17.10 | 56.42 |
| HT4 | 11.33 | 16.80 | 15.62 | 43.75 |

Note: The annual economic return from timber was EAI. The WTP per deer was $12.96 in a hunting site where respondents observed on an average of 8 deer per visit. The annual return of cattle forage was obtained after reducing pasture management cost as $20 acre$^{-1}$ ($49.42 ha$^{-1}$) (Halich et al., 2021). The cost of deer management was $0.

The annual economic return from Control and HT, if managed for timber, cattle, and deer, were negative due to significantly negative return from forages and thus considered as unfavorable management combination. However, HT ($12.85) and HT4 ($11.33) generated relatively better annual return per hectare from timber production compared to Control, HT2 and HT3. HT4 could be a better management options for landowners interested in reducing the risk of economic losses by diversifying their management objectives and also generating forest in their property (Adhikari et al., 2021a). The uneven aged forest with the biodiversity could generate additional benefit beyond timber, cattle, and deer management through USDA conservation program which is out of scope of this paper.

3.4: Impact of Rainfall Variation on Timber Production

The sawlog weight increased with the increase in rainfall in all treatments however the sensitivity towards rainfall changes varied in each treatment (Table 6). The HT and HT3 treatments were benefitted by the increase in rainfall for sawlog production compared to other treatment. The rate of increase in sawlog weight with the change in rainfall was relatively larger in HT and HT3 compared to Control, HT2 and HT4. The sawlog volume decreased by 48% and 41% with the 50% decrease in rainfall in HT and HT3. However, with the same amount of decrease in rainfall, sawlog volume only decreased in 9%, 17%, and 21% in HT4, Control, and HT2. With the 50% increase in rainfall, sawlog timber volume increased by 73% and 64% in HT3 and HT in comparison to 18%, 15%, and 11% in HT2, Control, and HT4. The HT4 is relatively stable with the change in changing rainfall. In the Control stand the production of sawlog was limited by competition and mortality (Hennessey et al., 2004; Jokela et al., 2004; Masters et al., 2006) which could have limited the growth regardless of the rainfall status. Such competition were removed in HT and HT3 due to hardwood thinning, pine harvesting, and burning which provide ample space and resources for existing trees to grow as sawlog timber.

The impact of change in rainfall on pulpwood production under various treatments varied by treatments (Table 6). The pulpwood volume in the control plot was almost non-responsive to changes in rainfall. The growth of new plants were limited in the Control stands due to competition in light and space thus limiting the growth of pulpwood. The percentage change in pulpwood timber remain comparatively stable in HT and HT4 as the rainfall increased indicating relatively less sensitive to change in rainfall. The increase in rainfall might have favored the conversion of pulpwood into sawlog timber in these stands. The growth of pulpwood decreases with the increase in sawlog volume in HT plots indicating the maturation and growth of pulpwood into sawlog.

The pulpwood growth in HT3 was very small but highly sensitive to change in rainfall. The increase in rainfall in combination with the reduced competition, ample light, and space created a favorable environment for small trees to grow as pulpwood timber. The percentage change in pulpwood growth decreased with increase in rainfall in HT2. The frequent disturbance by fire might have limited the growth of pulpwood timber because frequent fire kills newly growing woody understory vegetation, which otherwise would convert to pulpwood after a few years of growth.

The impact of change in rainfall in various treatments for pulpwood production was not as distinctly visible as in the sawlog production. The sensitivity of various treatment varied with fire return interval, competition, availability of space, and sunlight. The sawlog timber growth was favored by the increase in rainfall in comparison to pulpwood growth in the study site. Higher sensitivity of pulpwood growth with the change in rainfall was observed in stands such as HT2 and HT3 which were frequently treated with fire; these plots are also good producer of cattle and deer forages.

Table 6: Change in sawlog, and pulpwood production with the Change in Rainfall in the south-central transitional region of the USA

| **Rainfall Change (%)** | **Saw Log [Ton ha$^{-1}$ (% Change) ]** | | | | | **Pulp Wood [Ton ha$^{-1}$ (% Change) ]** | | | | |
|---|---|---|---|---|---|---|---|---|---|---|
| | Control | HT | HT2 | HT3 | HT4 | Control | HT | HT2 | HT3 | HT4 |
| **-50%** | 28.17 (- 0.17 %) | 24.24 (- 0.48 %) | 19.82 (- 0.21 %) | 13.20 (- 0.41 %) | 84.24 (- 0.09 %) | 8.99 (0.00 %) | 92.86 (- 0.06 %) | - 0.27 (0.57 %) | 0.22 (- 0.63 %) | 4.35 (- 0.12 %) |
| **-30%** | 30.62 (- 0.10 %) | 25.99 (- 0.44 %) | 21.79 (- 0.13 %) | 16.63 (- 0.25 %) | 86.53 (- 0.06 %) | 8.90 (- 0.01 %) | 103.02 (0.05 %) | - 0.22 (0.29 %) | 0.37 (- 038 %) | 4.92 (- 0.01 %) |
| **-10%** | 33.06 (- 0.03 %) | 37.19 (- 0.20 %) | 23.57 (- 0.06 %) | 20.73 (- 0.07 %) | 91.45 (- 0.01 %) | 8.82 (- 0.02 %) | 102.84 (0.05 %) | - 0.25 (0.43 %) | 0.52 (- 0.13 %) | 4.84 (- 0.02 %) |
| **0%** | 34.05 | 46.38 | 25.06 | 22.31 | 92.37 | 9.02 | 98.37 | - 0.17 | 0.59 | 4.97 |
| **10%** | 35.26 (0.04 %) | 54.39 (0.17 %) | 25.67 (0.02 %) | 25.95 (0.16 %) | 94.34 (0.02 %) | 9.19 (0.02 %) | 96.57 (- 0.02 %) | - 0.22 (0.29 %) | 0.67 (0.13 %) | 4.99 (0.00 %) |
| **30%** | 37.21 (0.09 %) | 67.16 (0.45 %) | 27.85 (0.11 %) | 31.90 (0.43 %) | 96.84 (0.05 %) | 9.39 (0.04 %) | 94.57 (- 0.04 %) | - 0.17 (0.00 %) | 0.82 (0.38 %) | 5.07 (0.02 %) |
| **50%** | 28.17 (0.15 %) | 24.24 (0. 64%) | 19.82 (0.18 %) | 13.20 (0.73 %) | 84.24 (0.11 %) | 8.99 (0.08 %) | 92.86 ( - 0.10 %) | - 0.27 (0.14 %) | 0.22 (0.67 %) | 4.35 (0.05 %) |

Note: This table represent timber produced in each stands after treatments were applied. The values in the tables were obtained by deducting weight of pre-existing timber in 1984 from total timber weight in 2024 after 40 years.

3.5: Impact of Rainfall Variation on Economic Return from Timber

The impact of rainfall variation had mixed effect in the economic return from timber (Table 7). The NFV of Control, HT, and HT4 increased with the increase in rainfall. The NFV of HT2 and HT3, however, decreased with the increase in rainfall. The NFV from the Control and HT were relatively less sensitive to changing rainfall. The NFV decreased by 34.88% with the 50% decrease in rainfall. The change in NFV was relatively smaller in Control (20.58%) and HT4 (13.80%) with 50% decrease in rainfall. The increase in NFV is also highest (38.92%) in HT compared to Control and HT4. This indicates that HT4 was more sensitive to change in rainfall. The weight of merchantable timber produced was comparatively higher in HT (144.75 ton $ha^{-1}$) compared to 97.33 ton $ha^{-1}$ and 43.09 ton $ha^{-1}$ in HT4 and Control resulting into higher NFV.

The change in NFV from timber in HT2 and HT3 stands were negative. The NFV in HT3 decreased by 73.63% with the 50% increase in rainfall and increased by 41.19% with the 50% increase in rainfall. Whereas the NFV increased by 14.30% with 50% decrease in rainfall and decreased by 12.32% with the 50% increase in rainfall. The results demonstrated that frequently burned stands are more vulnerable to change in rainfall and less frequently burned and thinned stands are suitable for timber production under increasing rainfall scenario.

Table 7: Change in net future value (NFV) ($ ha$^{-1}$) from timber in 40 years cycle with change in Rainfall.

| Rainfall Change (%) | Change in NFV [$ ha$^{-1}$ (%)] | | | | |
|---|---|---|---|---|---|
| | Control | HT | HT2 | HT3 | HT4 |
| **-50%** | 583.78 (- 20.58 %) | 1,142.71 (- 34.88 %) | - 1,081.91 (14.30 %) | -812.90 (41.19 %) | 1,334.63 (- 13.80 %) |
| **-30%** | 645.84 (- 12.14 %) | 1,267.68 (- 27.76 %) | - 1,030.74 (8.89 %) | - 723.51 (25.66 %) | 1,398.13 (- 9.70 %) |
| **-10%** | 708.10 (- 3.67 %) | 1,553.86 (- 11.45 %) | - 985.23 (4.09 %) | - 616.97 (7.16 %) | 1,523.86 (- 1.58 %) |
| **0%** | 735.04 | 1,754.80 | - 946.57 | - 575.76 | 1,548.32 |
| **10%** | 767.51 (4.42 %) | 1,946.27 (10.91 %) | - 931.09 (- 1.64 %) | - 481.87 (- 16.31 %) | 1,599.30 (3.29 %) |
| **30%** | 819.21 (11.45 %) | 2258.69 (28.71 %) | - 874.84 (- 7.58 %) | - 327.73 (- 43.08 %) | 1,663.99 (7.47 %) |
| **50%** | 875.05 (19.05 %) | 2437.76 (38.92 %) | - 829.97 (- 12.32 %) | - 151.81 (- 73.63 %) | 1,809.88 (16.89 %) |

4. Management Implications

These results provide important management implication. First, our study results suggest that active management tools such as prescribed fire and thinning are important to enhance the growth of timber and forage. Less frequently burned stands such as HT and Control are suitable for timber production. For example, harvested and thinned stands in this research provided highest annual economic return. But more frequently burned stands such as HT2 and HT3 generated highest economic return because of higher cattle and deer forage production. The HT3 stand though have negative return from timber, generated highest annual return compared to other treatments because of higher cattle forage production. The stand with harvested pine, thinned hardwood, and burned every four year generated ample amount of timber and forage. This management regime can be applied to reduce financial risk by diversifying management goals and generate additional income by enrolling into conservation programs. This result has important implications in south central USA because timber production, wildlife management, and cattle grazing are important land management objectives of landowners in this region (Bulter et al., 2021).

Second, landowners represent a diverse group with multiple management objectives. Therefore, the outreach needs to focus on clientele needs (Joshi and Mehmood, 2011). This research finding suggests that single management regime does not meet diverse management objectives of landowners. Developing a sustainable land management plan for multiple objectives requires landowners to make an informed decision on livestock production and wildlife habitat management (Herrero et al., 2009; Hines et al., 2021). On this front, this research provides valuable insight for landowners to make informed decisions to estimate their economic return before making a land management decision.

Third, the impact of rainfall change varies by treatment and types of targeted timber. For example, In HT3 type management regime sawlog and pulpwood production are highly sensitive to change in rainfall. However, Sawlog production is highly sensitive to change in rainfall in management regime such as HT. The result suggests that sawlog timber producers will benefit from the increase in rainfall. But the pulpwood production varies with the change in rainfall under various treatments. The findings of this research helps landowners to make informed decision about the rainfall sensitivity of the management regime to optimize their management objectives.

A critical limitation of this study is price involvement in calculating NFV and LEV from timber production. The timber prices used in the economic analysis have been averaged over several years, which might have ignored the structural changes in the timber market over the years. The timber ban in the pacific northwest (Perez-Gracia et al., 2007), covid, inflation, and international tariffs (Zhang et al., 2020) have changed the price and structure of the timber market in the past 40 years, which has affected the timber price and economic return from timber. These stochastic events were not considered in our financial analysis though we acknowledge that these changes impact the financial health of a long-term investment such as timber management.

## 5. Conclusion

This research quantified economic return from actively managed forest, savanna, and grassland for about 40 years using prescribed fire, harvesting, and thinning in the south-central ecoregion of the USA. We further simulated change in rainfall patterns to observe its impact on timber production. The production of sawlog increased with the increase in rainfall pattern in all treatments. The production of pulpwood varied with the stand treatments. Pulpwood volume shows less variation in the control stand with pre-existing mature trees and more variation in stands such as HT, and HT4, which have the potential to add timber volume.

The HT2 and HT3 stands were more frequently treated with fire and have virtually negligible volume accumulation of timber. However, they have a higher volume of annual forage production for cattle and deer compared to areas that are less frequently treated with prescribed fire. The Control and HT plots produce a smaller volume of cattle and deer forage than HT2, HT3, and HT4 stands, thus requiring a larger area to support cattle and deer. But control and HT plots could generate higher revenue through timber production. The economic return and impact of rainfall in timber production varied by treatment. Management regimes, thus, should be selected based on the management objective of landowners. Landowners can benefit by prioritizing land management goal before selecting a management regime that is suitable to optimize their economic return.

**APPENDIX II.A**

(Images of research fields were removed to reduce file size.)

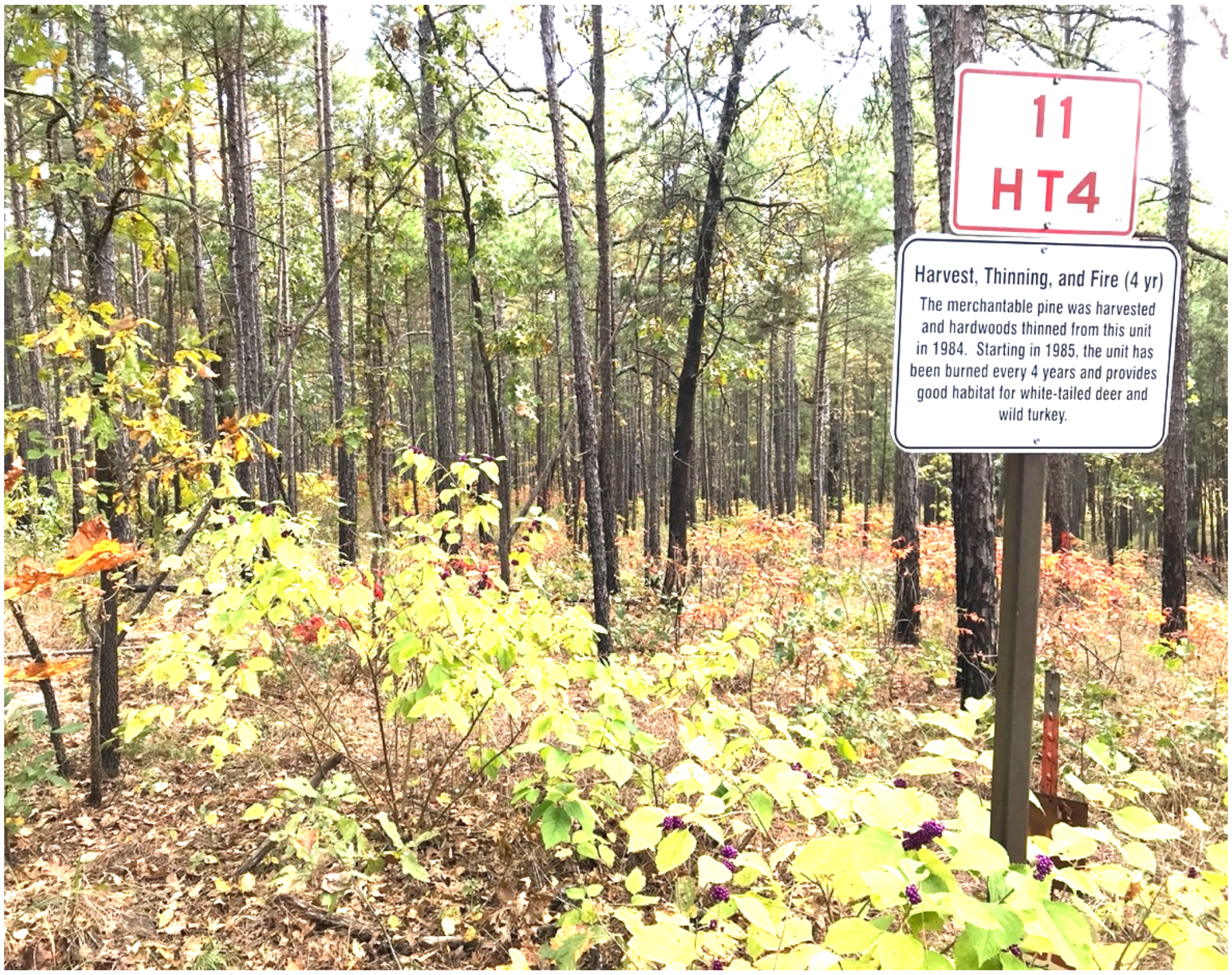

Figure 4: Diverse ecosystems in forest habitat research station, Pushmataha wildlife management area, Oklahoma resulting from various combinations of hardwood thinning, pine harvesting and burning for about 40 years.

# CHAPTER III: WILLINGNESS TO PAY FOR THE DEER HUNTING BASED ON SITE ATTRIBUTES IN SOUTH CENTRAL ECOTONE OF THE USA

**Abstract**

The savannas and woodlands in the south-central ecotone of the USA are transitioning to closed-canopy forest in the absence of fire. More open forest structure and its associated ecosystem its services can be restored using active management such as prescribed fire. Active management can create deer habitat desired by deer hunters. However, active management brings additional financial burden. Deer (*Odocoileus virginianus*) hunting can be an important source of revenue for landowners in this region which can compensate the cost of active management. To address this, I used best worst choice model to rank deer habitat characteristic such as seer sanctuary, food plots, and forest canopy cover for eastern Oklahoma. I further calculated marginal difference in willingness to pay (WTP) for habitat characteristics. We found that deer hunters are willing to pay more for a site that provides better deer hunting experience and opportunity to observe a greater number of deer which increase change of hunting success. The difference in marginal WTP to observe ten and six deer per visit instead of one deer per visit were $11 and $9 respectively. The willingness to pay for deer habitat with food plot and deer sanctuary was higher in comparison to deer habitat without them. Forest canopy types had non-significant impact on WTP which provide flexibility for landowners to freely change forest canopy cover to meet their multiple land management objectives in addition to deer hunting to optimize their revenue.

## 1. Introduction

The south-central transitional ecoregion of the USA is a dynamic mosaic of forest, grassland, and savanna. This ecotone is sandwiched between eastern forest and western Great Plains extending from southern Missouri to northern Texas (Hallgren et al., 2012; Joshi et al., 2019b). The heterogenic and dynamic ecosystem in this region provides an opportunity for multiple-use management involving objectives such as wildlife management, livestock grazing, and timber management, among others. This region, however, is drifting towards a closed-canopy forest due to exclusion of fire, encroachment of native but invasive species such as Eastern Redcedar (ERC, *Juniperus virginiana*) (henceforth, ERC), drought, and climate change (Joshi et al., 2019b).

Active management using prescribed fire, thinning, and harvesting can restore the characteristic feature of this ecotone (Hallgren et al., 2012). These practices, however, bring additional financial burdens (Starr et al., 2019). For example, the average costs of prescribed fire are over \$20 per acres (\$50 per ha) and chemical applications can cost above \$50 per acres (\$123.50 per ha) (Maggard and Barlow, 2018). Although some productive forestlands within the ecotone have commercial timber potential (Shephard et al., 2021), active management costs in the majority areas need to be compensated through other income streams such as deer hunting lease (Martínez-Jauregui et al., 2016). Given the relatively low quality timber and the lack of strong timber markets (Clark et al., 2007; Kaur et al., 2020; Riddle, 2019), managing land for deer hunting can serve as a primary cost offset tool in this region. Understanding the attributes of deer hunting sites and their preference among deer hunters are important in designing quality deer habitat to meet the need of deer hunters, optimize landowners invest in their land, and meet the habitat need of deer. The major objective of this paper was to quantify landowners' willingness to pay (WTP) for characteristics of deer hunting sites and rank their preferences for the specific characteristics.

The size and health of deer population depends upon the quality and quantity of available food and cover (Fulbright and Ortega-S, 2013; Yarrow, 2009). In addition, the growth, development, and the quality of individual deer is affected by availability of food (Yarrow, 2009). Habitat use (Leslie Jr. et al., 1996) and the dispersal distances of deer (Long et al., 2005) are affected by the canopy cover and forage production. Open-canopy created by fire increases the abundance of deer and alters their food preferences (Yarrow, 2009). An interspersion of forested, woodland, and non-forested areas and the transitional region between cover types (also called as edge), creates diversity in the food and cover for deer (Leslie Jr. et al., 1996; Yarrow, 2009).

Deer biologists often have emphasized importance of deer habitat improvement to maintain healthy deer populations (Fulbright and Ortega-S, 2013; Yarrow, 2009). Prescribed fire, thinning, food plot planting, as well as timber harvesting play crucial role in maintaining quality deer habitat and healthy deer population (Yarrow, 2009). For example, timber harvest, controlled burning, and plantings food plots provides quality food and cover (Yarrow, 2009). Timber harvest often creates "edge effects" which is heavily used by deer (Alverson et al., 1988; Yarrow, 2009). Prescribed burning increase browse yields and improve the palatability and nutrient of understory vegetation (Yarrow, 2009). Food plots can be designed to provide continuous food supply to maintain healthy deer populations (Fulbright and Ortega-S, 2013; Yarrow, 2009). Rotating open spaces, legume, cereals and grass, and clover (*Trifolium spp.*) in food plots increase diversity in food availability for deer throughout the year (Yarrow, 2009).

There is abundant research on the economic aspect of deer hunting [e.g. (Hussain et al., 2016; Hussain et al., 2007; Hussain et al., 2010; Mingie et al., 2017a; Mingie et al., 2017b; Serenari et al., 2019)]. The focus of economic research has been deer hunting lease revenues (Hussain et al., 2007), hunting club attributes (Mingie et al., 2017a) and membership dues (Mingie et al., 2017b), and regulatory changes (Serenari et al., 2019), among others. Despite the long history of economic research in deer hunting and ecological research emphasizing importance of habitat management to maintain quality deer population, the impact of landowners' preference on deer hunting site characteristics and their WTP for deer hunting is not well known.

To complement previous research, our goal was to determine the WTP of deer hunters to improve hunting site sand associated characteristics. Active management of ecosystem using tools such as prescribed fire, thinning, and herbicide increases heterogeneity (Adhikari et al., 2021a) thus providing diverse ecosystem services. The diversity within ecosystem can be leveraged by managing land for multiple objectives such as deer hunting, cattle grazing, and timber management. The accurate economic quantification of hunting site attributes provides opportunity for landowners to optimize revenue from their land in combination with other objectives.

Existing literature on lease hunting portrays landowners as primary supplier of the land for lease hunting (Hussain et al., 2007) and hunters as the consumer (Hussain et al., 2010; Hussain et al., 2004). This framework has two limitations. First, hunting based wildlife management is one of the most common land management objectives of landowners. Many landowners are avid hunters themselves thus may be interested in improving wildlife habitat on their own property. Second, lack of active management has led to marginal wildlife habitat in the south-central ecoregion of the USA. Many landowners in this region are likely to consider leasing hunting land for themselves if they find improved wildlife habitat in their proximity.

To address these issues, this research studies relative preference of deer hunting sites characteristics among deer hunters and their WTP as in the form of hunting site lease fee in for landowners in south central USA. Specifically, this research conducted a study to measure hunter's stated preference on important attributes such as deer sanctuary, food plots, and forest canopy cover, which are crucial in maintaining good deer habitat (Fulbright and Ortega-S, 2013; Yarrow, 2009). landowners will be able to make informed decisions by calculating economic tradeoff among various attributes of deer hunting sites to meet the objective of landowners, deer hunters, and maintain good deer habitat quality, while managing the land for multiple objectives.

## 2. Methods

### 2.1: Study Area

This study was conducted in eastern Oklahoma, USA which. This region covers Arkansas valley, Cross Timbers, Central Plains, Flint Hills, Alley, and Ouachita Mountains in Oklahoma lying roughly east to Interstate-35 (Figure 5). Historically, the transitional ecoregion was well known for savanna, woodlands, and forest dominated by post oak (*Quercus stellata*), blackjack oak (*Quercus marilandica*), shortleaf pine (*Pinus* *echinata*), and hickory (*Carya spp.*) (Johnson et al., 2010; Joshi et al., 2019b). There were expanses of tallgrass prairie vegetation intermixed. The diversity in the region the range of hunting sites defined the choice sets provides in our survey. Besides, the diversity in the region provides an opportunity for landowners to manage for multiple objectives such as wildlife management, cattle grazing, and timber management.

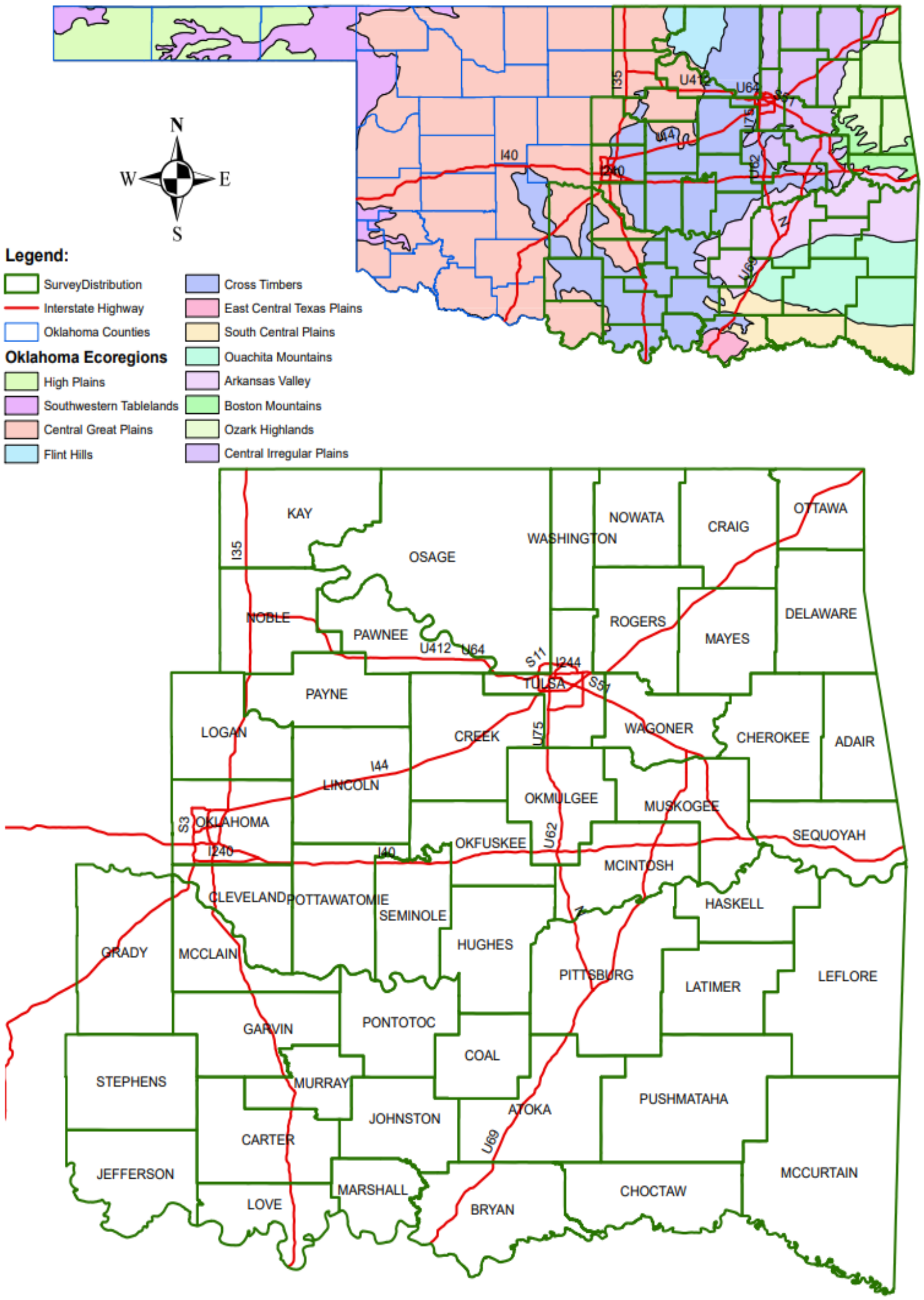

Figure 5: Study region: map of Oklahoma representing various ecoregions (top) and counties receiving surveys (bottom).

### 2.2: Survey Design

#### 2.2.1: Best Worst Choice (BWC) Modeling

The BWC is rooted in random utility theory (RUT) (Flynn et al., 2007) which states that the satisfaction derived by consumers depends upon the characteristics of the available alternatives (Hanley et al., 2001; Lancaster, 1996; Rakotonarivo et al., 2016). During the utility maximization process, consumers choose an alternative that provide them with the highest level of utility (Lancaster, 1996).

This research used profile case (Case 2) best-worst choice method (Louviere et al., 2015). The BWC proposes a hypothetical product or scenario explained by a set of characteristics and the price associated with these attributes provides economic value of the hypothetical product. BWC has two parts: 1) best worst scaling (BWS) and 2) binary choice question (Soto et al., 2016; Soto et al., 2018). BWS ask consumer to select one best and one worst item from a given set of characteristics in the choice set (Louviere et al., 2015; Soto et al., 2016; Soto et al., 2018). The BWS generates more robust information (Soto et al., 2018) on attribute impact, i.e., mean utility of an attribute across all levels, and level scale values (LSV) (Soto et al., 2018). The binary choice question of BWC, also known as discrete choice experiment (DCE), ask consumers whether they want to purchase the product (Yes) or not (No). The DCE data from BWC can be used to estimate unconditional demand and WTP (Soto et al., 2018). The BWC, in addition to unconditional demand and WTP, similar to traditional DCE, generates best-worst ranking data with less cognitive burden (Flynn et al., 2007; Louviere et al., 2015; Soto et al., 2016).

#### 2.2.2: Attributes, Levels, and Choice Sets

Deer sanctuary, food plots, forest canopy cover, number of deer observed per visit, and lease fee per acre per year were used as attributes. Since each attribute had three levels, a balanced orthogonal main effect full factorial design resulted into 81 unique choice sets, each characterizing a deer hunting site. Eighteen out of 81 choice sets were selected and divided into three blocks (Mingie et al., 2017a) using SAS OPTEX Procedure (SAS Institute Inc., 2016). Each respondent observed six choices sets and can make best and worst choices among four levels in each choice set. A binary question, whether the respondent would lease the hunting site or not, was asked immediately following each choice set. The description of attributes, and the example choice set presented to responders in survey is provided in Table 8. Attributes, levels of each attribute and their respective keys are presented in the first column of Table 9.

Table 8: Attributes, their descriptions, and example choice set presented to responders in survey.

a) Deer sanctuary is a small area within a hunting site maintained to attract more deer. This small area is an ideal deer habitat and hunting is permitted but other disturbances are not.
b) Food plots are plots maintained inside deer hunting sites where deer preferred forages are planted and maintained to attract deer.
c) Forest canopy condition is how open the forest top cover (canopy) is. This can vary from closed canopy with little understory to open canopy with abundant understory.
d) Total number of deer observed per visit.
e) Hunting lease rate ($per acres per year).

Assume the current market lease rate is approximately $10 per acre per year and each hunting site size is about 160 acres or about a quarter the size of one square mile.

Q: In your opinion, what are the best attribute and the worst attribute of this hunting site?

| Best attribute of site (Check only one box) | Hunting site attributes | Worst attribute of site (Check only one box) |
|---|---|---|
| ☐ | Deer sanctuary with food plots | ☐ |
| ☐ | Closed canopy forest | ☐ |
| ☒ | 10 deer observed/visit | ☐ |
| ☐ | $16 hunting lease/acre/year | ☒ |

➔ Would you lease this hunting site? ☒ Yes ☐ No

2.3: Survey Administration

The mailed survey was conducted following the tailored design methods suggested by Dillman et al. (2014). A mailing list of landowners in Oklahoma owning about 160 acres (about 65 Hectare) or more land was obtained from a commercial vendor, Dynata (https://www.dynata.com/). From this group, the survey was bulk-mailed to 2,500 randomly selected Oklahoma landowners from forest-grassland transitional ecoregion. Of the 2500, 16 were unable to participate due to lack of accurate addresses, death, or out of business. The survey package included a personalized cover letter, questionnaire, and prepaid return envelope.

Two rounds of survey packages two months apart were mailed, and each was followed by reminder postcards after a month of mailing. A second round of survey and postcard were sent to only those landowners who failed to respond to first mailing. A total of 508 completed surveys were received resulting in a final response rate of 20.5%. The demographics of the landowners were compared with National Woodland Owner's Survey database (Caputo and Butler, 2021). Early (responses received before mailing second round of survey mailing) and late (responses received after mailing second round of survey) responses were compared to gauge potential non-response biases using chi-square tests on respondents' age, gender, income, education, and race.

2.4: Data Analysis

Best-worst scaling and DCE components of BWC were analyzed separately (Soto et al., 2016; Soto et al., 2018). They are discussed separately after describing coding of attributes and levels.

2.4.1: Coding of Attributes and Levels

The coding of independent variable does not depend upon the choices of respondent. They depend upon assignment of best and worst attributes in the pair among all available best-worst pairs in each choice set. Levels and attributes were effect coded (*Table 13*, Appendix A). The effect coded level receive -1, if the reference level is present in a choice set, it receives 1 if the effect coded level is present in the choice set, and 0 otherwise (Flynn et al., 2007; Soto et al., 2018). The effect of coding for attributes is different in comparison to levels of an attribute. The attribute, (Say *K*), receives 1 and -1 for all pairs in which the attribute (*K)* is picked as best and worst respectively, and 0 otherwise (Flynn et al., 2007; Soto et al., 2018). Effect coding for level variables was performed in two steps: 1) dummy variables were created for all levels of an attribute *K* and 2) the reference level was subtracted from non-reference dummy levels of the attribute *K* to obtain effect coded levels. The effect coded variables were correlated within the attribute but uncorrelated with the population mean (Flynn et al., 2007).

2.4.2: Best worst Scaling

Best Worst Scaling data can be analyzed using various models based on the underlying assumption regarding consumer's decision-making process on selecting the best and worst item in the given choice set (Flynn et al., 2007; Louviere et al., 2015). The paired model has lower standard error compared to other models (Flynn et al., 2007).

Paired model assumes that a consumer selects one best-worst pair among all available J(J-1) best and worst pairs. The selected pair represented the maximum difference in the latent scale utility deriving from the given choices (Flynn et al., 2007; Soto et al., 2016; Soto et al., 2018). The distance between attribute levels on a latent scale was proportional to the relative choice probability of a given pair (Flynn et al., 2007; Marley and Louviere, 2005).

The attribute impact was estimated by estimating all attributes levels in the same scale except for the base attribute (Flynn et al., 2007). The coefficient of effect coded attribute impact variable was interpreted as the mean utility of attribute across all levels (Flynn et al., 2007; Soto et al., 2018). The constant term in the model with effect coded attribute represents population mean, which is an average utility across the population (Flynn et al., 2007). The negative sign of a coefficient in the model signifies its relative position in comparison to the reference but it does not mean a negative relationship with the dependent variable (Soto et al., 2016). The coefficient of omitted level was negative sum of non-omitted levels of an attribute *K* (Flynn et al., 2007).

This paper used a paired model to analyze the data. There were 4*(4-1) = 12 pairs available for a respondent to choose in a choice set in this study. The choice variable, with values 1 for best-worst pair and 0 otherwise, is used as dependent variable in the BWS model. The BWS model was estimated using a random parameter logit model using maximizing simulated log likelihood function (Hole, 2007; Lusk and Briggeman, 2009; Soto et al., 2018) using *mixlogit* (Hole, 2007) in Stata 15.1. Standard errors of coefficients were estimated in the form of robust standard errors to avoid potential issue of heteroscedasticity (StataCorp, 2017). An attribute with least impact was used as reference attribute and coefficient value was assigned as zero (Lusk and Briggeman, 2009; Soto et al., 2016; Soto et al., 2018). The reference or base serve as the reference point in utility latent scale (Lusk and Briggeman, 2009; Soto et al., 2018).

2.4.3: Empirical Model for BWS:

The total utility of respondent (i) for selecting item $x_{tj}$ as best and $x_{tk}$ as worst pair, $(j, k \in M; j \neq k)$ in a choice set X of BWS profile t, is given by following utility equation:

$$U^{i}_{jkt} = \beta_j X^{i}_{tj} - \beta_k X^{i}_{tj} + \epsilon_{i_{tj}} - \epsilon^{i}_{tk} \quad \text{(Eq. 1)}$$

Where β are coefficients to estimate and ϵ are random errors.

The probability of choosing attribute levels $x_{tj}$ as best and $x_{tk}$ as worst fits into the multinomial logit (MNL) form with the following probability:

$$P_{BW}(j, k|X) = \frac{\exp(\beta_j x_{tj} - \beta_k x_{tk})}{\sum_{l,m \in M; l \neq m} \exp(\beta_l x_{tl} - \beta_m x_{tm})} \quad \text{(Eq. 2)}$$

The MNL logit assumes that all individual same level of utility on each level and thus derive utility based on given choices (Louviere et al., 2015; Lusk and Briggeman, 2009; Train, 2009). This is a stronger assumption as the choice of alternatives differs between individual and available substitutes in the choice set (McFadden and Train, 2000; Train, 2009). Respondent's selection of best and worst attribute and levels in a choice set is limited by the availability of attributes or levels in the choice set. This individual specific assumption is called as independent of irrelevant alternatives (IIA) assumptions. This research used mixed logit model to overcome two shortcomings of conditional logit: 1) conditional logit does not account preference heterogeneity among individuals for available choices and 2) it does not account for the IIA assumptions (Train, 2009) . The IIA assumption and preference of individual is taken into consideration by mixed logit model (Train, 2009) which is given by:

$$\beta_{ij} = \beta_j + \sigma_j u_{ij} \qquad \text{(Eq. 3)}$$

Where, $\beta_j$ is parameter coefficient with random deviation $\sigma_j u_{ij}$; $u_{ij} \sim N(0, \sigma_j)$ is an error term (McFadden and Train, 2000).

This research used mixed logit (also called random parameter logit) model (Train, 2009), mixed logit model with maximum simulated log likelihood to analyze the data (McFadden and Train, 2000; Soto et al., 2018; Train, 2009).

2.4.4: Discrete Choice (Binary) Task

The binary question in BWC gives conditional demand by eliciting information on preference of respondent on given profile compared to respondent's status quo situation (Flynn et al., 2007). The DCE part of BWC allows to compute WTP (Flynn et al., 2007). The binary part of BWC is analyzed using random effect Probit (REP) (Louviere et al., 2015) and random effect Logit (REL) models using maximum likelihood estimator and mean-variance adaptive Gauss-Hermite Quadrature integration method (StataCorp, 2017). These models assume that individual heterogeneity is uncorrelated with attributes of choices (Soto et al., 2018).

The binary question in the BWC is used as dependent variable. The dependent variable is coded as 1 for "yes" and 0 for "no" responses. Two REP and two REL models—each using effect coded dummy variables, and quantitative price variable (*chlease*) to compute marginal WTP—were estimated. Control variables were included in the model to control potential unobserved individual heterogeneity arising due to socioeconomic characteristics (Coast et al., 2006; Soto et al., 2016). Control variables used were whether responders have hunted in last five years or not (*a1hunted* = 1 if yes, 0 otherwise), is deer often their primary game (*a3pgame* = 1 if yes, 0 otherwise), do you lease land for hunting (*a4lease* = 1 if yes, 0 otherwise), those who rented their land (*c4rent* = 1 if yes, 0 otherwise), farmer/rancher (*c5farmer* = 1 if yes, 0 otherwise) and business (*c5business* = 1 if yes, 0 otherwise) as primary occupation. I used probit and logit models because the probit and logit models assume normal and logistic distribution of error terms (StataCorp, 2017). The results based on both RLP and REL revealed trivial differences. Therefore, the empirical findings based on REL model are only elaborated further Standard errors of coefficients are cluster-robust standard errors nested within respondent.

2.4.5: Empirical Model for DCE

The REP utility ($U_{it}$) equation for an individual i, for binary choice task t:

$$U_{it} = \beta' * X_{it} + \alpha_i + \epsilon_{it} \qquad \text{(Eq. 4)}$$

$$Y_{it} = 1\ [U_{it} > 0] \qquad \text{(Eq. 5)}$$

where, $\beta'$is a vector of estimated coefficients (marginal utilities) of attributes and levels of choice set ($X_{it}$), $\alpha_i$ is individual specific error, $\epsilon_{it}$ is uncorrelated error term, $Y_{it}$ is binary dependent variable with value 1 if condition in the equation 5 is met and )
otherwise (Soto et al., 2018).

The marginal WTP is calculated by taking the negative ratio of coefficient of non-price variable to that of price (lease) variable in the model. Negative marginal WTPs can be interpreted as discount in price for an attribute level to be acceptable (Habb and McConnell, 1997; Soto et al., 2018). The standard error of WTP is robust standard error calculated using delta method (Mingie et al., 2017a). The trade-off among levels within an attribute are calculated by taking the difference in WTPs between levels. The absolute value impact of levels within a attribute is determined by taking absolute value of the difference in WTPs between levels (Lancsar et al., 2007; Soto et al., 2016). The ranking of absolute value impact is proportional to the absolute difference of in WTPs (Soto et al., 2016).

## 3. Results and Discussion

### 3.1: Best Worst Scaling (BWS)

#### 3.1.1: Best Worst Count and Differences

Deer observed (44%) and lease fee (47%) received highest percentage of best and worst counts respectively (Table *9*). The best-worst difference score for deer observed per visit was 26% and that for lease fee was -37% indicating most preferred and least preferred attributes of deer hunting sites among survey responders. Among all available 12 levels, 10 deer observed was selected as best level by 21% and $16 lease fee was selected as the worst level by 21% of responders. The selection of highest number of deer observed and highest lease fee as the best and the worst attributes of deer hunting site

respectively from given choices were important indication that the low-cost (Mingie et al., 2017a) and high-quality (Eliason, 2020; Hammit et al., 1989; Lebel et al., 2012) deer hunting experience was preferred by our respondents.

Table 9: Deer hunting site characteristics (attributes and levels of choice set) preference among landowners.

| Attributes and Levels | Best (B) (%) | Worst (W) (%) | B -W (%) |
|---|---|---|---|
| **Deer sanctuary and food plot (*Sanctuary*)** | **28.22** | **23.31** | **4.91** |
| **No deer sanctuary no food plot (*Nsnfp*)** | 3.20 | 14.21 | - 11.01 |
| **Sanctuary without food plot (*Snfp*)** | 9.74 | 6.33 | 3.41 |
| **Sanctuary with food plots (*Sfp*)** | 15.28 | 2.77 | 12.51 |
| **Forest canopy cover (*Canopy*)** | **18.48** | **12.58** | **5.9** |
| **Open (*Open*)** | 4.76 | 2.91 | 1.85 |
| **Moderate (*Moderate*)** | 7.46 | 4.48 | 2.98 |
| **Close (*Close*)** | 6.26 | 5.19 | 1.07 |
| **Deer observed/visit (*Deer*)** | **43.57** | **17.41** | **26.16** |
| **1 (*Deer1*)** | 4.90 | 14.14 | - 9.24 |
| **6 (*Deer6*)** | 17.21 | 2.06 | 15.15 |
| **10 (*Deer10*)** | 21.46 | 1.21 | 20.25 |
| **Lease fee/year/acre (*Lease*)** | **9.73** | **46.70** | **- 36.97** |
| **$6 (*Lease6*)** | 5.97 | 11.02 | - 5.05 |
| **$10 (*Lease10*)** | 2.13 | 15.07 | - 12.94 |
| **$16 (*Lease16*)** | 1.63 | 20.61 | - 18.98 |
| **Total (N = 1407)** | **100** | **100** | **0** |

Note: Attributes and their respective keys are aligned to the left of the first column in the table and are bold-faced. Levels and their keys are aligned to the right of the first column in the table. Key for each attribute and level are *italicized* inside parenthesis. N represent total number of choice sets with selected best-worst pair.

Although less than one-third respondents chose deer sanctuary and food plot as a best choice, it had the lowest best-worst difference score (5%) indicating that responders are less divided on whether sanctuary is the best or the worst attribute. Among levels within deer sanctuary and food plots, however, deer hunting site with food plot was selected as the best characteristic by 15% and no deer sanctuary and no food plot was selected as the worst characteristic by about 14% responders. This means respondents favored a better-quality deer hunting site with the deer sanctuary and food plot.

All presented forest canopy cover in the level scale also received low counts as best and worst levels (6%) indicating canopy cover had less important characteristics of a deer hunting site. The best-worst difference is also smallest among available levels for all canopy covers. This result, however, does not account for heterogeneity in individual's preference. The result accounting individual's preference heterogeneity is further discussed based on the result of mixed logit model (Flynn et al., 2007; Soto et al., 2018).

3.1.2: Best Worst Scaling using Mixed Logit Model

The attribute impact and ranking result of BWS paired model analyzed using random parameter mixed logit model is summarized in Table 10. Deer sanctuary (*Sanctuary*) has the smallest impact thus selected as reference attribute and has zero as coefficient following BWS convention (Lusk and Briggeman, 2009; Soto et al., 2016; Soto et al., 2018). Number of deer observed per visit (*Deer*) and lease fee (*Lease*) were significant respectively at 1% significance level. Canopy cover (*Canopy*) was not significant. The sign of coefficients signifies relative position of impact in the latent scale with respect to base category (Soto et al., 2016). The number of deer observed per visit and lease fee ranked highest and lowest in the latent scale of utility respectively. These results, consistent with the previous results (Eliason, 2020; Hammit et al., 1989), suggest that deer hunters often desire reasonable chances to view, pursue, and hunt deer to have high quality hunting experience. Hunters prefer deer habitat with open and low vegetative cover with high visibility for hunting (Brinkman et al., 2009). Previous research suggested that number of deer observed and smaller membership dues have positive impact in the deer hunting experience (Gruntorad et al., 2020; Hammit et al., 1989; Mingie et al., 2017a) and satisfaction (Lebel et al., 2012; Miller and Graefe, 2001).

Table 10: Best worst scaling of best worst choice model using mixed logit model and standard deviation of random parameters in the model.

| BWS Pair | Coeff. (Std. Err.)[Rank] | St. Dev. (Std. Err.)[Rank] |
|---|---|---|
| **Attributes Impact and Ranking** | | |
| **Deer sanctuary** [a] | 0.00 [2] | Reference |
| **Canopy cover** | 0.09 (0.09) [3] | 0.53 *** (0.16) |
| **Deer observed** | 0.60 *** (0.14) [4] | 1.31 *** (0.14) |
| **Lease fee** | - 1.46 *** (0.21) [1] | 1.65 *** (0.16) |
| **Levels Scale Impact Values and Ranking** | | |
| **No sanctuary no food plot** [a] | - 1.52 *** (0.13) [2] | Reference Level |
| **Sanctuary without food plot** | 0.23 ** (0.09) [8] | 0.44 ** (0.17) |
| **Sanctuary with food plot** | 1.29 *** (0.13) [11] | 1.10 *** (0.19) |
| **Open canopy [a]** | - 0.01 (0.07) [5] | Reference |
| **Moderate canopy** | 0.19 ** (0.08) [7] | 0.11 (0.07) |
| **Close canopy** | - 0.18 ** (0.08) [4] | 0.07 (0.15) |
| **1 deer observed [a]** | - 2.06 *** (0.13) [1] | Reference |
| **6 deer observed** | 0.66 *** (0.09) [9] | 0.13 (0.19) |
| **10 deer observed** | 1.40 *** (0.11) [12] | 0.90 *** (0.20) |
| **$6 lease fee [a]** | 0.83 *** (0.11) [10] | Reference |
| **$10 lease fee** | 0.02 (0.08) [6] | 0.06 (0.11) |
| **$16 lease fee** | - 0.84 *** (0.11) [3] | 0.90 *** (0.157) |
| ***Number of choices*** | 1407 | |
| ***Log likelihood*** | - 2545.6 | |
| ***Wald chi2(11)*** | 376.0 *** | |
| ***N*** | 16884 | |

**Notes:**

Mean Coeff. = $\beta$ coefficient from regression models.

St. Er. = Standard Error. SD = Standard Deviation of Mean

$^{*}$ *p-value* < 0.10, $^{**}$ *p-value* < 0.05, $^{***}$ *p-value* < 0.001.

N denotes total number of observations used in the model but not the number of respondents.

a = Base levels and attributes in regression models. The mean coefficient or additional utility of base level from the mean is equal to the negative sum of non-omitted levels of an attribute (Flynn et al., 2007).

The number inside [ ] bracket represent ranking of attributes and levels. They are ranked separately. Attributes are ranked from 1 (lowest) to 4 (highest) and levels are ranked 1 (lowest) to 12 (highest).

The level scale impact value results from mixed logit model was also summarized in Table 10. The coefficient of reference level was the negative sum of non-omitted levels of an attribute (Flynn et al., 2007). Deer sanctuary with food plots (*Sfp*), deer sanctuary without food plots (*Snfp*), 6 deer observed per visit (*Deer6*), and 10 deer observed per visit (*Deer10*) was significant at 99% confidence level. Moderate canopy cover was significant at 95% confidence level with positive sign. Close canopy cover was significant at 95% confidence level with negative sign. $16 lease fee was significant at 99% confidence interval but with negative sign.

Ten deer observed per visit was ranked as the best characteristic of deer hunting site followed by deer sanctuary with food plot, and $6 lease fee per acres. One deer observed per visit was ranked as the worst characteristics of deer hunting site followed by no deer sanctuary no food plots, and $16 lease fee per acre. After considering individual heterogeneity $16 lease fee was no longer the worst characteristic of deer hunting site, but one deer observed per visit was the worst characteristics. As pervious research suggests (Hammit, et al., 1989, Lebel, et al., 2012), probability of observing large number of deer provides better hunting experience and higher satisfaction to deer hunters (Hammit et al., 1989; Lebel et al., 2012). Therefore, deer hunters want management agencies to increase deer population for quality hunting experience (Hammit et al., 1989). Likewise, deer hunters were willing to pay slightly higher in lease fee with the quality deer hunting experience. Nonetheless, hunters carefully select their deer hunting site to increase harvest success (Hammit et al., 1989).

### 3.2: Binary Choice Model and Willingness to Pay

#### 3.2.1: DCE and Marginal WTP

The result of logit and probit models with quantitative lease were the same in terms of significant variables and retained signs (Table 11) suggesting the reliability of statistical analysis. In both effect coded REL and REP models, compared to one deer observed per year, six and 10 deer observed per visit are positive and significantly at 99% confidence level determine landowner's decision to lease a hunting site. Likewise, compared to the base category of $6 lease fee, $16 lease fee is negative but significant at 99% confidence level, meaning that higher lease fee will deter landowner's inclination towards hunting lease. Deer Sanctuary with food plot, compared to no deer sanctuary and no food plot is positive and significant. This means presence of deer sanctuary with food plot will motivate landowner to lease the hunting site. Canopies are not significant in both models.

Demographic control variables included in the random effect logit and probit models except *a1hunted* were significant and positive; *a1hunted* was non-significant and negative. Though the past hunting experience in this research was not significant, the negative sign could be because of lack of good quality game (Hussain et al., 2004) The respondents who rented out their land has also the highest WTP followed by those who leased their land for hunting and targeting deer as their primary game species.

Table 11: Model summary of binary (DCE) part of best-worst choice.

| Binary Choice | Random Effect Logit Model | | | Random Effect Probit Model | | |
|---|---|---|---|---|---|---|
| | Effect Coded Attributes | Quantitative Lease (Price) | | Effect Coded Attributes | Quantitative Lease (Price) | |
| | Coeff. (Std Err.) | Coeff. (Std Err.) | WTP (Std Err.) | Coeff. (Std Err.) | Coeff. (Std Err.) | WTP (Std Err.) |
| **_cons** | -5.08 *** (1.32) | -2.38 * (1.31) | N/A | -2.82 *** (0.74) | -1.33 * (0.73) | N/A |
| **Nsnfp** [a] | - 0.39 | - 0.37 | -1.48 (0.73) | - 0.23 | - 0.22 | -1.55 (0.73) |
| **Snfp** | -0.13 (0.15) | -0.12 (0.15) | -0.45 (0.59) | -0.07 (0.08) | -0.06 (0.08) | -0.43 (0.60) |
| **Sfp** | 0.52 ** (0.16) | 0.49 ** (0.16) | 1.93 (0.70) | 0.29 *** (0.09) | 0.27 ** (0.09) | 1.97 (0.70) |
| **Open** [a] | 0.073 | 0.06 | 0.25 (0.60) | 0.05 | 0.04 | 0.32 (0.60) |
| **Moderate** | 0.04 (0.14) | 0.03 (0.14) | 0.12 (0.57) | 0.02 (0.08) | 0.01 (0.08) | 0.06 (0.57) |
| **Close** | -0.11 (0.15) | -0.09 (0.14) | -0.37 (0.58) | -0.06 (0.08) | -0.05 (0.08) | -0.38 (0.59) |
| **Deer1** [a] | - 1.72 | - 1.70 | -6.75 (1.20) | - 0.96 | - 0.94 | -6.81 (1.20) |
| **Deer6** | 0.64 *** (0.18) | 0.62 *** (0.18) | 2.47 (0.83) | 0.35 *** (0.10) | 0.34 *** (0.10) | 2.48 (0.83) |
| **Deer10** | 1.09 *** (0.20) | 1.07 *** (0.20) | 4.28 (0.79) | 0.61 *** (0.11) | 0.60 *** (0.11) | 4.33 (0.80) |
| **Lease6** [a] | 1.26 | N/A | N/A | 0.70 | N/A | N/A |
| **Lease10** | -0.000390 (0.16) | N/A | N/A | - 0.01 (0.09) | N/A | N/A |
| **Lease16** | -1.26 *** (0.225) | N/A | N/A | - 0.69 *** (0.12) | N/A | N/A |
| **A1hunted** | - 0.92 (1.35) | -0.89 (1.34) | -3.53 (5.33) | - 0.52 (0.76) | - 0.51 (0.75) | -3.66 (5.44) |
| **A3pgame** | 2.29*** (0.68) | 2.27 *** (0.67) | 9.03 (2.94) | 1.28 *** (0.38) | 1.27 *** (0.38) | 9.16 (3.014) |
| **A4lease** | 3.81*** (0.73) | 3.80 *** (0.72) | 15.13 (3.27) | 2.11*** (0.40) | 2.11*** (0.40) | 15.22 (3.30) |
| **C4rent** | 4.65** (1.57) | 4.67** (1.57) | 18.58 (6.02) | 2.52** (0.89) | 2.53** (0.89) | 18.25 (6.21) |
| **F5farmer** | 1.38** (0.57) | 1.37** (0.57) | 5.46 (2.22) | 5.470.76 ** (0.32) | 0.76 ** (0.32) | 5.45 (2.26) |
| **F5business** | 1.34* (0.72) | 1.31* (0.71) | 5.22 (2.85) | 0.72 * (0.399) | 0.71* (0.40) | 5.11 (2.89) |
| **lnsig2u** | 2.08*** (0.23) | 2.07 *** (0.22) | | 0.91 *** (0.22) | 0.89 *** (0.22) | |
| ***AIC*** | 827.1 | 826.2 | | 827.5 | 826.8 | |
| ***BIC*** | 907.2 | 901.3 | | 907.7 | 901.9 | |
| **ll** | -397.6 | -398.1 | | - 397.8 | -398.4 | |
| **ll_c** | -495.7 | -495.9 | | - 495.0 | -495.4 | |
| **chi2** | 99.58 | 95.10 | | 108.7 | 103.8 | |
| **chi2_c** | 196.2 | 195.6 | | 194.4 | 193.9 | |
| **sigma_u** | 2.828 | 2.809 | | 1.575 | 1.564 | |
| **rho** | 0.709 | 0.706 | | 0.713 | 0.710 | |
| ***N*** | 1105 | 1105 | 1105 | 1105 | 1105 | 1105 |

Willingness to pay was calculated from REL and REP model with the quantitative lease variable (Table 11). The highest WTP is for 10 deer observed per visit followed by six deer observed per visit and deer sanctuary with food plots in both models. The negative sign of some WTP estimates reveal that landowners indeed expect compensation to accept those outcomes. To this end, highest negative WTP is for 1 deer observed per visit and no deer sanctuary and no food plot in both models. Sign of WTP values consistent in both models though WTP values vary slightly.

Results show that the number of deer and availability of deer food at the site are valuable characteristics of deer hunting sites. Higher number of deer observed and better-quality hunting site provides quality deer hunting experience and better deer hunting opportunities (Hammit et al., 1989). The deer number can be increased by maintaining quality deer habitat. Deer sanctuary with food plots is an example of a good deer habitat due to availability of deer food and shelter (Fulbright and Ortega-S, 2013).

3.2.2: Difference in Marginal WTP from Binary Models

The absolute value impact ranking, also called the absolute value of difference in marginal WTP between two levels (Soto et al., 2016), is summarized in Table 12. The absolute value ranking is same for REP and REL models. The absolute value impact ranking shows that 1 deer to 10 deer has the highest marginal WTP followed by 1 deer to 6 deer and no deer sanctuary no food plot to deer sanctuary with food plot. This result is consistently aligned with better satisfaction received from high quality deer hunting experience and possibility of observing higher number of deer, which increases the possibility of hunting a deer and thus higher WTP. Canopy cover types ranked low in marginal WTP differences indicating that they have smaller influence in comparison to other attributes in a quality deer hunting experience.

Table 12: Difference in marginal WTP from DCE binary models and absolute value impact ranking

| Marginal WTP Differences (“From” to “To”) | RP Logit Model WTP ($) (Std. Err.) | RP Probit Model WTP ($) (Std. Err.) | Absolute Value Impact Ranking |
|---|---|---|---|
| **Nsnfp to Snfp** | 1.04 (1.13) | 1.13 (1.14) | 4 |
| **Nsnfp to Sfp** | 3.41 $^{**}$ (1.30) | 3.52 $^{**}$ (1.30) | 7 |
| **Snfp to Sfp** | 2.38 $^{**}$ (1.07) | 2.399 $^{**}$ (1.08) | 6 |
| **Open to Medium Canopy** | - 0.12 (1.01) | - 0.259 (1.01) | 1 |
| **Open to Close Canopy** | - 0.612(1.03) | - 0.70 (1.05) | 3 |
| **Medium to Close Canopy** | - 0.50 (0.98) | - 0.44 (0.99) | 2 |
| **1 to 6 Deer** | 9.22 $^{***}$ (1.91) | 9.29 $^{***}$ (1.91) | 8 |
| **1 to 10 Deer** | 11.02 $^{***}$ (1.85) | 11.15 $^{***}$ (1.86) | 9 |
| **6 to 10 Deer** | 1.80 $^{*}$ (1.08) | 1.86 $^{*}$ (1.10) | 5 |

**Note:** Absolute value impact ranking is based on absolute positive value of levels from binary random Logit and Probit models (DCE part of BWC). Rank: 9 = Highest and 1 = Lowest Ranked Levels.

The difference in marginal WTPs suggests the trade-off in WTP between two levels of an attribute (Soto et al., 2016). Landowners’ WTP compensation to switch from lower to higher level of an attribute and order of impact from the models with the quantitative lease are summarized in Table 12. The WTP compensation to go from deer hunting site with deer sanctuary but without food plot to that with deer sanctuary and food plot (*Snfp to Sfp*) is higher than going from hunting site without deer sanctuary and food plot to that with deer sanctuary without food plot (*Nsnfp  to Nsfp*). This indicates that food plots might have added more value to deer hunting site in comparison to deer sanctuary. However, going from deer hunting site without deer sanctuary and food plot to site with deer sanctuary and food plot (*Nsnfp  to Sfp*) yields the highest WTP. So, having deer sanctuary and food plots in deer hunting sites can provide higher perceived value to landowners rather than having either of them.

The WTP compensation to go from 1 deer to 6 deer (~$9) is much higher than going from 6 deer to 10 deer (~$2) suggesting a diminishing marginal utility (Table 12). The number of  deer observed contributes towards satisfaction in deer hunting and deer hunting experiences (Gruntorad et al., 2020; Hammit et al., 1989), however the degree of satisfaction seems to diminish.  Based on the finding of this research, constructing food plots and deer sanctuary to attract deer in the hunting site and provide an opportunity for deer hunters to observe between 6 to 10 deer per visit could be the best management strategy for landowners to optimize their economic return.

The difference in marginal WTP for going from open canopy to medium canopy, open canopy to close canopy, and medium canopy to close canopy are negative (Table 12). Alternatively, going from closed and medium canopies forest to open canopy generate higher WTP from deer hunting. The habitat types with less visual obstruction gives better opportunities for hunter to observe deer (Lebel et al., 2012). In particular, the open canopy provides a better visibility and an opportunity for the herbaceous forage growth in comparison to medium and closed canopy forests.  The dispersal distance of juvenile male deer is also greater with less forest cover (Long et al., 2005).

Therefore, an opportunity to observing more deer would have contributed to the satisfaction of deer hunters (Lebel et al., 2012; Miller and Graefe, 2001). Accordingly, this would have motivated hunters to express higher WTP for open canopy forests.

The trade-offs between switching forest canopy, as revealed from foregone marginal WTP and the statistical insignificance, is relatively small. These findings, consistent with the previous landowner research, suggest that landowners generally manage forestland for multiple objectives (Bulter et al., 2021). Open canopy cover will less woody plants and more forbs and other browse is considered as a better habitat for deer (Fulbright and Ortega-S, 2013). Open to medium canopy forest are considered as a good habitat for Turkey (Holbrook et al., 1987; Pollentier et al., 2017). Since canopy cover has non-significant and small influence on WTP, integration of cattle grazing after careful planning (DelCurto et al., 2005; Vavra, 2005), timber production, and managing land for multi-species hunting could generate additional revenue beside deer hunting.

More frequent deer sightings means extensive use of habitat by deer, which is more likely happen in a property having better habitat quality (Fulbright and Ortega-S, 2013; Stewart et al., 2000). This would increase with the increase in high quality and nutritional food, increased forage availability (Lebel et al., 2012), and would decrease as plants mature (Stewart et al., 2000). Therefore, creating openings in mature and regenerating forest could increase hunting efficiency of deer hunters (Lebel et al., 2012).

Study results reveal some important management implications. For example, landowners are willing to pay higher premium for the habitat that provides higher deer observations. As previous research suggests (Adhikari et al., 2021a; Masters et al., 2006), prescribed fire is a cost-effective active management tool that reduces canopy cover and increases forb and woody browse in the understory which is better habitat for deer (Fulbright and Ortega-S, 2013). This suggests lack of awareness among landowners on the management prescriptions that are likely to provide the preferred management outcomes. Therefore, education and outreach on active forest management, overall costs, and cost effectiveness can help landowners to achieve their management goals.
Economic incentives with active forest management are limited in the study region (Joshi et al. 2019). Since food plots and deer sanctuary can improve deer hunting habitat, which can serve as a mechanism to generate additional revenues by increasing demand for lease hunting. Food plots are often expensive to establish and maintain. However, forage produced in the food plots often has significantly more benefit compared to food supplement (Pittman, 2019) and is preferred by deer (McQueen, 2020). Food plots maintained using prescribed fire are more cost effective and bring additional positive ecological benefits (Pittman, 2019). Food plots reduces the cost of production by reducing the money spent in food palates (McQueen, 2020). Landowners can further reduce the cost of management by reducing plantation frequency and relying on previous year plantation to voluntarily supply deer forage yet, receive same amount of forage production (Pittman, 2019). Food plots need to be carefully established to ensure that anthropogenic provision of food does not harbor mycotoxin contaminants, which may cause severe negative effects on wildlife health (Murray et al., 2016).

**4. Conclusions**

This research found that the number of deer observed is the most important and valued characteristics of deer hunting site followed by site quality. The respondents identified 10 deer observed per visit as the best attribute and 1 deer observed per visit as the worst attribute of the deer hunting site. A deer habitat with deer sanctuary and food plots is regarded as better deer habitat in comparison to the one without sanctuary and food plots. Actively managing land for habitat improvement, increase in forage production and its quality, improved visibility for deer hunters by maintaining diverse canopy cover are vital in improving revenue from deer hunting.

Study results have important management implications and suggest the need for appropriate outreach. The wildlife management services are continuously trying to improve the deer hunting site quality to maintain quality deer population and attract more deer. This finding of this research is also beneficial for wildlife agencies in south central USA to optimize deer habitat quality by adding hunting site characteristics that generate higher return and removing site attributes that does not adds to the economic return. For example, adding food plots could enhance the deer hunting site quality in which could attract more deer hunters in public land.

**Appendix III.A**

Table 13: Effect coding of variables

| Attribute and Level | Effect Coding | | |
|---|---|---|---|
| **Deer Sanctuary** | **Nsnfp (Base)** | **Snfp** | **Sfp** |
| Nsnfp | 1 | -1 | -1 |
| Snfp | 0 | 1 | 0 |
| Sfp | 0 | 0 | 1 |
| **Forest Canopy** | **Open (Base)** | **Moderate** | **Close** |
| Open | 1 | -1 | -1 |
| Moderate | 0 | 1 | 0 |
| Close | 0 | 0 | 1 |
| **Number of Deer** | **1 (Base)** | **6** | **10** |
| 1 | 1 | -1 | -1 |
| 6 | 0 | 1 | 0 |
| 10 | 0 | 0 | 1 |
| **Lease** | **6 (Base)** | **10** | **16** |
| $6 | 1 | -1 | -1 |
| $10 | 0 | 1 | 0 |
| $16 | 0 | 0 | 1 |

Table 14: Attributes and levels used in choice set

| Attributes | Levels |
|---|---|
| **Deer Sanctuary and food plots (Sanctuary)** | No deer sanctuary no food plots (Nsnfp)<br>Deer sanctuary without food plots (Snfp)<br>Deer Sanctuary with food plots (Sfp) |
| **Forest canopy cover (Canopy)** | Open canopy cover (Open)<br>Moderate canopy cover (Moderate)<br>Closed canopy cover (Close) |
| **Number of Deer observed per visit (Deer)** | 1 (Deer1)<br>6 (Deer6)<br>10 (Deer10) |
| **Hunting lease per acres per year (Lease)** | $6 (Lease6)<br>$10 (Lease10)<br>$16 (Lease16) |

# CHAPTER IV: INTENTIONS OF LANDOWNERS TOWARDS ACTIVE MANAGEMENT FOR WHITE-TAILED DEER HUNTING IN THE FOREST-GRASSLAND TRANSITIONAL ECOREGION OF THE SOUTH-CENTRAL USA

**Abstract**

The forest-grassland ecotone of the south-central USA is a mixture of forest, savanna, and tallgrass prairie. The ecotone is increasing in woody plant dominance due to the exclusion of fire and other anthropogenic factors. Active management, such as prescribed fire and thinning, can restore savanna and prairie ecosystem to maintain a full suite of ecosystem services and improves suitable habitat for wildlife such as white-tailed deer (*Odocoileus virginianus*). Active management, however, comes with the cost of management and acceptance of management tools. Deer hunting is a vital source of revenue generation to offset the landowner's management cost in the region. We studied Oklahoma landowners' perceptions regarding active and sustainable management of forest and rangeland for deer habitat using two established theories of reasoned action and planned behavior as well as expanded theories adding moral norms. We analyzed mailed survey data using structural equation modeling. We found that subjective norms and perceived behavior control significantly affected deer hunting intention when moral norms were introduced into the model. Attitudes independently and significantly affected intentions of deer hunting but were negatively correlated with the intentions. This study suggests that landowners have positive social pressure and are interested towards active management but associated financial burden and risk could be shaping negative attitudes.

## 1. Introduction

The forest-grassland ecotone of the south-central USA, originally spanning from southern Illinois to Texas, consists of a mosaic of upland forests, savanna, and tallgrass prairie (Hallgren et al., 2012; Joshi et al., 2019b). The ecotone regularly experiences disturbances such as drought (Clark et al., 2007; Hallgren et al., 2012), invasive species (Joshi et al., 2019b; Starr et al., 2019), anthropogenic factors such as urbanization (Joshi et al., 2019b), and climate change (Joshi et al., 2019b; Starr et al., 2019). The transitional nature of the ecoregion makes it more vulnerable to climate change (Füssel, 2007) and changing management regimes can cause large changes in land cover as evidenced by shifting vegetation dynamics towards closed-canopy forests (Hoff et al., 2018a; Hoff et al., 2018b; Joshi et al., 2019b). Fire exclusion leads to an increase in forest cover, abundance of mesophotic, fire-sensitive hardwoods species, and fire-sensitive eastern redcedar (ERC; (*Juniperus virginiana*) which potentially increase wildfire risk (Hoff et al., 2018b).

Active management using prescribed fire and thinning are important tools to restore this region's characteristic features, increase resiliency against changing climate, and build self-sustaining ecosystem (Clark et al., 2007; Joshi et al., 2019a; Starr et al., 2019). Management costs and potential liabilities from using fire, however, have restricted the application of active management tools (Starr et al., 2019). Previous research in this region suggests that healthy and resilient forests provide an opportunity to increase revenue which in turn drives active management (Joshi et al., 2019b; Starr et al., 2019). For example, prescribed burning in combination with herbicides can be used to create open forest structure and more diverse seasonal forage for white-tailed deer (*Odocoileus virginianus*) (Jones et al., 2009; Leslie Jr. et al., 1996). Given the context of declined production of quality timber due to encroachment of less valuable timber species (Riddle, 2019) such as ERC (Clark et al., 2007; Kaur et al., 2020) and large economic benefit of wildlife management at the local and state level (Poudyal et al., 2020), wildlife management can be an important motivation for active management of forested and non-forested ecosystems in this region.

Among wildlife management activities, deer hunting is an important cultural tradition transferred from generation to generation (Byrd et al., 2017; Denevan, 1992; Lovell et al., 2004; Mann, 2002). Deer hunting is used to harvest game (Byrd et al., 2017; Hrubes et al., 2001) and provides psychological (Hrubes et al., 2001), social (Byrd et al., 2017; Hrubes et al., 2001), emotional, mental, and physical (Hrubes et al., 2001) benefits to the hunters. Deer hunting also provides important economic benefits and serves as a vital wildlife management tool (Byrd et al., 2017; Peterson, 2004). This study, thus, addresses the question: how do landowners' beliefs, attitudes, and norms about sustainable management of forest and rangeland for deer habitat impact their intentions for the active management of ecosystems for deer hunting.

Theory of reasoned action (TRA) and theory of planned behavior (TPB) were used for this research. TRA and TPB are widely used in comparative study (Daigle et al., 2002; Hrubes et al., 2001; Rossi and Armstrong, 1999) and were specifically applied to deer hunting (Daigle et al., 2002), willingness to pay (Lopez-Mosquera et al., 2014), and several other purposes summarized by (Ajzen, 2011). No previous study, however, to the best of our knowledge, used TRA and TPB to study landowners' intentions towards active management of forest and rangeland for deer habitat management and expanded these theories by including moral norms.

This research contributes to existing knowledge in three ways. First, this research studied landowners' intentions of applying active management for deer hunting because landowners can generate revenue from hunting leases by actively managing their land to improve deer habitat. Second, limited social science research explored the interrelationship between values, attitudes, norms, and behavior intentions in this grassland-forestland tension zone of the United States; TRA and TPB were used to study inter-relationships between above mentioned human psychology. Third, TRA and TPB were expanded by adding moral norms into both theories as suggested by Ajzen (1991). Past researchers highlighted the importance of improvement, refinement, and modifications (Lopez-Mosquera et al., 2014; Miller, 2017) of these theories by adding new predictors, testing concepts and models, and merging theories with additional attributes (Miller, 2017). Moral norms affect subjective norms and perceived behavior control of an individual (Heidari et al., 2018; Lopez-Mosquera et al., 2014). This paper, thus, tested four models—two theories with, and without moral norms— to study intentions towards active management of forest, and rangeland for deer habitat management.

## 2. Methods

2.1 Theoretical framework: theory of reasoned action (TRA) and theory of planned behavior (TPB)

Theory of reasoned action proposes that human intention is an immediate precursor to an action. The action originates from a belief that performing an activity leads to the intended outcome (Madden et al., 1992), assuming the action is under the volitional control of an individual. The theory, however, did not account for an action that the individual intends to perform but is not under the actual control (i.e., volitional control) of the individual (Ajzen, 2002). This limitation involving volitional control is addressed in TBA by adding perceived behavioral control as one of the factors affecting the behavioral intention of an individual (Fishbein and Ajzen, 1975; Madden et al., 1992; Rossi and Armstrong, 1999). Theory of planned behavior, thus, can be understood as the addition of perceived behavioral control to TRA such that TPB reduces to TRA when the behavior is under the volitional control (Ajzen, 2020). The theoretical models (Figure 6) for this paper were adopted from (Ajzen, 1991) and (Madden et al., 1992). The solid and dotted line referred to as direct and indirect links between two components the theory.

The theory of planned behavior assumes that a belief towards an action shapes a person's attitude and norms which further shapes intentions toward the action. Positive beliefs, attitudes, norms, and intentions towards an action motivates an individual to perform a given action (Ajzen, 1991, 2002, 2011). The beliefs can be categorized into behavioral beliefs, normative beliefs, and control beliefs. Behavioral beliefs originate from the experience of an individual while performing an action which shapes a person's attitude towards an action. Normative beliefs originate from social standards, values, norms, and pressure which shapes subjective norms of an individual. Control beliefs shape perceived behavioral control which is a perception of an individual that action is under the volitional control of the individual (Ajzen, 2002). This research was designed assuming that landowners used their beliefs to form attitudes, subjective norms, perceived behavior, and intentions while responding to respective survey questions.

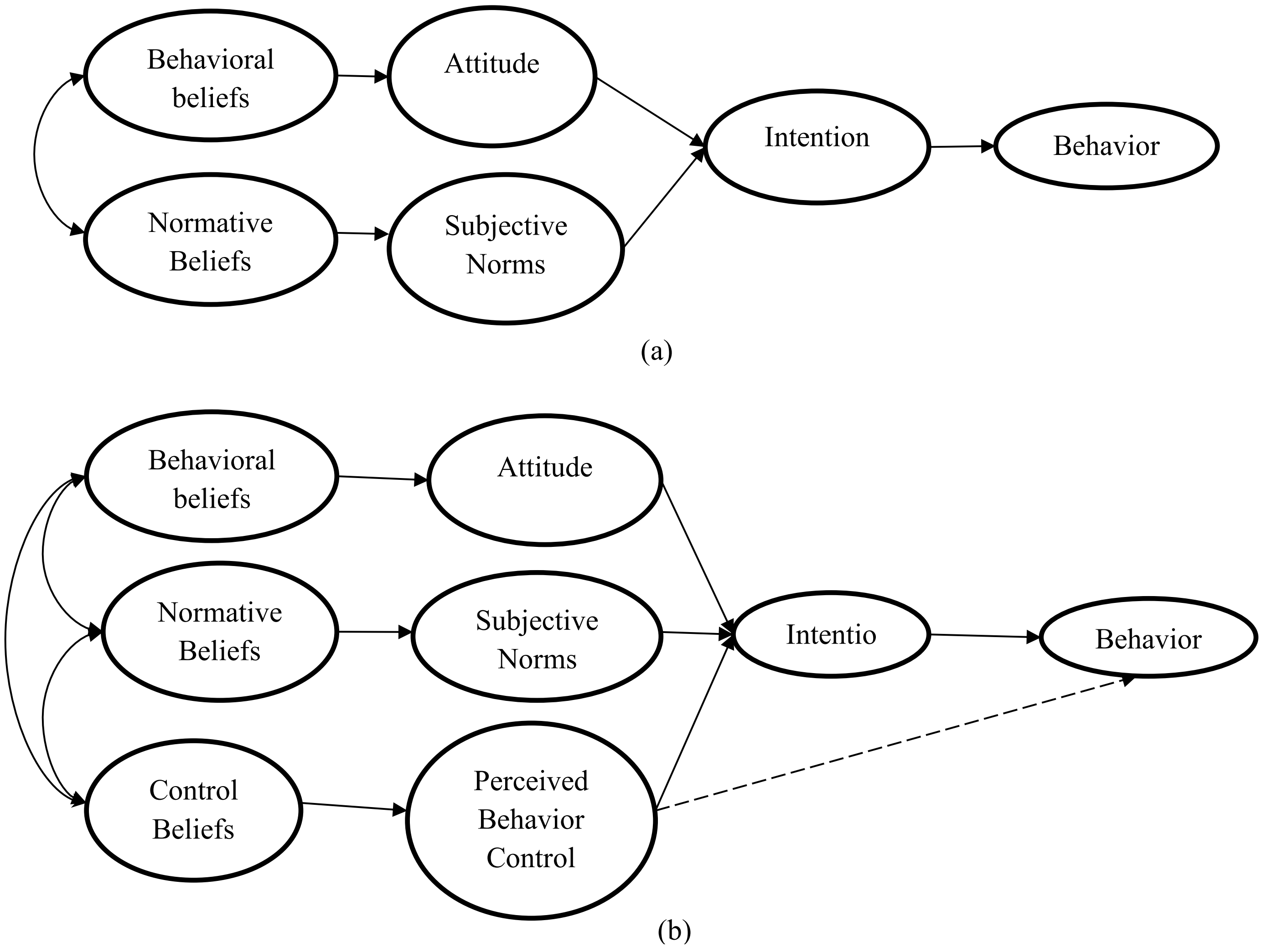

Figure 6: (a) Theory of reasoned action and (b) Theory of planned behavior.

## 2.2 Survey Design and Administration

The mailed survey was conducted following the method suggested by Dillman et al. (2014). The study area represented a portion of the transitional ecoregion of south-central USA in Oklahoma (Figure 7). A mailing list of landowners in Oklahoma owning 160 acres (65 ha) or more land with forest and rangeland was obtained from a commercial vendor, Dynata ([https://www.dynata.com/](https://www.dynata.com/)). The survey was bulk-mailed to 2,500 randomly selected Oklahoma landowners from forest-grassland transitional ecoregion. The survey package included a personalized cover letter, questionnaire, and prepaid return envelope. Out of selected landowner, 16 did not participate because of various reasons such as deceased, refused to participate, no longer in business, and inaccurate location identifier, reducing total sample to 2,484.

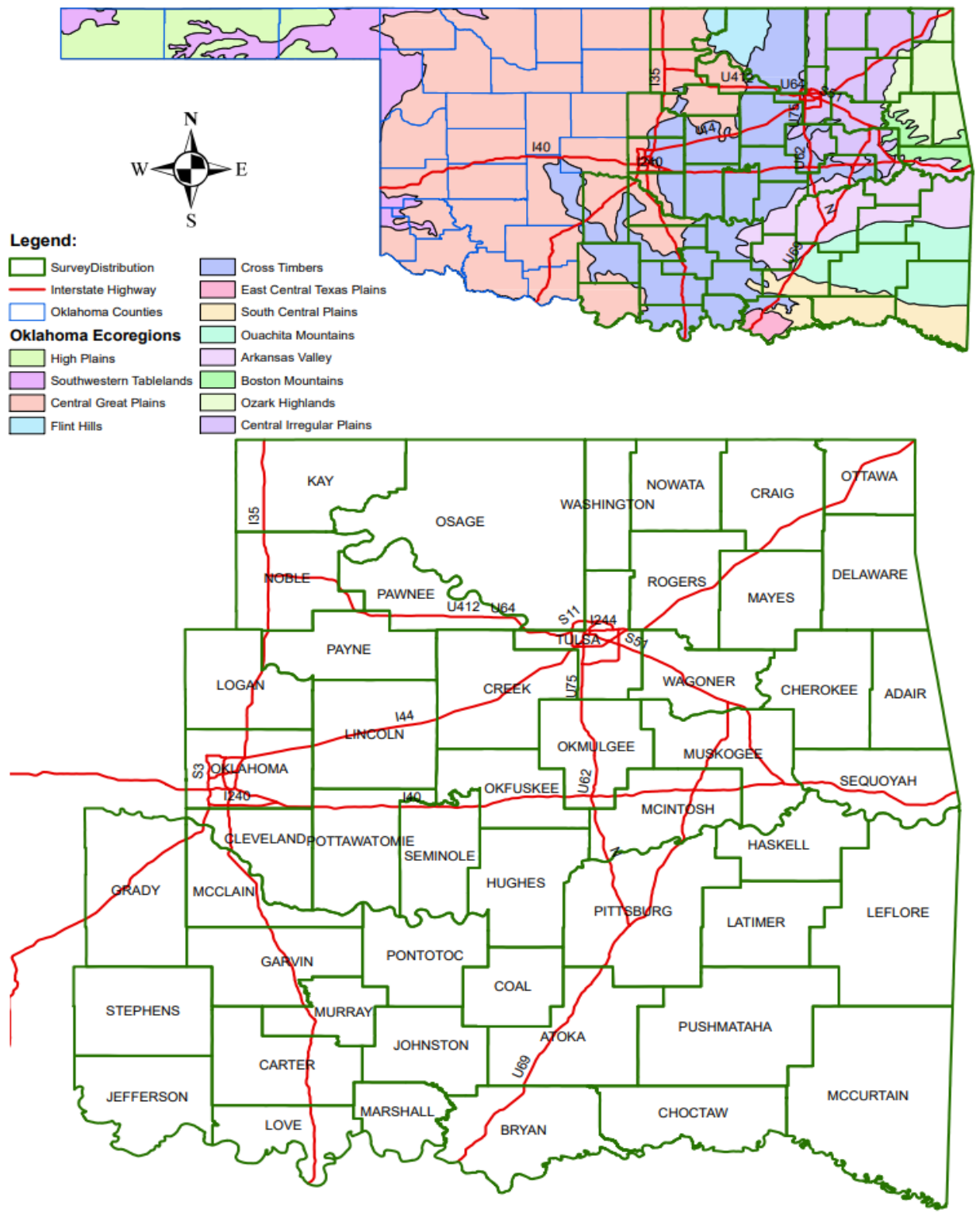

Figure 7: Study region (counties) of Oklahoma

Two rounds of surveys with a gap of about two months, each followed by reminder postcards after about a month of survey mailing, were sent to the randomly selected landowners. A second round of survey and postcards were sent to only those landowners who did not respond during the first round of mailing. A total of 508 completed surveys were received from the two lots of mailing and postcard reminders with the final response rate of 20.45%. The demographics of the landowners were compared with National Woodland Owner's Survey database (Caputo and Butler, 2021). Early and late response biases were conducted using chi-square tests on age, gender, income, education, and race among landowners' response received after first and second lots of survey and postcards.

The questions were asked as 5-point Likert scale (1 as strongly disagree to 5 as strongly agree) for all variables except those representing intentions. Intentions were asked as landowners' willingness to pay (USD), travel distance (miles) to alternate hunting site with similar quality, and interest (yes/no) in active management of their land. The observed variables loaded as intentions in the model were normalized by dividing the difference of the mean and observed value for each observation by the standard deviation of the variable (Urbano et al., 2019) because of difference in measurement scale. Mean and standard deviation before standardization were reported for all standardized and non-standardized variables. Cronbach alpha values were obtained after standardization for standardized variables because standardized variables are used in structural equation models (SEM). Missing observations were removed on a list-wise basis resulting into 177 observations to use in the SEM models, calculating fit indices, factor loadings, Cronbach alpha values, mean, and standard deviation of observed variables.

## 2.3 Hypothesis

We tested the following hypotheses related to the TRA, TPB, and moral norms regarding active management of forest and rangeland for deer habitat:

Hypothesis 1 ($H_1$): Positive subjective norms shape positive intentions.

Hypothesis 2 ($H_2$): Positive attitude shape positive intentions.

Hypothesis 3 ($H_3$): Positive attitudes shape positive moral norms.

Hypothesis 4 ($H_4$): Positive perceived behavior control shape positive intentions.

Hypothesis 5 ($H_5$): Positive subjective norms shape positive moral norms.

Hypothesis 6 ($H_6$): Positive perceived behavior control shape positive moral norms.

Hypothesis 7 ($H_7$): Positive moral norms shape positive intentions.

## 2.4 Structural Equation Model (SEM)

### 2.4.1 Model Fit Indices and Internal Validity

The internal validity of measurement variables was determined using Cronbach alpha. A Cronbach alpha value above 0.60 (Coon et al., 2020; Cronbach, 1951) was used as an indicator of internal consistency of variables loading in the latent constructs. The model fit indicators were determined by using several model fit indicators such as the root mean squared error or approximation (RMSEA, < 0.05) (Schreiber, 2017; StataCorp, 2017), (pclose, > 0.05) (StataCorp, 2017), standardized root mean squared residual (SRMR, ≤ 0.08) (StataCorp, 2017), Comparative fit index (CFI, ≥ 0.95) (Schreiber, 2017; StataCorp, 2017), and coefficient of determination (CD, ≥ 0.95) (StataCorp, 2017). Akaike Information Criteria (AIC), the smaller the better, was used for model comparison (StataCorp, 2017). Model fit indicators were obtained after running each SEM model. RMSEA estimate population errors, CFI do baseline comparisons, and SRMR and CD compares size of residuals. CD is analogous to $R^2$ for the model (StataCorp, 2017).

### 2.4.2 Path analysis

Structural equation model (SEM) was used for the study. Structural equation model has endogenous or outcome variables (variables with arrows pointed towards them) and, exogenous or independent variables (variables with arrows pointed away from them) (Anderson and David, 1988; Gunzler and Morris, 2015). Structural equation model is a two-step modeling approach using a confirmatory factor analysis method and specifies the relationship of observed variables to their respective latent variables which can inter-correlate freely. Paths in SEM are structured based on underlying theories (Anderson and David, 1988). The linkage between endogenous and exogenous variables are shown by structural equations. The measurement error and the latent variables are modeled by measurement equations (Gunzler and Morris, 2015) which are:

$$y = \mu_y + \Delta_y \eta + \varepsilon$$

$$x = \mu_x + \Delta_x \xi + \delta$$

where, $\boldsymbol{\xi}$ represents a vector of *r* unobserved latent exogenous variables measured by the *q* observed variables **x**. $\boldsymbol{\eta}$ is a vector of *m* latent endogenous variables measured by the *p* observed variables $\boldsymbol{y}$. $\boldsymbol{\mu}_y$ and $\boldsymbol{\mu}_x$ are vectors of intercepts, $\boldsymbol{\Delta}_y$ and $\boldsymbol{\Delta}_x$ are matrices of slopes also referred to as loading matrices and, $\boldsymbol{\varepsilon}$ and $\boldsymbol{\delta}$ are residual terms in respective equations for **y** and **x**. $\boldsymbol{\mu}_y$ and $\boldsymbol{\varepsilon}$ have dimensions of $p$*1, $\boldsymbol{\mu}_x$ and $\delta$ have dimensions of $q$*1, and $\boldsymbol{\Delta}_y$ and $\boldsymbol{\Delta}_x$ have dimension of *q*r* (Gunzler and Morris, 2015).

The structural model that relates unobserved latent variables to each other can be expressed as:

$$\eta = \mu_\eta + B\eta + G\xi + z$$

where, $\mu_\eta$ is $m$*1 dimensional matrix of intercepts for the unobserved endogenous latent variables, $\boldsymbol{B}$ is an *m*m* matrix of slopes of the unobserved endogenous latent variables to each other, $\boldsymbol{G}$ is *m*n* matrix of slopes for the unobserved exogenous latent variables, and $\boldsymbol{z}$ is m*1 vector of random error for unobserved endogenous latent variables (Gunzler and Morris, 2015).

Four different models—TRA and TRA with moral norms (henceforth, TRA-moral) and TPB and TPB with moral norms (henceforth, TPB-moral) were fitted using SEM. To develop TRA-moral, TRA was extended by adding an additional path from subjective norms to intentions through moral norms. Similarly, TPB-moral was developed adding two additional paths from subjective norms and perceived behavior control intentions through moral norms resulting. Structural equation models were fit using "*sem*" command in STATA 15.1 which fits linear SEMs using maximum likelihood estimation method (StataCorp, 2017). Maximum likelihood estimators have asymptotic, unbiased, consistent, and efficient property under the normality assumption of observed variables (Anderson and David, 1988).

Structural equation models were fit after obtaining acceptable range of internal consistency and factor loadings in each latent variable for all four models. Observed variables were dropped if an acceptable range of internal consistency and factor loading were not obtained. The same set of observed variables (N = 177) were used in all four models to form latent constructs. Command *sem* assumes that observed endogenous, observed exogenous variables, latent endogenous, and latent exogenous variables were jointly distributed normally with mean ($\mu$) and variance-covariance matrix ($\sum$) (StataCorp, 2017). The coefficients reported are standardized coefficients which are correlation coefficients and can be interpreted as change in one variable given a change in another, both measured in standard deviation units (StataCorp, 2017).

## 3. Results

### 3.1 Demographics of Respondents

Most of the survey respondent were white males with formal education level of high school or above, 21-94 years old. Most were involved in farming or ranching, with annual income of $25,000 or higher. Participants in the survey were 86.62% male and 13.45% female, 82.62% white American, 11.80% native American, and 5.58% multiple races. About half, 48.36%, of the respondents reported their primary job as farmers/rancher, 25.60% as retired, 7.66% as business, 4.16% as working class (laborious) jobs, 2.63% as medical jobs, and 11.59% of the responders held other jobs unidentified in the survey. Average age of respondents was 67.44 (SD = 12.20) years. The percentage of respondents reported their highest level of education was 25.9% for high school or General Educational Development (GED) degree, 18.9% for some college experiences, 11.0% for associate or technical degree, 21.7% for bachelor's degree, and 20.4% for graduate degree. The age, gender, race, education, and income proportions from the survey data generally were similar to the National Woodland Owners Survey Database. The early and late response bias conducted using above mentioned demographics did not show response bias among responders from first and second lots of survey and postcards mailing.

### 3.2 Measurement and Structural Variables, and their Factor Loadings

Cronbach alpha and factor loading ($\lambda$) of observed variables in their respective latent variables, mean, and standard deviation are presented in Table 15. Measurement variables are variables observed in the survey. Structural variables are latent variables. Subjective norms (Cronbach $\alpha = 0.89$) consisted of observed variables *e1value, e1diverse, e1support,* and *e1livable* variables. Attitudes (Cronbach $\alpha = 0.87$) consisted of variables *e3manage*, *e3effort*, *e3wilder*, and *e3overall*. Moral norms (Cronbach $\alpha = 0.82$) were represented by variables *e2respect,*

*e2maintain* and *e2invest*. Cronbach alpha value of subjective norms, attitudes, and moral norms were above the value suggested by Cronbach (1951). Perceived behavior control (Cronbach α = 0.52) consisted of variables *e1resource* and *e1improve.* Lastly, intentions (Cronbach α = 0.38) consisted of *a7wtp, a9altdist*, and *c6interst*. Cronbach alpha values of perceived behavior and intentions were slightly below the suggested value of internal consistency.

Table 15: Validity of structural variables, descriptions, and descriptive statistics of measurement variables.

| Measurement Variables in SEM Models | Factor Loading ($\lambda$) (N = 177) | Mean (St. Dev.) (N = 177) |
|---|---|---|
| Subjective Norms (*SUBNORM*): Cronbach Alpha (α) = 0.89 | | |
| *e1value:* Sustainable management of forest, rangeland and deer habitat is important to the people I value most. | 0.77 | 3.81 (1.08) |
| *e1diverse*: My family and friends think that forest, rangeland, and deer habitat management could enhance plant and animal diversity. | 0.83 | 3.61 (1.13) |
| *e1support*: My family and friends are supportive of forest, rangeland, and deer habitat management activities. | 0.91 | 3.83 (1.04) |
| *e1livable*: My family and friends think that forest, rangeland, and deer habitat management would make our environment more livable. | 0.81 | 3.57<br>(1.12) |

Perceived Behavior Controls (*PBC*): Cronbach Alpha (α) = 0.52

| | | |
|---|---|---|
| *e1resource:* I have resource and opportunities to manage my land for forest, rangeland, and deer habitat management. | 0.46 | 3.49<br>(1.17) |
| *e1improve*: I think that I can improve forest, rangeland, and deer habitat on my property by actively managing them. | 0.72 | 3.96<br>(0.99) |

Moral Norms (*MORAL*): Cronbach Alpha (α) = 0.82

| | | |
|---|---|---|
| *e2respect:* I give respect and courtesy to people who are involved in forest, rangeland, and deer habitat management. | 0.68 | 4.25<br>(0.85) |
| *e2maintain:* I feel that I should actively manage forest, rangeland, and deer habitat on my property to maintain deer habitat for deer and wildlife. | 0.92 | 3.97<br>(0.99) |
| *e2invest:* I feel honored to invest money, time, and resources to manage forest, rangeland and deer habitat for deer and wildlife habitat. | 0.77 | 3.61 (1.17) |
| Attitudes (*ATTITUDE*): Cronbach Alpha (α) = 0.87 | | |
| *e3manage:* I am satisfied with the overall characteristics of forest, rangeland, and deer habitat that I maintain. | 0.68 | 3.64 (0.95) |
| *e3effort:* I am satisfied with the number of deer and wildlife that I observed with the management effort that I put in my property. | 0.85 | 3.71 (1.07) |
| *e3wilder:* I am satisfied with the wilderness of forest, rangeland, and deer habitat that I maintain. | 0.88 | 3.63 (1.02) |
| *e3overall:* I am satisfied with the overall benefits I am getting from forest, rangeland, and deer habitat that I manage. | 0.78 | 3.57 (1.04) |
| Intentions (*INTENT*): Cronbach Alpha (α) = 0.38 | | |
| *a7wtp:* Assume that you do not observe any deer in your regular hunting site. How many dollars/acres are you willing to spend to maintain the deer population you generally observe in that site to receive desired hunting experience? (USD) | 0.43 | 63.56<br>(109.43) |
| *a9altdist:* If you could not go to the site that you regularly hunt deer, how far would you drive one way to go to another deer hunting site of about the same quality? (miles) | 0.24 | 35.04<br>(93.58) |
| *c6interest:* Are you interested in knowing more about active forest or rangeland management in Oklahoma? | 0.61 | 0.61<br>(0.50) |

At least 60% of landowners agreed or strongly agreed (henceforth, agree) towards variables related to subjective norms, perceived behavior control, and moral norms. An exception is that only about 51% and 33% of landowners agreed and remained neutral respectively to the statement that they should actively manage their property to maintain habitat for deer and wildlife (*e2maintain*). About 85% of landowners also agreed that sustainable management of forest and rangeland is important to people they value most (*e1value*). About 66% of landowners agreed that their family and friends are supportive of sustainable management of forest, rangeland, and deer habitat management activities (*e1support*) and about 75% of landowners agreed that their family and friends think that active management could enhance biodiversity (*e1diverse*). About 72% landowners also agreed that their family and friends think that management of forest and rangeland would make our environment more livable (*e1livable*). Despite a relatively smaller portion of landowners feeling that they should actively manage their lands, about 73% of landowners agreed that they have resources and opportunities to actively manage their land (*e1resource*). About 68% landowners feel honored to invest money, time, and resources to actively manage their land for wildlife and deer habitat management (*e2invest*). Like previous statement, about 67% of landowners agreed that they give respect to people involved in active management of ecosystem (*e2respect*).

The satisfaction level from management, however, was relatively low for landowners given the level of management effort they input. Though about 61% of landowners were satisfied with the overall characteristics of their forest, rangeland, and deer habitat they manage on their property (*e3manage*), only about 40% of landowners were satisfied with the number of deer and wildlife they observed with the management effort they put in their property (*e3effort*). Only about 42% of landowners were satisfied with the wilderness they maintain (*e3wilder*) and about 37% remained neutral on the statement. Similarly, only about 23% of landowners agreed and about 31% remained neutral with the statement that they were satisfied with the overall benefits they are receiving from the land they manage. The detailed summary of the results is presented in Table 16. The variables that are not included in the SEM models but asked in the survey are presented in Table 20, Appendix A.

Table 16: Distribution of landowners' responses to observed variables used in SEM models.

| Variables | Strongly Disagree (%) | Disagree (%) | Neutral (%) | Agree (%) | Strongly Agree (%) | N |
|---|---|---|---|---|---|---|
| E1value | 1.90 | 0.95 | 11.64 | 44.42 | 41.09 | 421 |
| E1diverse | 5.21 | 4.27 | 15.88 | 34.60 | 40.05 | 422 |
| E1support | 5.23 | 4.51 | 24.70 | 36.58 | 28.98 | 421 |
| E1livable | 4.05 | 5.24 | 19.05 | 43.10 | 28.57 | 420 |
| E1resource | 2.87 | 4.78 | 18.66 | 45.69 | 27.99 | 418 |
| E1improve | 3.84 | 5.76 | 26.38 | 38.85 | 25.18 | 417 |
| E2respect | 3.11 | 4.78 | 24.64 | 43.54 | 23.92 | 418 |
| E2maintain | 6.21 | 10.26 | 32.70 | 29.36 | 21.48 | 419 |
| E2invest | 2.63 | 10.05 | 19.38 | 47.13 | 20.81 | 418 |
| E3manage | 3.83 | 6.94 | 27.99 | 45.45 | 15.79 | 418 |
| E3effort | 19.48 | 15.68 | 28.03 | 23.28 | 13.54 | 421 |
| E3wilder | 7.19 | 13.91 | 36.69 | 29.50 | 12.71 | 417 |
| E3overall | 27.01 | 18.96 | 30.81 | 18.48 | 4.74 | 422 |

Notes: The variables are defined in Table 1.

3.3 Model Results

The model fit statistics exhibited good fit for the four models representing TRA, TRA-moral, TPB, and TPB-moral (Table 17). SEM model results to test TRA, TRA-moral, TPB, TPB-moral are presented in Figure 8 and Figure 9. Models were presented in the figures using structural variables only; measurement variables were excluded in the figures to simplify presentation.

Table 17: SEM model fit statistics for all four models along with the sample size used in each model.

| **Fit Indicators/Models** | **TRA** | **TRA-moral** | **TPB** | **TPB-moral** |
|---|---|---|---|---|
| **Sample Size (N)** | **177** | **177** | **177** | **177** |
| Root mean Squared Error of Approximation **(RMSEA)** | 0.014 | 0.034 | 0.036 | 0.042 |
| 90% CI, Lower Bound | 0.000 | 0.000 | 0.000 | 0.016 |
| Upper Bound | 0.054 | 0.058 | 0.061 | 0.061 |
| Comparative fit index **(CFI)** | 0.998 | 0.988 | 0.985 | 0.978 |
| Standardized root mean squared residuals **(SRMR)** | 0.037 | 0.040 | 0.047 | 0.048 |
| Coefficient of determination **(CD)** | 0.990 | 0.991 | 0.994 | 0.995 |
| Akaike's information criterion **(AIC)** | 4908.249 | 6081.001 | 5892.968 | 7054.775 |

Model summary statistics for TRA and TRA-moral are given in Table 18. In the TRA model (Figure (8a)), the subjective norm ($\beta = 0.31, p < 0.001$) significantly affected intentions for active management of forest, rangeland, and deer habitat management for deer hunting (henceforth, intentions) supporting $H_1$. However, attitude ($\beta = -0.20, p < 0.05$) significantly affected intentions but showed a negative correlation in the TRA model thus partially supporting $H_2$. In the TRA-moral model (Figure (8b)), subjective norms ($\beta = 0.02, p = 0.10$) did not directly affect intentions, rejecting $H_1$ but indirectly affected intentions through moral norms. Subjective norms ($\beta = 0.53, p < 0.001$) significantly affected moral norms, and moral norms ($\beta = 0.49, p < 0.001$) significantly affected intentions and were positively correlated, thus supporting $H_5$ and $H_7$ respectively. Attitude significantly affected ($\beta = -0.27,\ p < 0.05$) intentions but was negatively correlated thus, partially supporting $H_2$. Attitude did not significantly affect moral norms, thus rejecting $H_3$.

In the TPB model (Figure (9a)), subjective norms ($\beta = 0.21, p > 0.10$) and perceived behavior control ($\beta = 0.24, p > 0.10$) did not affect intentions, thus not supporting $H_1$ and $H_4$. Attitude ($\beta = -0.27, p < 0.05$), similar to previous models, significantly affected intentions and having a negative correlation, thus, partially supporting $H_2$. In TPB-moral (Figure (9b)), subjective norm ($\beta = 0.02, p > 0.10$) did not directly affect intentions again failing to support $H_1$. Subjective norms, however, indirectly affected intentions through moral norms, similar to the TRA-moral model. Subjective norms ($\beta = 0.36, p < 0.001$) significantly affected moral norms and moral norms ($\beta = 0.50, p < 0.001$) significantly affected intentions with positive correlation, again supporting $H_5$ and $H_7$ respectively. Attitude again directly affected ($\beta = -0.21$, $p < 0.05$) intentions and retained a negative correlation, but did not affect moral norms; thus, $H_2$ was partially supported and $H_3$ was not supported. Model summary statistics for TPB and TPB-moral are also given in Table 18.

Table 18: Model summary of four SEM models (TRA, TRA-moral, TPB, and TPB-moral).

| Structural/ Latent Variables | TRA | | TRA-moral | | TPB | | TPB-moral | |
|---|---|---|---|---|---|---|---|---|
| | **$\beta$- Coef. (Std. Err.)** | **Z-Statistics** | **$\beta$- Coef. (Std. Err.)** | **Z-Statistics** | **$\beta$- Coef. (Std. Err.)** | **Z-Statistics** | **$\beta$- Coef. (Std. Err.)** | **Z-Statistics** |
| SUBNORM ➔ INTENT | 0.31 *** (0.092) | 3.41 | 0.02 (0.10) | 0.17 | 0.21 (0.13) | 1.64 | 0.02 (0.11) | 0.21 |
| ATTITUDE ➔ INTENT | -0.20 ** (0.10) | -1.96 | -0.21 ** (0.09) | -2.28 | -0.27 ** (0.13) | -2.04 | -0.21 ** (0.10) | -2.03 |
| MORAL ➔ INTENT | - | - | 0.49 *** (0.172) | 2.86 | - | - | 0.50 ** (0.20) | 2.55 |
| PBC ➔ INTENT | - | - | - | - | 0.24 (0.24) | 0.96 | -0.01 (0.21) | -0.02 |
| SUBNORM ➔ MORAL | - | - | 0.53 *** (0.72) | 7.31 | - | - | 0.36 *** (0.09) | 4.03 |
| ATTITUDE ➔ MORAL | - | - | 0.08 (0.061) | 1.33 | - | - | - | - |
| PBC ➔ MORAL | - | - | - | - | - | - | 0.38 ** (0.147) | 2.57 |

Note: β-Coef. = βeta coefficients are correlation coefficients, Std. Err. = Standard Error of βeta coefficients. SUBNORM ➔ INTENT: subjective norms (SUBNORM) impact Intentions (INTENT) and so on. All arrows in the table are in accordance with arrows in respective models. Dashes (-) indicate irrelevant variable in the model. *** = $p < 0.001$, ** = $p < 0.05$ and * = $p < 0.10$

Unlike stated hypotheses, attitude consistently showed a negative sign in all four models. However, the pairwise coefficients among subjective norms and attitudes are positive and significant in all four models. Also, subjective norms and perceived behavior control, and subjective norms and attitudes were also positive and significant in TPB and TPB-moral models (Table 19).

Table 19: Standardized correlation coefficients of latent variables in four SEM models (TRA, TRA-moral, TPB, and TPB-moral)

| **Components of Theories** | **TRA** | | **TRA-moral** | | **TPB** | | **TPB-moral** | |
|---|---|---|---|---|---|---|---|---|
| | **Cor. Coef. (Std. Er.)** | **Z-Statistics** | **Cor. Coef. (Std. Er.)** | **Z-Statistics** | **Cor. Coef. (Std. Er.)** | **Z-Statistics** | **Cor. Coef. (Std. Er.)** | **Z-Statistics** |
| **SUBNORM*ATTITUDE** | 0.35 *** (0.07) | 4.76 | 0.35 *** (0.074) | 4.77 | 0.35 *** (0.07) | 4.75 | 0.35 *** (0.07) | 4.75 |
| **SUBNORM*PBC** | - | - | - | - | 0.71 *** (0.08) | 8.73 | 0.70 *** (0.08) | 8.87 |
| **PBC*ATTITUDE** | - | - | - | - | 0.52 *** (0.10) | 5.04 | 0.48 **** (0.09) | 5.21 |

Note: SUBNORM*ATTITUDE: Bivariate correlation between subjective norms (SUBNORM) and attitudes (ATTITUDE). Dashes (-) indicate irrelevant relationship in the model. *** = $p < 0.001$, ** = $p < 0.05$ and * = $p < 0.10$.

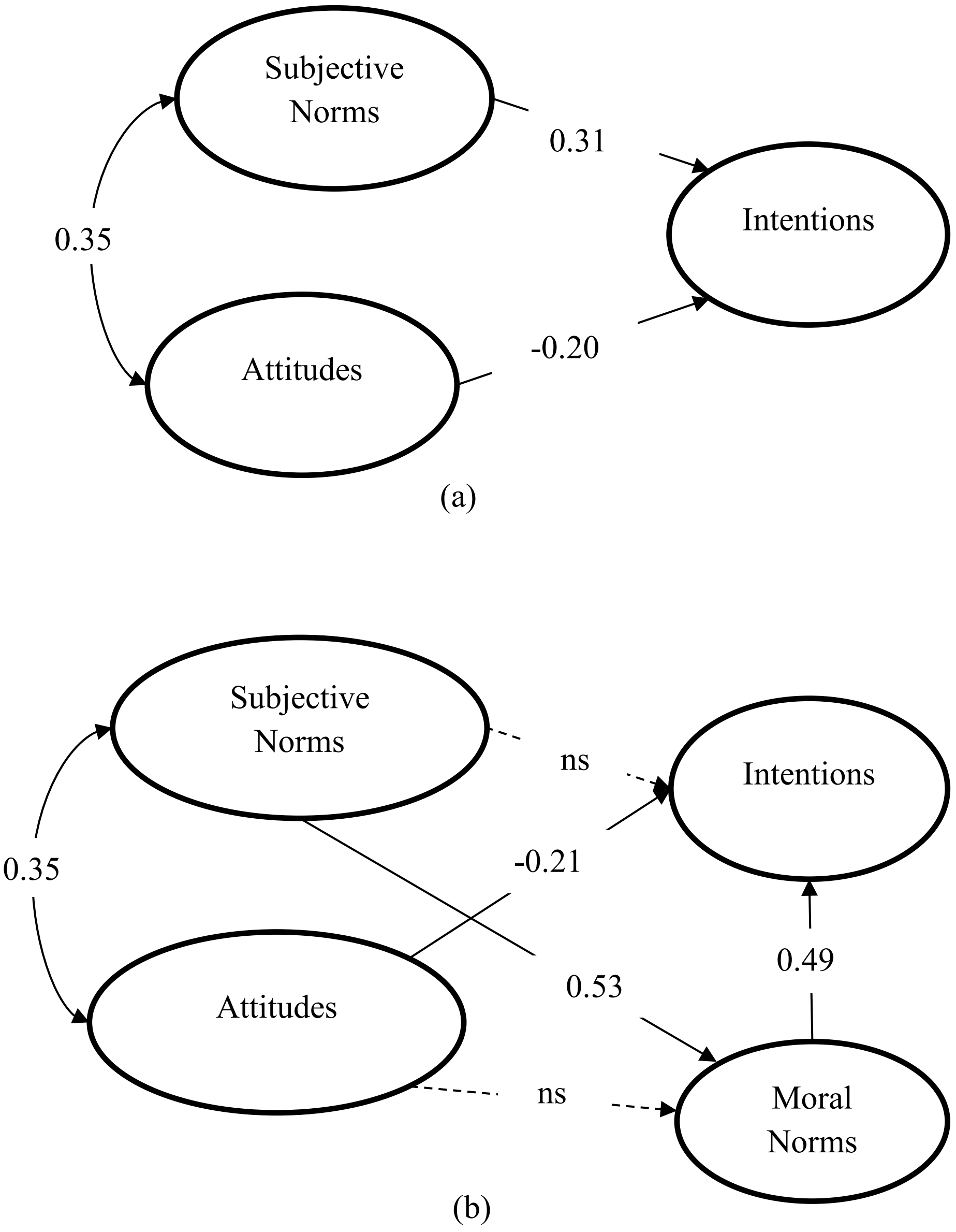

Figure 8: (a) Theory of reasoned action (TRA) and (b) Theory of reasoned action with moral norms (TRA-moral). Values on the arrow and "ns" indicate coefficients and "non-significant" relationships, respectively.

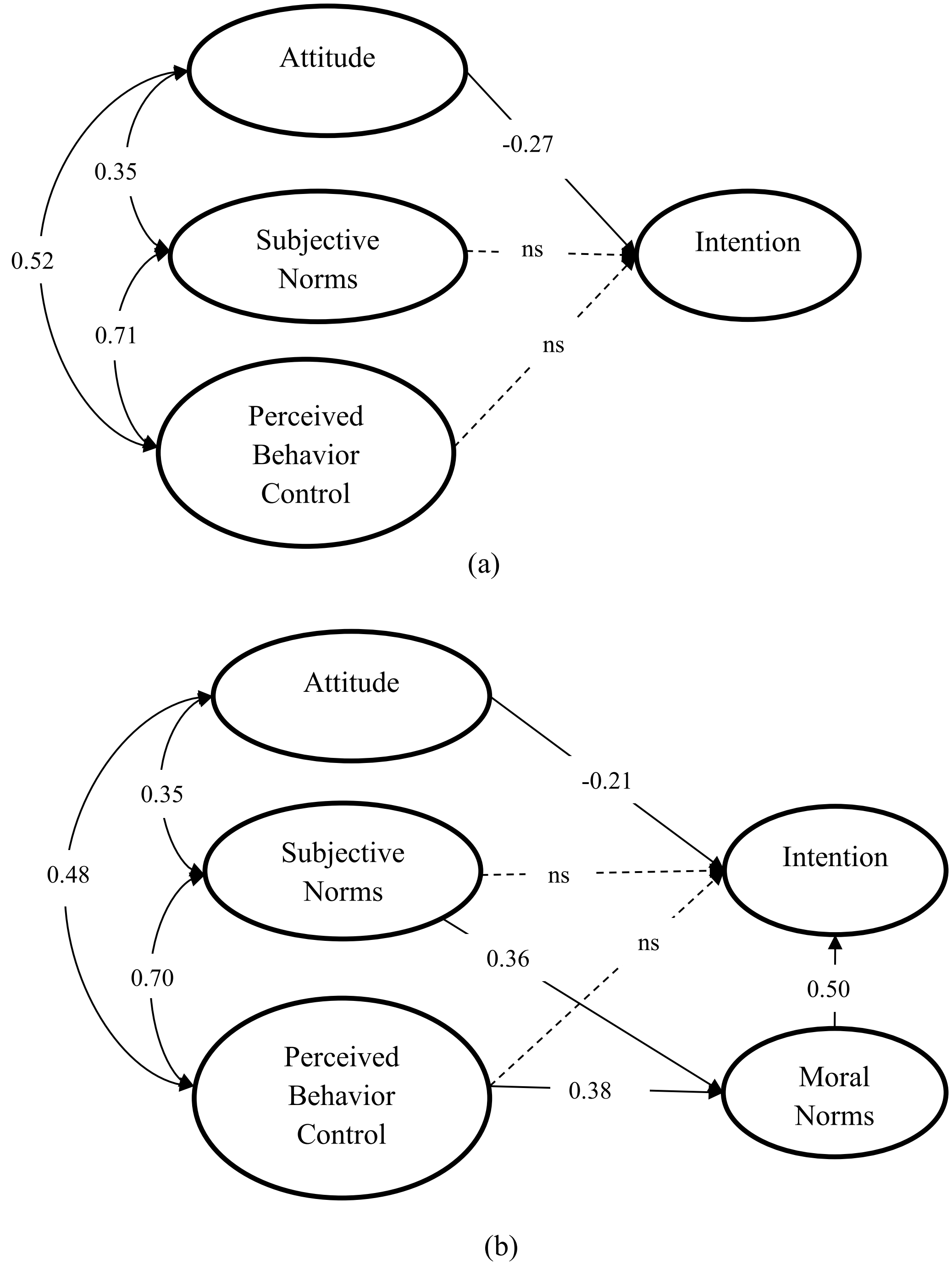

Figure 9: (a) Theory of planned behavior (TPB) and (b) Theory of planned behavior with moral norms (TPB-moral). Values on the arrow and “ns” indicate coefficients and “non-significant” respectively.

## 4. Discussion

Previous research indicated that Oklahoma landowners were supportive of using active management tools such as prescribed burning (Elmore et al., 2010) but concerned about the fire liabilities (Elmore et al., 2010; Kaur et al., 2020; Starr et al., 2019) and associated financial risk (Kaur et al., 2020; Starr et al., 2019). Fire suppression and exclusion since the mid 1900's has reduced grasslands, savannas, and open woodlands and increased closed-canopy forests (Hoff et al., 2018a; Joshi et al., 2019b). Thus, active management is needed to restore the full suite of ecosystem services along the south-central ecotone. In this context, our research studied how landowners' attitude, perceived behavior control, moral norms, and subjective norms influence active management on forest and rangeland to improve deer habitat deer hunting revenue.

Our results showed that landowners have positive social pressure (subjective norms) and positive perception regarding their ability (perceived behavior control) to actively manage their land. Landowners in the study region also expressed strong positive moral support towards people and activities related to active management of forest, rangeland, and deer habitat management. Attitudes, however, have negative impacts towards active management. Attitude is the reflection of the behavioral belief that is originated from an individual's experience of performing an action (Ajzen, 2002). Despite having positive attitude, the negative relationship of attitude in the model might be from past unpleasant experience related to active management. Several researchers suggested financial burden (Kaur et al., 2020; Starr et al., 2019) and fire liabilities (Elmore et al., 2010; Joshi et al., 2019a; Starr et al., 2019) as major demotivating factors for landowners to actively manage their forested land.

The pairwise correlations among subjective norms, perceived behavior control and attitude were positively correlated signifying that the landowners with the positive subjective norms and perceived behavior control tend to have positive attitude towards active management. Active management tools such as prescribed fire has high support, but associated risk and liability issues should be addressed (Elmore et al., 2010). Similarly, financial burden associated with the active management is one of the major barriers for the active management in this ecoregion (Starr et al., 2019). Addressing associated risk and liabilities issues as well as financial burden could change the attitude of landowners and thus positively affecting the intentions for the active management.

Among four different models developed and discussed, TRA was best supported by our data, as reflected by AIC value, to explain the intentions for active management of forest, rangeland, and deer habitat management for deer hunting. The behavior, managing land for deer hunting, is under the volitional control of our study population, who are landowners owning at least 160 acres of land, which is why the behavior is better explained by TRA (Madden et al., 1992).

Perceived behavior control and intentions in SEM models have lower Cronbach alpha values. However low Cronbach alpha values are not uncommon in SEM models. Lopez-Mosquera and Sanchez (2012) also reported lower than suggested Cronbach alpha value for perceived behavior control and intentions. Latent constructs often have lower Cronbach alpha coefficients because of random error, even with meticulously planned variables (Ajzen, 2011).

## 5. Management Implications and Conclusion

This paper sheds light on the role of attitude, perceived behavior control, and social and subjective norms on shaping intentions of landowners towards active management of forest, rangeland, and deer habitat in south-central ecotone of the USA. Exclusion of fire in this region has increased closed canopy forest and recruitment of fire susceptible species such as ERC. Meanwhile, the declining recruitment and retention and of deer hunters has raised concerns regarding the future of hunting (Peterson, 2004) and the associated tradition, cultural (Byrd et al., 2017), economic, and wildlife management strategies (Byrd et al., 2017; Peterson, 2004). This creates an opportunity to actively manage land to maintain deer habitat as a source of revenue as well as restore ecosystem services. Our research suggests that landowners are positive about actively managing their land and have positive pressure from family and friends. Landowners also show respect to stakeholders involved in active management of ecosystems and feel respectful towards those who actively manage their land. Landowners, however, are not satisfied with the management outcomes, ecosystem services, and benefits they received from their land, thus, providing an opportunity to improve ecosystem services by actively manage their land. Landowners are supportive of actively managing their land for deer hunting by maintaining a good deer habitat; motivating landowners towards stewardship of natural resources and respect towards stakeholders involved in active management motivates them to actively manage their land.

Fire related risk and liabilities can be offset by adopting prescribed burning and application of alternative management techniques such as overstory and understory thinning, herbicide application, and brush management. The perceived risk and liabilities of fire decrease with the increase in knowledge and experience associated with prescribed burning (Joshi et al., 2019a), suggesting a need for outreach activities among landowners to increase awareness about prescribed fire.

The management cost associated with the active management can be in part be offset through hunting leases; improved deer habitat through active management could motivates deer hunters to pay more money per acres as lease fee. Based on our research we suggest extension specialists and policy makers to focus on educating landowners to make them aware about the cost and benefit associated with the active management. This will give more confidence to landowners in adoption of active management tools and helps landowners realize financial benefits which can outweigh associated cost. The realization of reduced risk and added financial benefits could motivate landowners to adopt management tools in their forest and rangeland.

## 6. Future Research Consideration

This paper studied landowners' attitudes, perceptions, social and peer pressure related to active management of ecosystem using SEM and broadened the scope of wildlife management research through the inclusion of moral norms in TRA and TPB models. This dimension of wildlife management research further can be expanded to other game species. Furthermore, the use of TPB and TRA for other species, and expansion of these theories using moral norms in the hunting research are yet to be understood fully. Also, TBP is criticized for ignoring human emotions, identity, and moral values (Miller, 2017) which is addressed by this paper by expanding theories by adding moral norms as suggested by (Ajzen, 1991). The recent development of theory of planned behavior is more suggestive towards mediating role of perceived behavior control between attitude-intention and subjective norms-attitude and inclusion of sociodemographic variables into the model (Ajzen, 2020; Sok et al., 2020) which is out of scope of this paper but something to consider for future research.

**Appendix IV.A**

Table 20: Distribution of landowners' responses to variables presented in same section of survey but not included in SEM.

| Variables | Strongly Disagree (%) | Disagree (%) | Neutral (%) | Agree (%) | Strongly Agree (%) | N |
|---|---|---|---|---|---|---|
| e1govt: It would be difficult to conduct forest, rangeland, and deer habitat management without government support. | 4.73 | 3.78 | 22.70 | 41.61 | 27.19 | 423 |
| e1commun: It would be difficult to conduct forest, rangeland, and deer habitat management without support from the community. | 6.35 | 4.94 | 27.53 | 34.35 | 26.82 | 425 |
| e2harvest: Excessive harvesting of natural resource may limit their use for the future generation. | 4.03 | 7.82 | 24.64 | 39.34 | 24.17 | 422 |
| e3benefit: Active management of forest, rangeland, and deer habitat can bring economic as well as environmental benefits. | 7.88 | 10.98 | 33.65 | 27.92 | 19.57 | 419 |
| e3human: The primary use of forest, rangeland, and deer habitat management should be to benefit human beings. | 4.06 | 8.59 | 31.50 | 36.99 | 18.85 | 419 |
| e3restrict: Restricting excessive use of forest, rangeland, and deer habitat can enhance recreational opportunities. | 6.65 | 9.74 | 27.79 | 37.29 | 18.53 | 421 |
| e3time: It is important to spend time managing forest, rangeland, and deer habitat. | 2.39 | 9.07 | 25.30 | 45.11 | 18.14 | 419 |
| e3balance: Sustainable management of forest, rangeland, and deer habitat is important to maintain balance and diversity in the natural environment. | 9.50 | 14.01 | 27.55 | 31.59 | 17.34 | 421 |
| e3connect: I feel connected with nature when I get involved in forest, rangeland, and deer habitat management. | 0.96 | 10.29 | 22.73 | 49.52 | 16.51 | 418 |
| e3environ: The primary use of forest, rangeland, and deer habitat management should be to benefit the environment. | 6.92 | 16.23 | 29.36 | 31.50 | 15.99 | 419 |
| e3noneed: There is no need for active, forest, rangeland, and der habitat management. | 41.73 | 28.78 | 19.66 | 6.00 | 3.84 | 417 |

# CHAPTER V: CONCLUSION

This research answered three important questions about the active management of ecosystem to develop sustainable ecosystem resilient under changing climate in the south-central transitional ecoregions. This research quantified ecosystem benefit from timber production, deer habitat management, and cattle management under changing rainfall condition in forest, savanna, and grassland of south-central transitional ecoregion of USA resulted from about 40 years of active management using prescribed fire and thinning. This research studied WTP for deer hunting site attributes and preferences ranking from deer hunter's perspective. This research further studied intentions of landowner's to actively manage their land for deer habitat management.

This research found that the active management of ecosystem using prescribed fire has often positive economic return. The return however varies because of varying productivity of management regimes and cost of management. For example, the highest growth of sawlog and pulpwood volume were found in HT4, and HT. The volumetric growth of sawlog volume increased but the pulpwood volume growth did not occur with the increase in rainfall. This could be because pulpwood is converted into sawlog as tree grew older. Forage availability was higher in stands burned every two to three year supporting highest number of cattle and deer. Control and HT stands are suitable for timber production.

The variation in productivity and economic return will helps landowners in the south-central US to make an informed decision to optimize their economic return and prioritize their land management goals such as deer habitat management, cattle grazing, and timber production. Landowners should choose management regime such as HT to maximize return from timber and HT2 and HT3 to maximize return from forage production. Management regime such as HT4 support multiple objectives which could reduce the risk of financial risk and improve biodiversity in the forest. This research found that landowners can maximize their economic return from the HT3 stands even though the annual return from timber was negative.

This research found that maintaining deer habitat to provide quality deer hunting experience would generate higher economic return. Deer hunters are willing to a pay higher price for a site that provides better deer hunting experience and opportunity to observe higher number of deer thus, increasing the change of hunting deer. Landowners might benefit by maintaining food plots and deer sanctuaries inside their deer hunting sites; the willingness to pay for deer habitat with food plot and deer sanctuary is higher in comparison to deer habitat without them. Forest canopy cover does not have significant impact in the WTP for deer hunting which provides flexibility for landowners to modify canopy cover in their property to manage their land for multiple objectives to optimize their revenue.

This research further conclude that the landowner in the south-central transitional ecoregion holds positive values and intensions towards active management of ecosystem. Landowners perceive that they are capable of and have positive support from friends and families to actively managing their land. However, the negative past experiences are limiting the active management of ecosystem using prescribed fire. This finding is supported by previous research.  found that Oklahoma farmers and public are supportive of prescribed fire. However, financial burden and liabilities major barriers of active management using prescribed fire (Starr et al., 2019). Our research quantified ecosystem service benefit of active management of ecosystem for timber production, deer habitat, and cattle management to reduce financial burden to landowners.Elmore et al. (2010) found that Oklahoma farmers and public are supportive of prescribed fire. However, financial burden and liabilities major barriers of active management using prescribed fire (Starr et al., 2019). Our research quantified ecosystem service benefit of active management of ecosystem for timber production, deer habitat, and cattle management to reduce financial burden to landowners.

VITA

Bijesh Mishra

Candidate for the Degree of

Doctor of Philosophy

Thesis:

ECONOMICS AND HUMAN DIMENSION OF ACTIVE MANAGEMENT OF FOREST-GRASSLAND ECOTONE IN SOUTH-CENTRAL USA UNDER CHANGING CLIMATE

Major Field: Natural Resource Ecology and Management
Concentration: Forest Resource Economics
Minors: Statistics and Agricultural Economics

Biographical:

Education:

Completed the requirements for the Doctor of Philosophy in Natural Resource Ecology and Management with Ph.D. Minor in Statistics and Ph.D. Minor in Agricultural Economics at Oklahoma State University, Stillwater, Oklahoma in July, 2022.

Completed the requirements for the Master of Science in Environmental Studies at Kentucky State University, Frankfort, Kentucky in May 2017.

Completed the requirements for the Bachelor of Science in (Agriculture) (Agricultural Economics) at Tribhuvan University, Kathmandu, Nepal in July 2013.

ECONOMICS AND HUMAN DIMENSION OF ACTIVE MANAGEMENT OF FOREST-GRASSLAND ECOTONE IN SOUTH-CENTRAL USA UNDER CHANGING CLIMATE

BY

BIJESH MISHRA

Supplement Material: These charts were not included in main dissertation. The charts were presented in tabulated format in the dissertation. This document translated tabulated information into visual format. Please see full dissertation for detailed information.

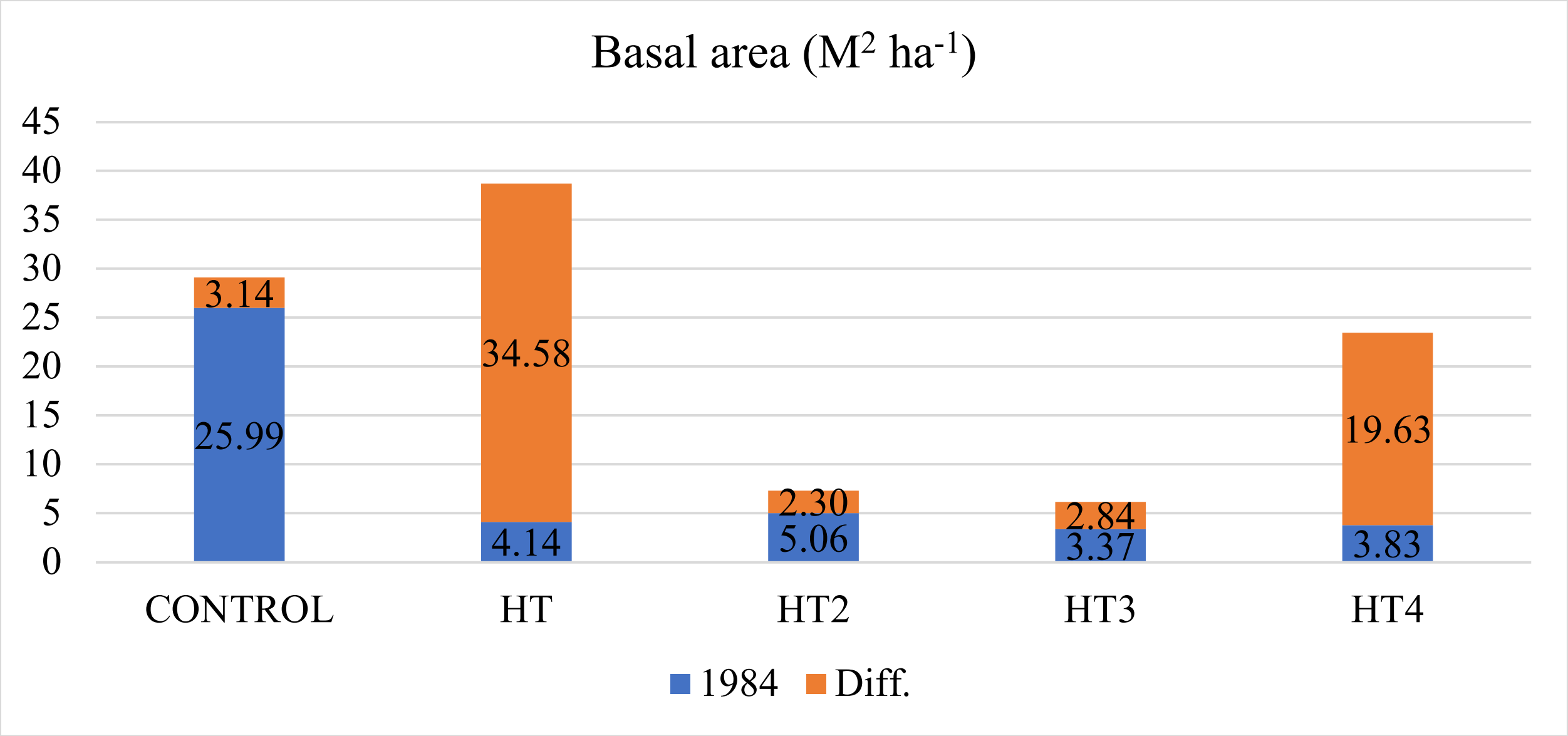

Figure 1: Basal area of timber from various treatments in the study region.

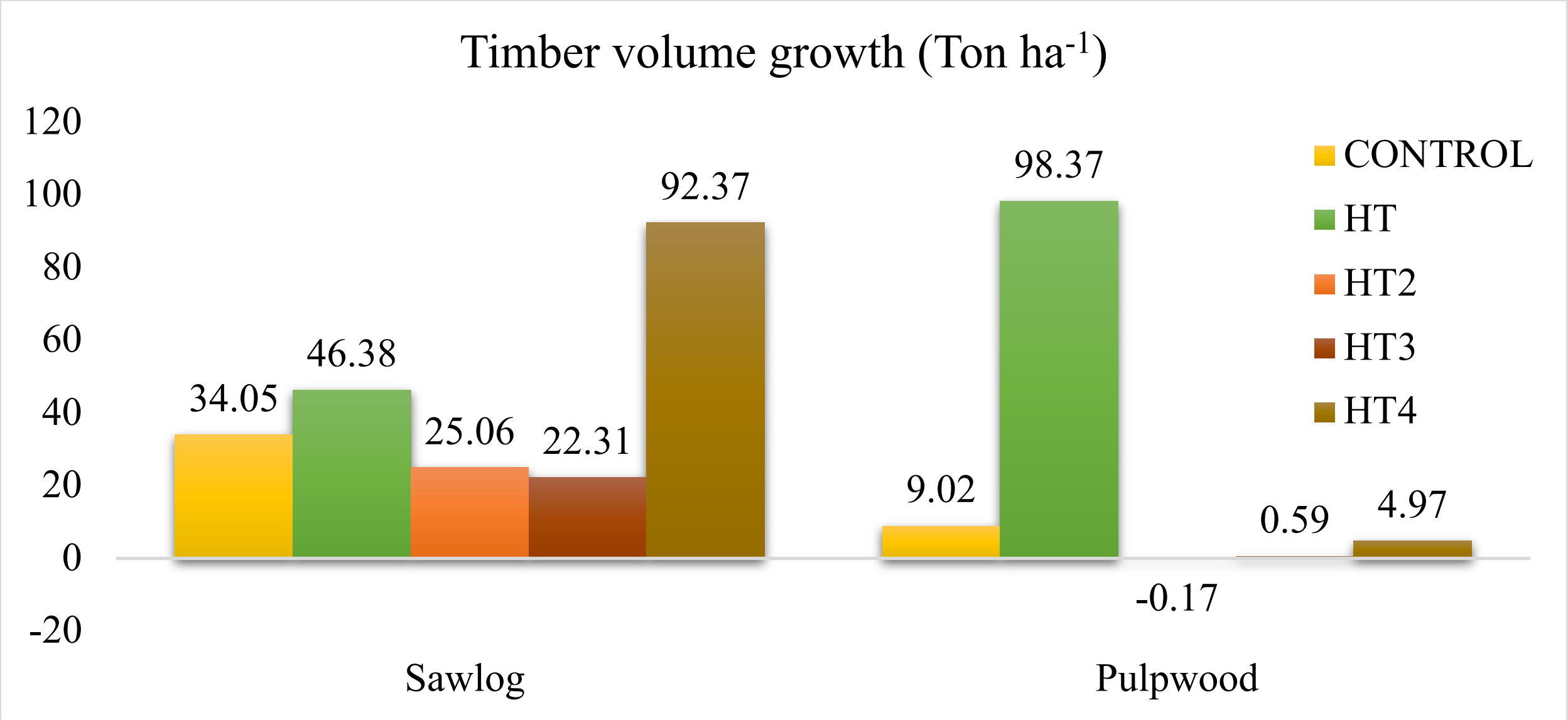

Figure 2: Timber volume growth in various treatments in the study region.

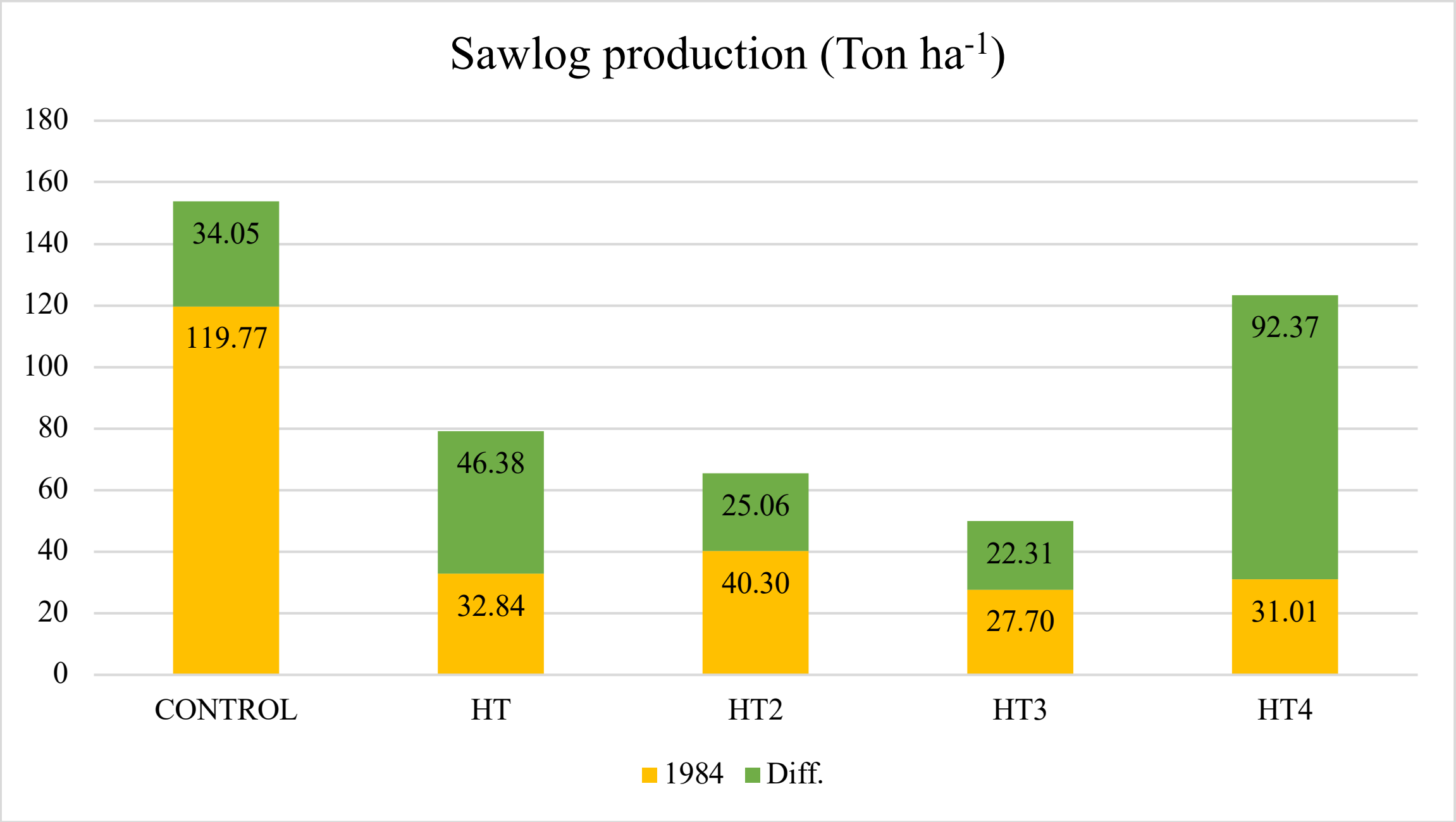

Figure 3: Sawlog production from various treatments in the study area.

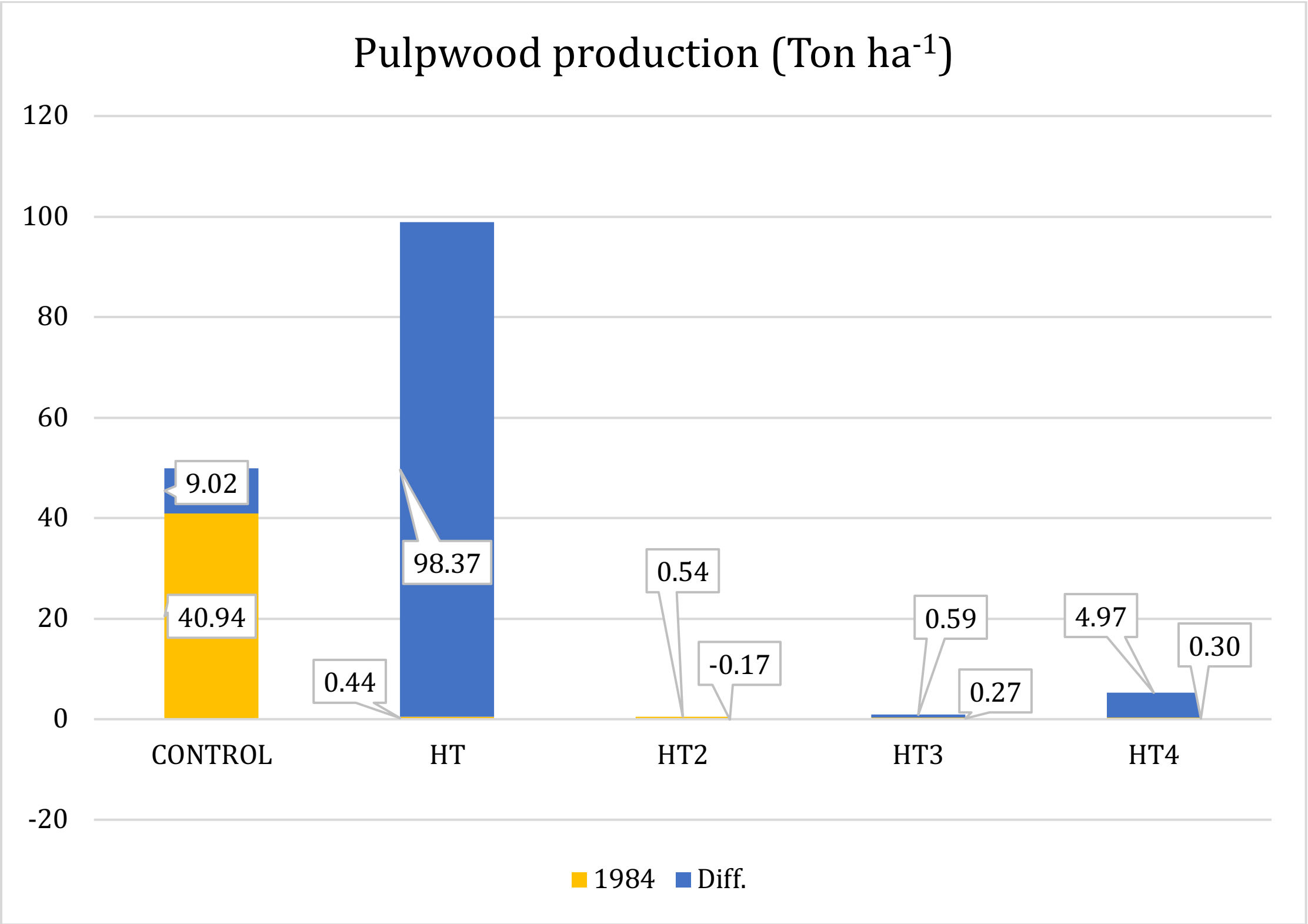

Figure 4: Pulpwood production from various treatments in the study area.

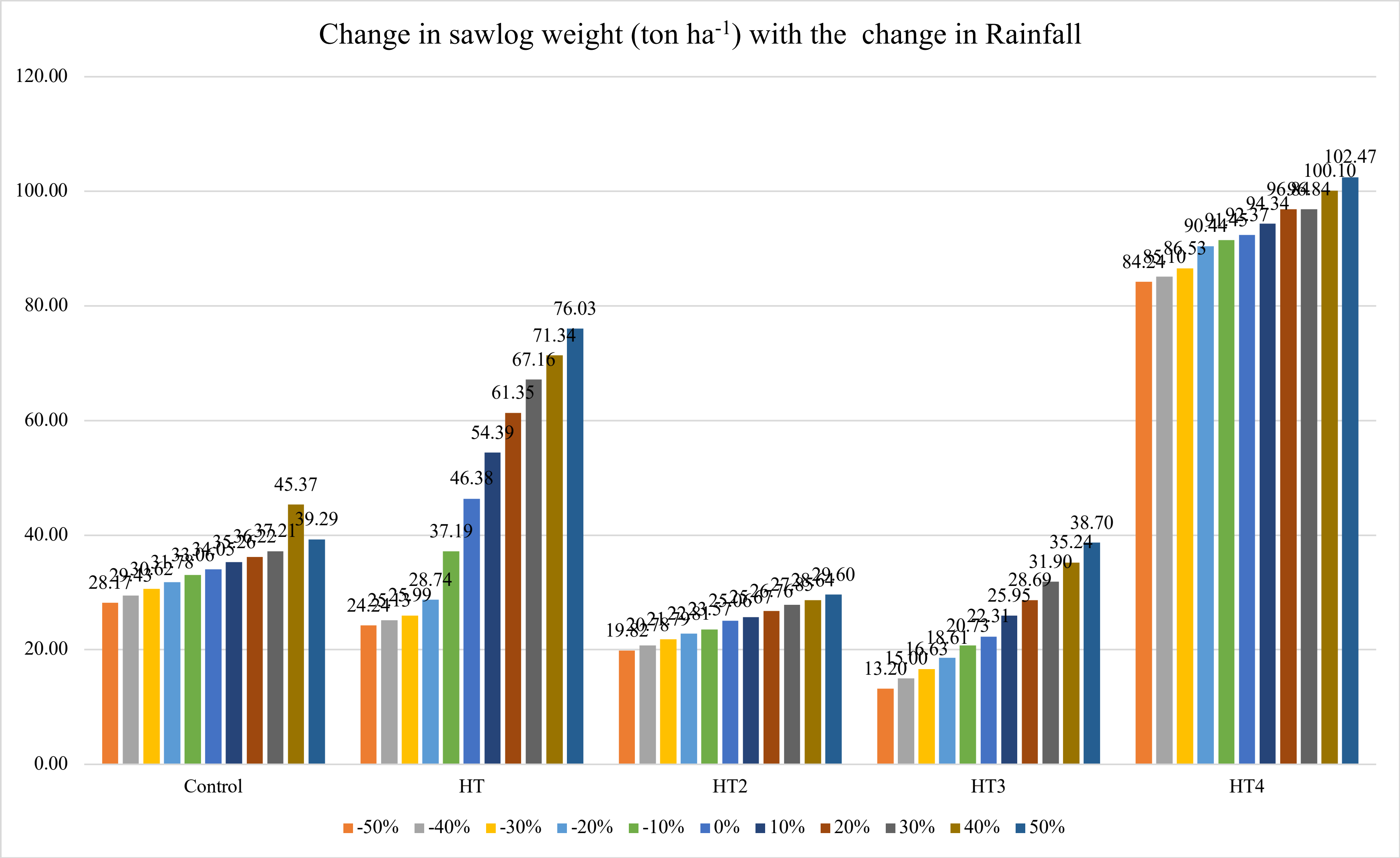

Figure 5: Change in sawlog production (ton ha$^{-1}$) with the change in rainfall (%) in various treatments n the study area.

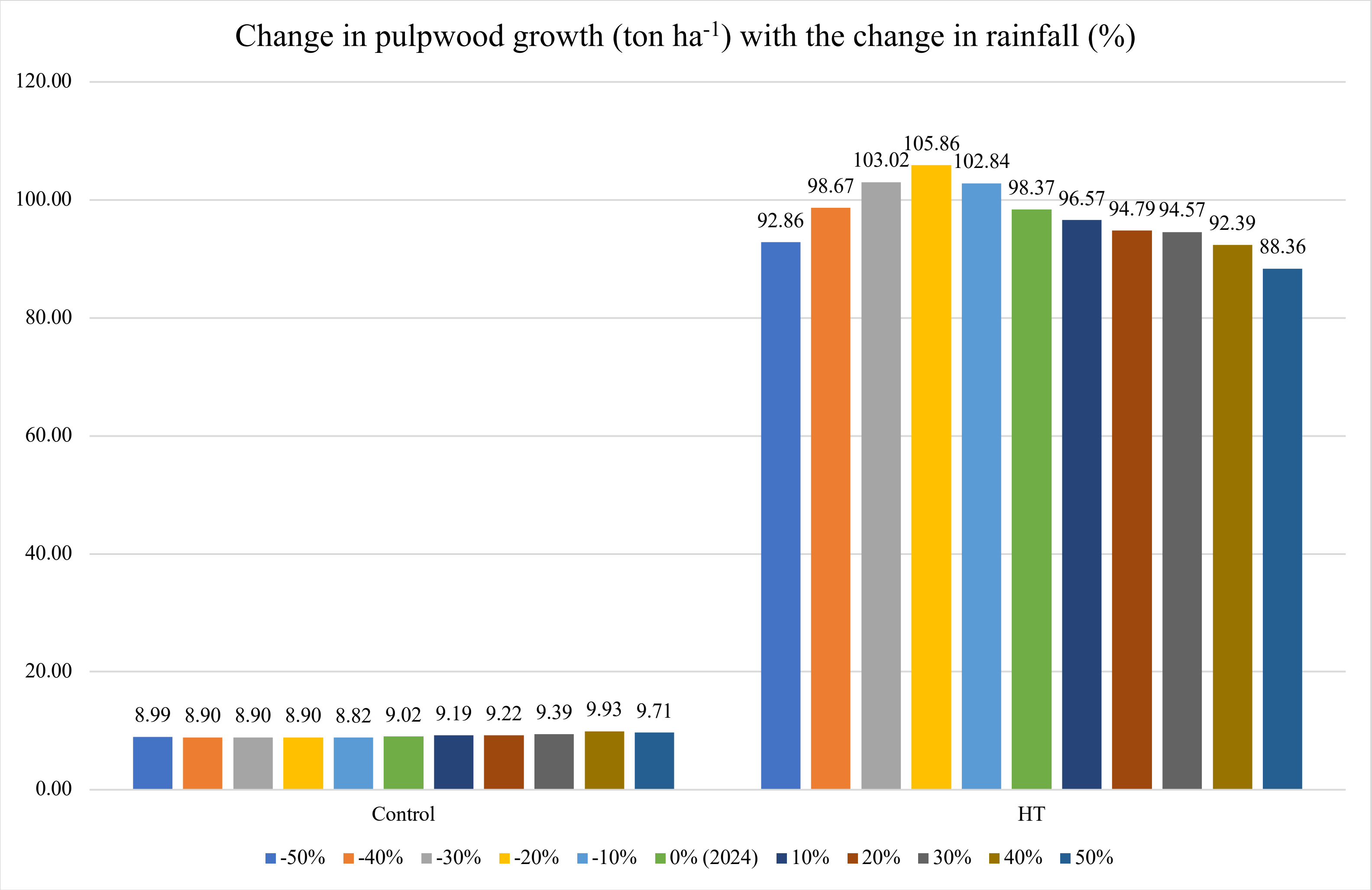

Figure 6: Change in pulpwood production (ton $ha^{-1}$) with change in rainfall (%) in non-burned stands in the study area.

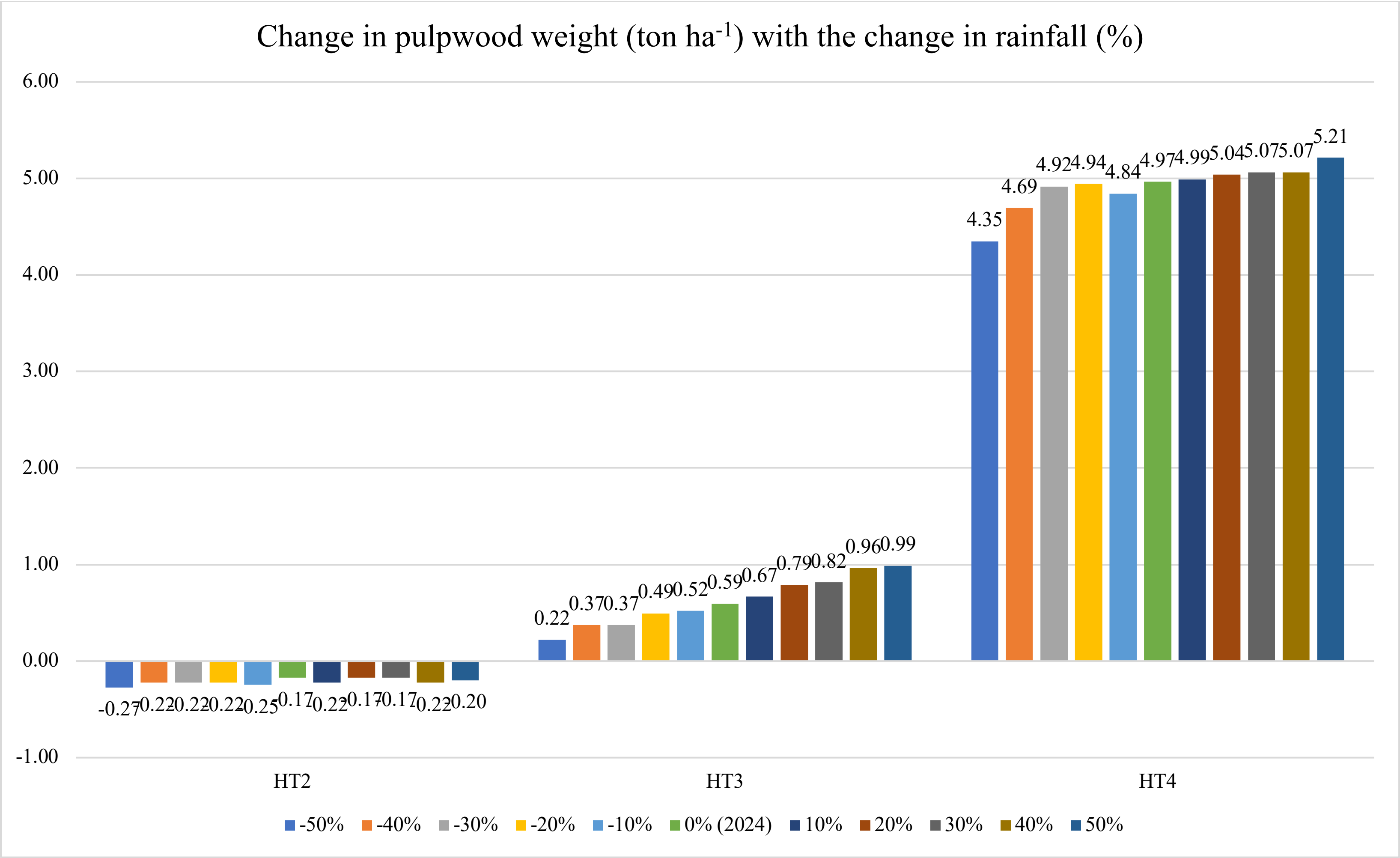

Figure 7: Change in pulpwood growth (Ton ha$^{-1}$) with the change in rainfall (%) in burned stands in the study area

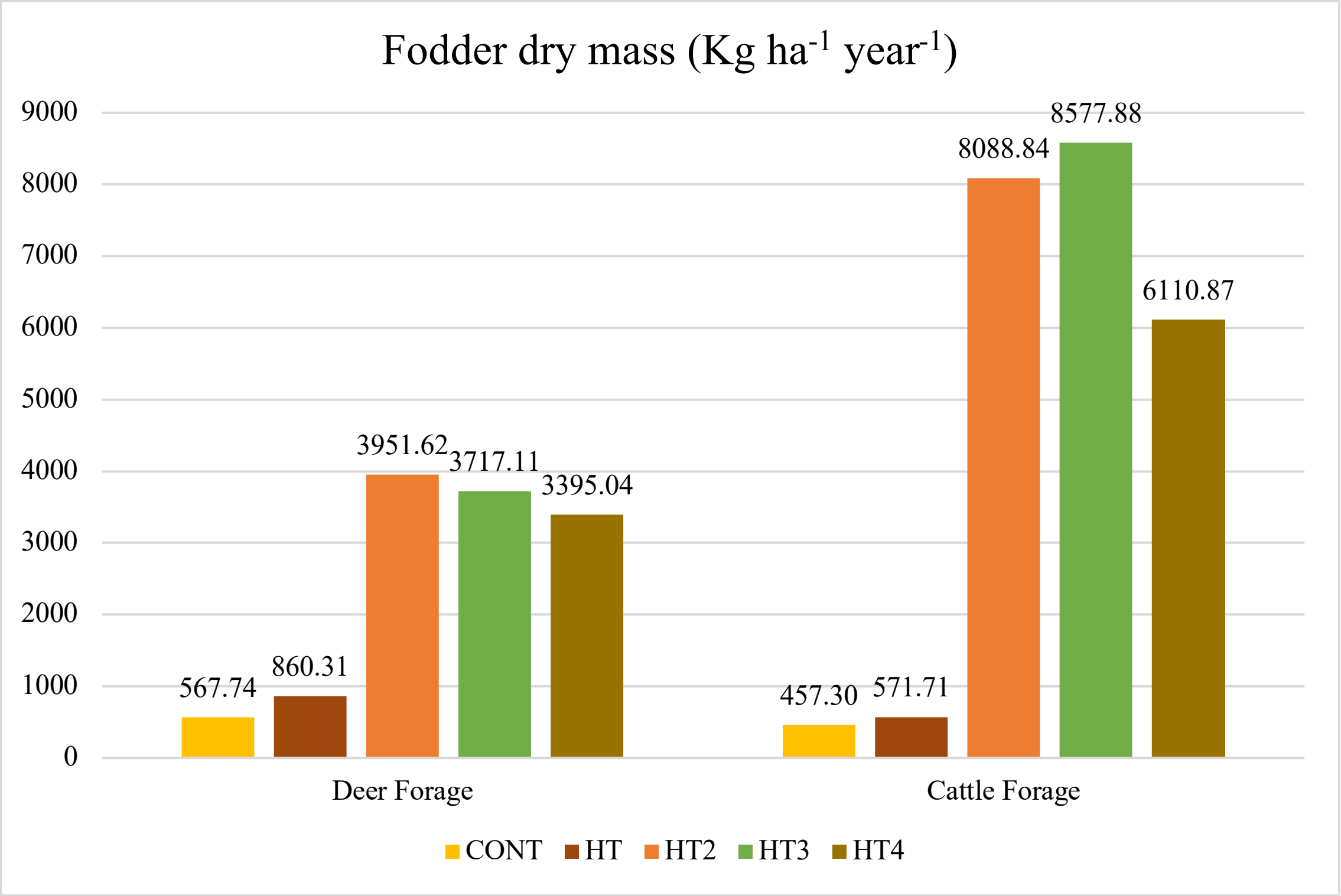

Figure 8: Dry weight of deer and cattle forage produced in various treatments in the study site.

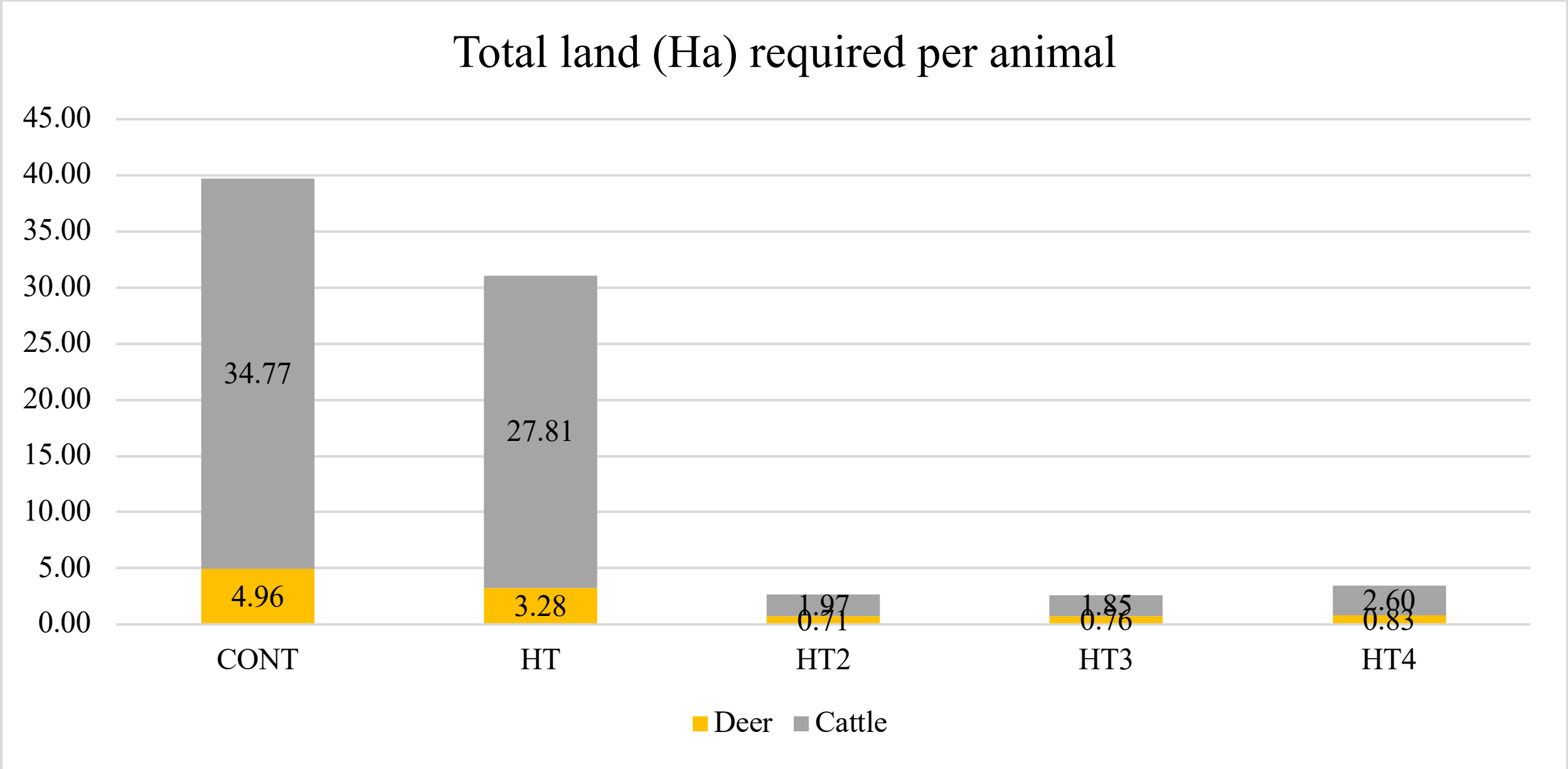

Figure 9: Land Requirement per cattle and deer in various treatments in the study region.

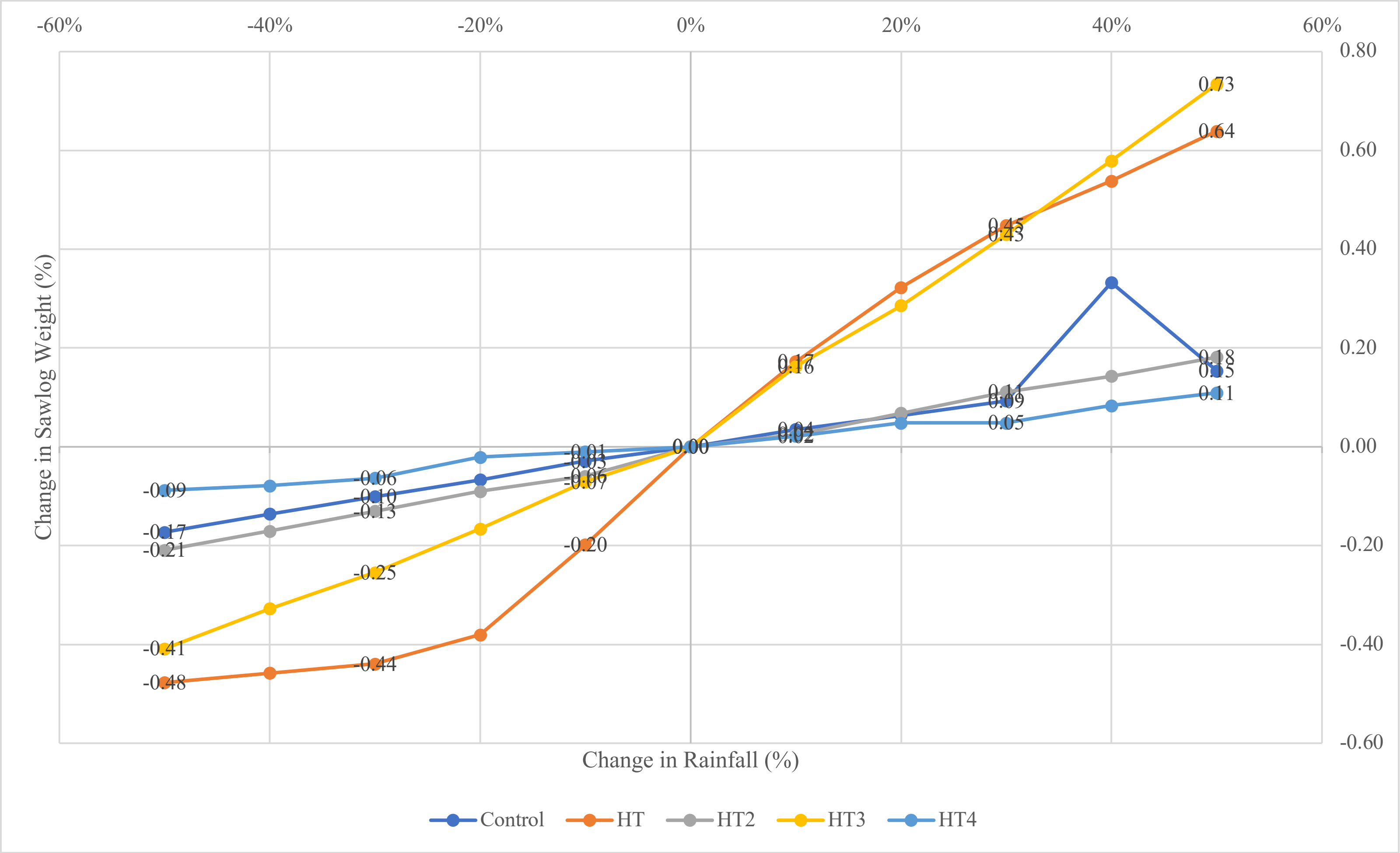

Figure 10: Change in sawlog weight (%) with the change in rainfall (%) in various treatments in the study area.

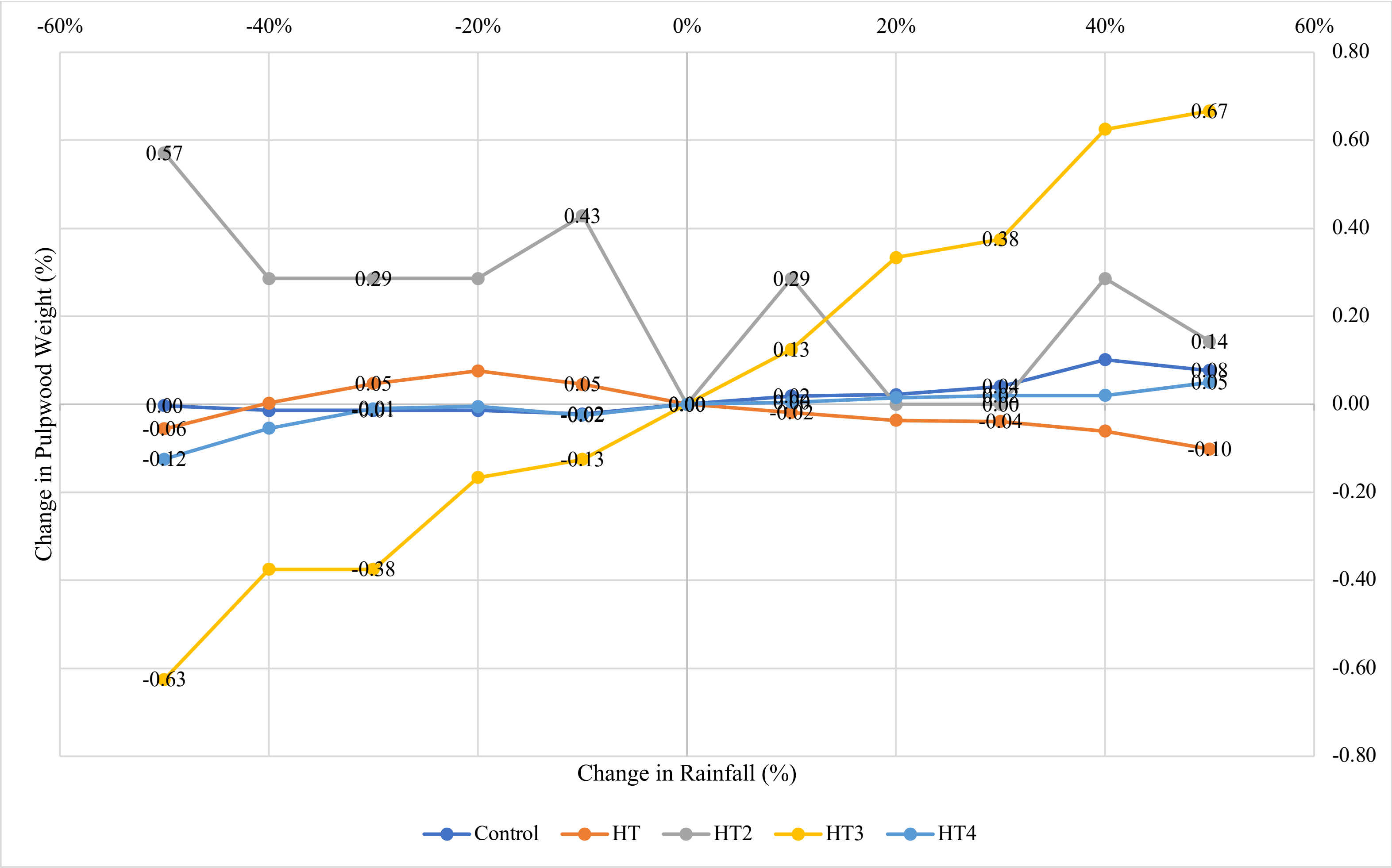

Figure 11: Change in pulpwood weight (%) with the change in rainfall (%) in various treatments in the study area.

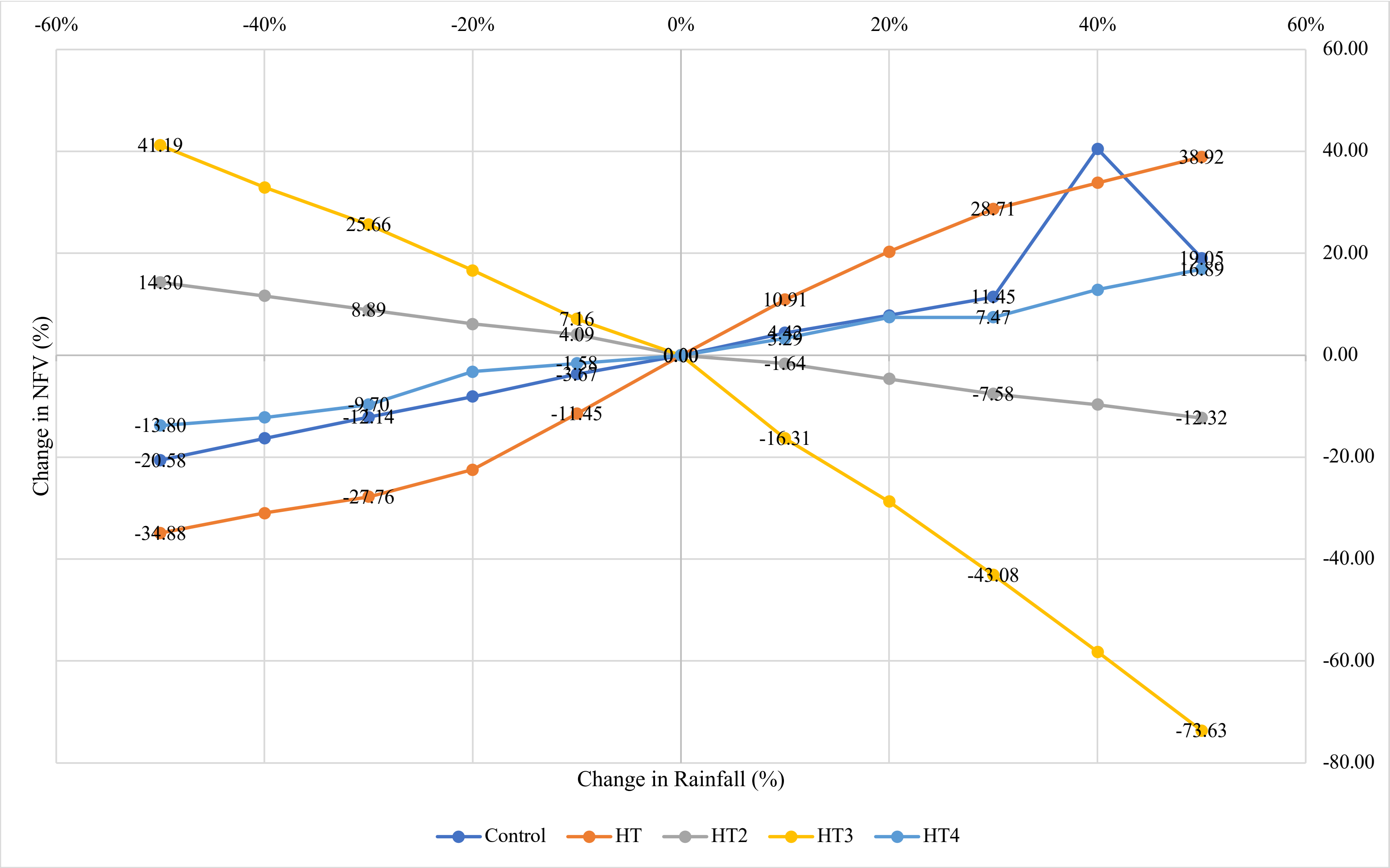

Figure 12: Change in net future value (NFV) (%) with the change in rainfall (%) from various stands in the study area.